# A simple stacked ensemble machine learning model to predict naturalized catchment hydrology and allocation status

Michael J. Friedel[1,2,*], Dave Stewart[3,4] , Xiao Feng Lu[4], Pete Stevenson[4], Helen Manly[4], Tom Dyer[4]

[1] University of Colorado, Denver, Colorado, United States (michael.friedel@ucdenver.edu)
[2] Earthquest Consulting Ltd, Auckland, New Zealand 1025 (mike@earthquestconsulting.com)
[3] RainEffects Ltd, Dunedin, New Zealand
[4] Otago Regional Council, Dunedin, New Zealand

ABSTRACT

New Zealand legislation requires that Regional Councils set limits for water resource usage to manage the effects of abstractions in over-allocated catchments. We propose a simple stacked ensemble machine learning model to predict the probable naturalized hydrology and allocation status across 317 anthropogenically stressed gauged catchments and across 18,612 ungauged river reaches in Otago. The training and testing of ensemble machine learning models provides unbiased results characterized as very good (R2 > 0.8) to extremely good (R2 > 0.9) when predicting naturalized mean annual low flow and Mean flow. Statistical 5-fold stacking identifies varying levels of risk for managing water-resource sustainability in over-allocated catchments; for example, at the respective 5th, 25th, 50th, 75th, and 95th percentiles the number of overallocated catchments are 73, 57, 44, 23, and 22. The proposed model can be applied to inform sustainable stream management in other regional catchments across New Zealand and worldwide.
*Keywords:* Machine Learning, Stacking model, Naturalized hydrology, Mean and MALF, Allocation status.

## 1 Introduction

The current demand for freshwater resources is threatening sustainable management and security of regional catchments worldwide (McManamay et al., 2022). Focus on stream water allocation (process of distributing in-stream water for various sector needs) and environmental flows (ideal state of river flow regimes required to promote the sustainability of aquatic ecosystems; Booker et al., 2022) in regional catchments is of increasing interest among the international community (Jain and Kumar, 2014; Hoekstra, 2014; McManamay, 2014; Richter, 2013; Tharme, 2003). In New Zealand, the National Policy Statement for Freshwater Management (NPS-FM; Ministry for the Environment, 2020) gives direction to water resource reforms that include development of regional water management plans with freshwater objectives involving out-of-stream water allocation and in-stream environmental biodiversity outcomes. According to the NPS-FM, these freshwater objectives need to describe desired water-resource outcomes that will be achieved at the sub-regional scale, called a freshwater management unit (FMU).

Important NPS-FM freshwater objectives are to limit the streamflow below which all abstractions must cease (minimum flow) and to limit the cumulative number of upstream abstractions above which the permitting of consented abstractions must cease (total allocation rate). Defining these limits on a catchment basis is considered important for quantifying the amount of freshwater resource that is available to out-of-stream users. In principle, comparing the difference between these limits provides a means for characterizing the catchment status as under-allocated or over-allocated. Knowledge of the catchment status is particularly important because the NPS-FM directs regional councils to reduce the allocation of water in over-allocated catchments. The NPS-FM further encourages councils to include desired water-resource outcomes in their regional plans, such as the use of predefined rules (allocation limits) for minimizing the potential cumulative effects of catchment abstraction on in-stream biodiversity through delivery of environmental flows while providing water for out-of-stream use (NPS-FM, 2020). In this way, the regional plans can better safeguard the water availability for public, industrial, and agricultural uses while ensuring a standard level of protection for cultural, social, and environmental values (Ministry for the Environment, 2015).

The NPS-FM provides impetus for councils to develop regional plans that manage the potential effects of in-stream abstractions as a freshwater objective, but there are challenges in in defining freshwater resource use limits associated with environmental flows and therefore catchment allocation status (Booker et al., 2018). In principle, there is a tradeoff in defining the freshwater resource use limits and, at the time of this study, there are no published guidelines describing how these limits should be set. In 2021, Hayes et al. presented evidence to the Environment Court on guidelines to help inform the Otago Regional Water Plan. These guidelines describe a method for determining the default allocation rate and the default minimum flow as a percentage of the naturalized 7-day Mean Annual Low Flow (MALF) based on knowledge of the naturalized mean daily flow (Mean). In doing so, these two freshwater limits can be expressed in units of flow at any location where the naturalized Mean and naturalized MALF has been determined. Unfortunately, many of the observed flows originating upstream of gauging stations reflect a combination of natural and human activities. For this reason, the natural flows cannot always be directly measured and therefore must be determined using a naturalization method.

Streamflow naturalization methods typically involve the use of models (Terrier, et al. 2021). Most published models use the water balance approach (Fantin-Cruz, et al., 2015; Jiongxin, 205; Yuan et al., 2017). Other reconstitution methods use spatially explicit process-based hydrology models that are data and computationally intensive but can predict streamflow at a daily time step (Barbarossa et al. 2017). In principle, these spatially explicit hydrological models can be developed and calibrated for regulated catchments and used to predict naturalized environmental streamflow following removal of the anthropogenic components (Gosain et al., 2005; Kim et al., 2012; Yin et al., 2017). In practice, these efforts are often challenged by uncertainty due to simplified process representation in the model structure, input data characterized by limited spatiotemporal measurements, and nonunique parameter estimates resulting from the calibration procedure (Ehlers et al., 2018; Gupta and Govindaraju, 2019; Jin et al., 2010; Moges et al., 2020; Setegn et al., 2009). As an alternative, recent applications include additional calibration constraints based on the regionalization of multiple hydrological models in data scarce and ungauged catchments (Garna et al., 2023; Golian et al., 2021; Mahapatra and Jha, 2022).

Regression based empirical models provide a practical alternative to the time-consuming, computationally intensive, and uncertain spatially explicit process-based hydrology models. In general, these models relate streamflow indices to explanatory catchment characteristics promoting scale-dependent understanding among hydrological processes and patterns in regional catchments (Farmer et al., 2015). These empirical models can be parametric with predictors based on equations (Barbarossa et al., 2017) or nonparametric with predictors based on information derived from data (Okkan and Serves, 2012; Wu et al., 2009). Despite the number and type of empirical approaches available, few studies compute naturalized environmental flow indices. In one related study by Booker and Woods (2014), the nonparametric Random Forest regression (ensemble machine learning) method was determined to outperform the process-based hydrological model when estimating environmental flow indices across ungauged catchments in New Zealand.

Despite the various approaches available for estimating naturalized streamflow, there are no studies that quantify the naturalized default limits and subsequent naturalized allocation status across regional catchments. Possible reasons may be attributed to the challenges in computing naturalized hydrological indices across regulated catchments, the tradeoff in computing catchment limits, and how to express management risk in terms catchment allocation with model uncertainty. The *aim* of this study is to develop and apply a novel regression-based workflow that informs the sustainable management of natural flow in regional catchments of Strahler order streams (Strahler 1964) from 1 to 7 across Otago, New Zealand. We hypothesize that the combination of natural hydrology and available physical catchment characteristics can provide mutual information suitable for machine learning-based model building and classifying the allocation status of Otago regional catchments. The objective is to use a simple stacked ensemble machine learning model to predict the probable naturalized hydrology and probable allocation status across 317 anthropogenically stressed gauged catchments and probable naturalized Mean and MALF across 18,612 ungauged river reaches in Otago New Zealand. This study extends the work of Booker et al (2014, 2018) whose studies on New Zealand stream catchments of Strahler stream order > 3 included using a random forest regressor to estimate deterministic indices of natural hydrology and using a weighting scheme to quantify the hydrological effect of permitted water abstractions.

## 2 Data and methods

The methodology used in this study involves the four-step workflow shown in Fig. 1. The four steps include the base model, simple stacking, limit setting, and allocation status briefly described next.

### *2.1 Base Model*

The base modelling task is used to predict naturalized hydrologic indices in the Otago Region. The 32,000 $km^2$ Otago Region includes anthropogenically stressed gauged catchments and ungauged river reaches**.** These catchments and river reaches span five FMUs: Catlins, Clutha (Mata-Au), North Otago, Taieri, Dunedin & Coast (Fig. 2). The Clutha (Mata-Au). These FMUs are further subdivided into five smaller water-management units called Rohe reflecting the specialized water-interests of different iwi tribes: Dunstan, Lower Clutha, Manuherekia, and Upper Lakes.

#### *2.1.1 Sources of data*

According to Rallo et al. (2002), one of the elements necessary for accurate ensemble machine learning predictions is base model diversity. Model diversity reflects the incorporation of training information (response and predictor variables also called target and features) characterizing mutually informative relations across different spatial and temporal sampling gradients. The types of regression data sourced for this study include information in natural hydrological indices (response variables) and catchment characteristics (predictor variables) associated with $1^{st}$ to $7^{th}$ order streams and catchment areas that range from 0.3 $km^2$ to 6,000 $km^2$ (Fig. 3.).

The natural hydrological indices, namely mean daily flow (Mean) and 7-day mean annual low flow (MALF), are computed from available daily streamflow time-series collated using the Hilltop software (2023, Hilltop) and Otago Regional Council (ORC) hydrology database (Table 1). From this database, a set of daily streamflow time-series are collected from gauging stations representing a range of hydrological conditions (natural to anthropogenically stressed) across the Otago region. Of these sites, only those sites with at least five years of continuous (> 11 months per year) daily flow records are identified for possible use. Additional filtering of time-series records is undertaken to remove those gauge stations affected by upstream engineering projects such as dams, diversions, or substantial abstractions. Lastly, selecting sites where the total consented upstream abstraction is less than 30% of the estimated median daily flow results in identifying 100 flow sites (Fig. 4) that approximate natural streamflow conditions for use in model building. In using empirically based regression methods, the differences among sites in hydrological regimes is assumed to exceed any differences in hydrological regimes due to differences in observation periods, which are different for each observed time-series. The reader is referred to Booker and Woods (2014) for more details on gauging station selection.

According to Booker and Snelder (2012), there are eight catchment characteristics (features) considered suitable for explaining variation in hydrological patterns across New Zealand (Table 2). The eight catchment characteristics include area, elevation, particle size, potential evapotranspiration (PET), rainfall variation, rain days, runoff volume. These physical characteristics represent median values obtained from the Freshwater Environments of New Zealand geo-database (Leathwick et al. 2011) sorted on reach numbers found in the River Environment Classification (Snelder and Biggs 2002). The catchment characteristics used in this study represent physical properties located upstream from gauged catchments and ungauged river reaches of mixed environmental conditions. For example, catchment characteristics acquired from the locations of 100 natural stream flow sites are presented in Fig. 4; whereas the location of 317 regulated (named) gauged streamflow sites are presented in Fig. 5; and the location of the 18612 ungauged river reaches (unnamed) are presented in Fig. 6.

*2.1.2 Ensemble machine learning predictions*

The base model predictions rely on four ensemble machine learning methods (Pedregosa et al., 2011), namely Random Forest Regressor (RFR), Gradient Boosting Regressor (GBR), Extreme Gradient Boosting Regressor (XGB), and Quantile Gradient Boosting Regressor (QGBR), to predict the Mean flow and the MALF. To do so, these ensemble methods learn relationships among response (hydrological indices) and predictor variables (catchment characteristics) without relying on statistical assumptions about the data (Dietterich 2000). The architecture of these methods is based on a regression tree that has a piece-wise constant surface where there is a region $R_m$ in input space *I* for each terminal node (that is, the hyper-rectangles induced by tree cuts). The constant associated with each region represents the estimated prediction $\hat{y}=c\hat{m}$ that the tree is making at each terminal node. Formally, the M−terminal node tree model is expressed as

$$\hat{y} = T(x) = \sum_{m=1}^{M} \quad \hat{c}_m I_{R_m}(x) \tag{1}$$

where $I_{R_m} = 1$, if $X \in R_m$ and 0 otherwise. The difference among ensemble methods is essentially related to how they minimize the prediction error (bias plus variance). The four ensemble machine learning methods are briefly described next.

*2.1.2.1 Random Forest Regressor*

The RFR method (Breiman, 2001) uses the bagging (bootstrap aggregating) procedure (Breiman, 2001) plus a perturbation procedure (subset splitting) to combine a set of base learners (tree models). The idea is to average many noisy but approximately unbiased trained tree models, and hence reduce the prediction variance. This variance reduction is achieved in the tree-growing process through random selection of the input variables. The collection of different classifiers overfit the data in different ways and through voting (e.g., using the mean, median, mode, or other statistical criteria) those differences are smoothed out (Breiman, 2001). The extension of bagging to RFR follows two steps (Hastie et al., 2009):

1. For b=1 to B: (a) Draw a bootstrap sample $Z^*$ of size N from the training data. (b) Grow a random-forest tree $T_b$ to the bootstrapped data, by recursively repeating the following steps for each terminal node of the tree, until the minimum node size $n_{min}$ is reached. i. Select *m* variables at random from the p variables. ii. Pick the best variable/split-point among the *m*. iii. Split the node into two daughter nodes.

2. Output the ensemble of trees $\{T_b\}^B$. To make a prediction at a new point x:

$\hat{f}_{rf}^{B}(x) = \frac{1}{B}\sum_{b=1}^{B} T_b\,(x)$ (2)

*2.1.2.2 Gradient Boosting Regressor*

The GBR method minimizes the predictive error using the boosting procedure (De'ath, 2007) to combine a set of weak learners (high bias, low variance) that generates a collectively strong model. In terms of decision trees, the weak learners are shallow trees, sometimes even as small as decision stumps (trees with two leaves). Boosting reduces error mainly by reducing bias by aggregating the output from many models as an additive model (Hastie et al., 2009) defined as:

$f(x) = \sum_{m=1}^{M} \beta_m\, b(x; \lambda_m)$ (3)

where $\beta_m$, m=1,2…M are the expansion coefficients corresponding to the M trees, x is the set of predictor variables, $\lambda_m$ parameterizes splitting variables at internal nodes and predictions at terminal nodes, and $b(x; \lambda_m)$ is a basis function that represents a single tree. This procedure uses stagewise gradient boosting to estimate $\beta_m$ and $\lambda_m$ sequentially from m=1 to M with each new tree fitted to the residuals of the previous tree. After an initial tree is trained, subsequent trees are fitted to the residuals of the previous tree rather than to the data directly. A stochastic gradient boosting and steepest-descent minimization is applied to estimate $\beta_m$to minimize the loss given by

$L(y, f(\boldsymbol{x})) = (y - f(\boldsymbol{x}))^2$ (4)

where y is the observed value of the response variable. In summary, the RFR uses bagging to minimize the variance and overfitting, while GBR uses boosting to minimize the bias and underfitting.

*2.1.2.3 Extreme Gradient Boosting Regressor*

The XGB regressor (Chen and Guestrin, 2016) is a specific implementation of GBR with regularization techniques that may improve on other decision tree methods, such as RFR and GBR. One difference is that this implementation follows a level-wise strategy, scanning across gradient values and using these partial sums to evaluate the quality of splits at every possible split in the training set. this method uses a sparsity-aware algorithm for sparse data and weighted quantile sketch for approximate tree learning summarized as follows (after Chen and Guestrin, 2016):

1. Given the input training set $\{x_i, y_i\}_{i=1}^{N}$, a differentiable loss function L($y$,F(x)), a number of weak learners M, and a learning rate $\alpha$.

2. Initialize model with a constant value: $\hat{f}_{(0)}(x) = arg_\theta\ min \sum_{i=1}^{N} L(y_i \theta)$

3. Compute gradient and hessian operators

   $\hat{g}_m(x_i) = \left[\frac{\partial L(y_i, f(x_i))}{\partial f(x_i)}\right]_{f(x)=\hat{f}_{(M-1)}(x)}$

   $\hat{h}_m(x_i) = \left[\frac{\partial^2 L(y_i, f(x_i))}{\partial f(x_i)^2}\right]_{f(x)=\hat{f}_{(M-1)}(x)}$

4. Fit a weak learner (e.g., tree model) using the training set $\left\{x_i - \frac{\hat{g}_m(x_i)}{\hat{h}_m(x_i)}\right\}_{i=1}^{N}$ by solving the optimization problem:

$$\hat{\phi}_m = arg_{\phi \epsilon \Phi} min \sum_{i=1}^{N} \frac{1}{2} \hat{h}_m(x_i) \left[ -\frac{\hat{g}_m(x_i)}{\hat{h}_m(x_i)} - \phi(x_i) \right]^2$$
$$\hat{f}_m\,(x) = \alpha \hat{\phi}_m(x).$$

5. Update the model: $\hat{f}_m\,(x) = \hat{f}_{m-1}(x) + \hat{f}_m\,(x)$.

6. Output: $\hat{f}_m\,(x) = \hat{f}_M\,(x) = \sum_{m=0}^{M} \hat{f}_m\,(x)$.

*2.1.2.4 Quantile Gradient Boosting Regressor*

The statistical concept of heteroscedasticity is said to be operating when the standard deviations of a predicted (target) response variable is non-constant with increasing values of the independent predictor variable(s). In this case, quantifying the prediction intervals (uncertainty) can be determined by computing the conditional percentiles of prediction models by application of quantile regression to the response variable (Koenker 2005). The QGBR algorithm used in this study performs gradient descent in functional space to minimize the objective function used by quantile regression. The following is a summary of the QGBR algorithm (after Zheng 2012):

1. Given the training data $\{x_i, Y_i\}$ , I=1, … n} the desired quantile value τ, and the total number of iterations M, initialize $f^0(\bullet) = 0$ or set $f^0(\bullet) =$ τ-th quantile of ($Y_i$, …, $Y_n$).
2. For m = 1 to M do:
3. Compute the negative gradient $-\frac{\partial}{\partial f}\rho_\tau(Y - f)$ and evaluate at $f^{(m-1)}(x_i)$: $U_i = I(Y_i - f^{(m-1)}(x_{i)} \geq 0) - (1 - \tau), i = 1, \ldots, n.$
4. Update the estimation by $f^{(m)}(\bullet) = f^{(m-1)}(\bullet) + \eta_m g$, where η is a step size factor.
5. End for
6. Output the τ-th quantile function $f^{(m)}(x)$.

*2.1.3 Training, testing, prediction*

Important base model tasks involve (standard practice) training and testing of the ensemble machine learning models. Several decisions are required during the model *training* phase of the study (see Fig. 1). First, a file with the naturalized catchment records is assigned. Second, a decision is made to assign either the natural Mean or the natural MALF as the response variable. Third, the number and type of catchment characteristics are assigned as independent predictor variables, e.g., area, elevation, potential evapotranspiration, particle size, rain days, rainfall, runoff, and slope. Fourth, an arbitrary random seed (also referred to as the random state number) is assigned to initialize the random number generator for shuffling of the catchment records. Fifth, a decision is made on the relative proportion of records assigned to the training and testing phases. Sixth, a decision is made to use default (or base) ensemble model parameters or invoke a hyperparameter tuning method to optimize the model parameter values Pedregosa et al., 2011.

The ensemble machine learning *testing* phase is undertaken by presenting the independent split fraction of the complete data set to the trained models. This phase is important for assessing the ability of models to generalize when presented independent catchment records. Testing scores are used to assess the R-Squared coefficient of determination (Lewis-Beck and Lewis-Beck, 2015) for each regression model as follows: 60-70% poor, 70-80% good, 80-90% very good, >90% excellent. Scatterplots (sometimes called calibration plots) of predicted values along the y-axis to observed values on the x-axis are inspected to visually identify prediction bias, where values with a 1:1 correspondence reveals an ideal unbiased model. Feature importance scores are reviewed to evaluate the relative influence that a feature may have on the model prediction process. Caution is exercised in the interpretation of these plots because highly correlated features result in splitting their importance giving the false impression that they have less importance. Lastly, deviance plots are inspected to ensure the model is not overfitting the set of training records, and summary statistics are provided to compare the original, training and testing processes. Once the training and testing phases are satisfactorily completed, the natural hydrological indices can be predicted for any catchment or stream segment by presenting the associated independent catchment characteristics to the ensemble machine learning models.

*2.2 Stacking*

Stacking generalization is the method of using a high-level (meta) model to combine lower-level (base) ensemble models with the aim of achieving greater predictive accuracy (Wolpert, 1992). For example, a meta model that is trained on k-fold predictions of lower-level ensemble models is then presented with independent data to make better informed predictions. However, there are sometimes stacked generalization issues in achieving improvements in performance using a meta model (Ting and Witten, 2011). For example, improvements using stacked generalization depends on the complexity of the problem and whether it is sufficiently well represented by the training data and complex enough that there is more to learn by combining predictions. There also is a dependency upon the choice of base models and whether they are sufficiently skillful and sufficiently uncorrelated in their predictions (or errors). For these reasons, this study embraces simple statistical stacking in which the results of multiple random subsamples of the field observations are presented to the ensemble models to improve accuracy and quantify and reduce the prediction interval and related uncertainty. Advantages in using this approach are to: (1) help prevent overfitting by providing a more robust estimate of the model's performance on unseen data, (2) compare different models and select those that perform best on average, (3) use of all the available data for both training and testing. Disadvantages in using this approach include the increase in computational time for training when considering multiple folds (randomly shuffled split sets), time consuming (cross-validation when multiple models need to be compared), and bias variance tradeoff (choice of the number of randomly shuffled split sets: too few folds may result in high variance, while too many folds may result in high bias).

Statistical stacking of ensemble model results are used to quantify the prediction uncertainty at predefined percentiles, e.g., 5$^{th}$, 10$^{th}$, 25$^{th}$, 50$^{th}$, average, 75$^{th}$, 90$^{th}$, and 95$^{th}$. The uncertainty in model predictions will vary and is likely due to (at least) three sources. *First*, catchments are assumed to be in a natural state when calculating the hydrological indices used in training the ensemble machine learning models. *Second*, there is a limited number of randomly selected catchment records used for k-fold training of the ensemble machine learning models. The limited number of catchment records available for training and testing underscore the challenges in identifying gauge stations reflective of natural conditions. *Third*, the upstream catchment characteristics are assumed to be optimal in type and number. In fact, modern feature selection techniques involving learn heuristics, such as the filter (Buscema et al., 2013; Friedel et al., 2020) or wrapper (Calvet et al., 2017) methods, may identify a smaller and/or different set of optimal catchment characteristics that satisfy the same hydrological index (response variable).

*2.3 Limit setting*

The third step in the stacked ensemble workflow involves application of the default limit guidelines proposed and adopted in the Environment Court of New Zealand for use by the Otago regional council. According to Hayes et al. (2021), the proposed limits (Table 3) serve two primary functions. First, these limits set the default allocation rates and the default minimum flows to avoid more than minor in-stream ecological effects. Second, these default limits define a threshold for more than minor instream effects. In the event the allocation rate is exceeded and/or the minimum flow is less than proposed, the ecological effects are likely to exert pressures that are considered more than minor. The possibility exists for the proposed instream values and NPS-FM objectives to be adjusted with alternative allocation rates and alternative minimum flows, but the assessment of ecological effects supporting these outcomes require the collection and incorporation of additional information (e.g., hydraulic-habitat modelling and/or invertebrate drift versus flow relationship) to properly assess the ecological effects supporting that outcome (Beca, 2008).

Minimum flow and allocation limits set as proportions of historical flow statistics, such as the default limits proposed by Hayes et al. (2021), assume spatially consistent reductions in habitat and/or ecological responses with flow reduction. However, the flow related ecological flow and habitat relationships often respond nonlinearly to spatiotemporal flows resulting in default minimum flows and default allocation limits that may result in different ecological and habitat protection levels for different size rivers and aquatic species (Snelder et al. 2011; Booker et al. 2014). The application of the so-called Hayes guidelines are simpler to apply than the methods of assessing environmental flows and habitat setting limits, and some guidance already exists on percentage flow alteration limits likely to pose low risk of adverse ecological effects (Richter et al., 2012). According to Hayes et al., (2021), the default limits for perennial rivers will also provide precautionary limits for permanently flowing segments of intermittent rivers, and the proposed method to calculate the limits for such reaches, based on percentage of MALF, is practical and environmentally conservative while allowing for modest levels of stream abstractions (also called takes). Lastly, the limits as proposed give effect to the NPS-FM directive of Te Mana o te Wai, whose translation means to put the health and wellbeing of waterbodies above other needs.

*2.4 Allocation status*

To determine the naturalized allocation status for the 317 gauged Otago catchments, the current allocation rate (i.e., the sum of consented catchment takes upstream from the gauge station; Fig. 7) is subtracted from the computed default allocation rate giving the default allocation rate available. If the default allocation rate available is positive, then the catchment status is deemed to be under-allocated with additional water available for future consents. Conversely, if the default allocation rate is negative, then the catchment status is deemed to be over-allocated with a net deficit of catchment water available for future consenting. This process is then repeated with default allocation rates computed at predetermined percentiles (e.g., 5$^{th}$ 25$^{th}$, 50$^{th}$, 75$^{th}$ 95$^{th}$ percentiles) and expected value providing results that collectively describe a cumulative distribution function for every regional catchment being evaluated. This approach provides added information over deterministic solutions resulting in flexibility in selecting the level of risk to manage based on the probable number of over-allocated catchments. Once chosen, the level of risk dictates the number and location of over-allocated catchments; for example, at the 5$^{th}$ (over conservative), 25$^{th}$, average (most likely or expected), 50$^{th}$ (median), 75$^{th}$, and 95$^{th}$ (over conservative). Maps of the spatial distribution of over-allocated catchments can then be developed for each percentile.

## 3 Results and discussion

The following sections briefly describe results of the base model, simple stacking, limit setting, and allocation status steps to inform the Otago regional land and water plan for sustainability of streams.

*3.1 Base model*

*3.1.1 Training and testing*

In this section, results are provided for model training and testing phases using catchment records acquired at natural streamflow sites (N=100) across Otago (see Fig. 4). The set of catchment records comprises hydrologic indices (Mean or MALF) each referred to as a target response and eight predictive features referred to as catchment characteristics (area, elevation, potential evapotranspiration, particle size, rain days, rainfall, runoff, and slope). These catchment records are randomly shuffled and split multiple times during the testing and training phase. In this study, the ratio used in shuffling and splitting records is 80% (N=80) for training and 20% (N=20) for testing. This ratio is a matter of choice, where alternative ratios also could be adopted, e.g., such as 50% training and 50% testing, 90% training and 10% testing, as part of the testing phase. A statistical summary of (dependent) hydrologic indices and (independent) catchment characteristics aggregated from natural streamflow sites is presented in Table 4.

The process used in selecting a subset of the catchment records is controlled by assigning a random seed number (called the random state) that initiates the record shuffling prior to splitting. This process is repeated to produce five different subsets (one per random state) of target hydrologic indices and feature catchment characteristics that are each presented to the suite of ensemble models. In this way, the shuffling process provides a means to evaluate the effect of different catchment characteristic subsets on the prediction bias and uncertainty of the ensemble models despite limited records. One side benefit in using this procedure is that each random number seed produces a single reproducible (deterministic) outcome that can be repeated using the same python script for review and/or use in other related analyses at any time. In this study, five different randomly shuffled sets are used to train each of the ensemble models along with variants in these models reflecting the application with and without hyperparameter tuning available as part of the scikit-learn toolbox (Pedregosa et al., 2011). Hyperparameter tuning includes random grid search and random grid search plus cross-validation methods available from this toolbox. In total there are 21 possible ensemble models that are evaluated as part of the training and testing phase. Given that 5 random states are applied to each model leaves the possibility of 105 model predictions for each hydrologic index at each location of interest.

A summary of results is presented for testing MALF and Mean flow predictions when using two different shuffled split sets, e.g., random state (seed) numbers 2 and 4 (Tables 5-8). This testing phase consists of presenting each independent (20%) spilt set following the random shuffling procedure to models trained with and without hyperparameter tuning. Examples are presented for predictions using models *without* hyperparameter turning, e.g., Random Forest Regressor (RFR), Gradient Boosting Regressor (GBR), Extreme Gradient Boosting (XGB), and Quantile Gradient Boosting Regressor at the $10^{th}$ (QGBR10), $50^{th}$ (QGBR50), and $90^{th}$ (QGBR90) percentiles, and predictions using models *with* hyperparameter tuning, e.g., Random Forest Regressor with random grid search (RFRgs), Random Forest Regressor with random grid search and cross-validation (RFRgscv), Gradient Boosting Regressor with random grid search and cross-validation (GBRgscv), and Extreme Gradient Boosting Regressor with random grid search and cross-validation (XGBgscv), and Quantile Gradient Boosting Regressor with grid search at $10^{th}$ (QGBR10gs), $20^{th}$ (QGBR20gs), $30^{th}$ (QGBR30gs), $40^{th}$ (QGBR40gs), $50^{th}$ (QGBR50gs), $60^{th}$ (QGBR60gs), $70^{th}$ (QGBR70gs), $80^{th}$ (QGBR80gs), and $90^{th}$ (QGBR90gs) percentiles.

In general, the different naturalized ensemble prediction results for untuned and tuned models generalize well as indicated by R-Squared values, where $0.8 > R^2 < 0.9$ is considered very good and $R^2 > 0.9$ is considered extremely good (Tables 5 and 6). These findings suggest that the proportion of variance in the dependent variable (hydrologic index) is predictable when using the independent variables (catchment characteristics). Those models that do not generalize well result in poor predictions for Mean flow prediction when using the GBRgscv model and negative prediction values (indicating the prediction is worse than using the mean value) for Mean flow and MALF when using the QGBR10 model. Similar test findings are observed when using other random state values (not shown here) to predict naturalized Mean and MALF. In most cases, the untuned and tuned parameter sets result in essentially the same (or very similar) testing results (see Tables 7-8). The predictions for models deemed as very good to extremely good are retained for use in forming a suite of model predictions for simple stacking.

Statistical results are presented for comparing the natural 7-day mean annual low flow (MALF) observations (known) to natural MALF predictions (testing phase) with and without hyperparameter tuning (Table 7). Similarly, statistical results are presented for the same models providing a comparison of the natural mean daily flow (Mean) observations (known) to natural daily Mean flow predictions (testing phase) with and without hyperparameter tuning (Table 8). The randomly shuffled spilt sets used in training and testing of the ensemble models shown here also reflects the assignment of random state numbers 2 and 4 to the *untuned models* (N=6)*:* RFR, GBR, XGB, QGBR10, QGBR50, QGBR90; and *tuned* models (N=15): RFRgs, RFRgscv, GBRgscv, XGBgscv, QGBR10gs, QGBR20gs, QGBR30gs, QGBR40gs, QGBR50gs, QGBR60gs, QGBR70gs, QGBR80gs, and QGBR90gs, where gs = grid search, rgcv = random grid search with cross-validation. Inspecting tables 7 and 8 provides a quantitative comparison of the ensemble model ability to predict natural MALF or natural Mean using a limited number of testing records (N=20).

In general, these model predictions approximate the statistical attributes associated with the observations despite the limited number of records and random state chosen. As anticipated, differences exist among statistical attributes ascribing their model predictability as weak learners when using limited number of records. For example, random states 2 and 4 produce randomly shuffled splits sets with different respective minimum and maximum values, e.g., MALF: 0.001 m3/s and 20.8 m3/s, and 0.002 m3/s and 16.8 m3/s; Mean flow: 0.02 and 82.0, and 0.02 and 82.0 (or 0.02 and 39.2 for random state 10 not shown here). Despite their reasonable model performance using limited number of records, these tables provide a means to identify underperforming models deemed inappropriate for use in the prediction of hydrologic indices. For example, QGBR10 is consistently underperforming when modeling natural MALF and natural Mean flow regardless of hyperparameter tuning as determined when comparing the observation to prediction statistics and during testing validation phase with coefficient of determination values that are typically negative but always below $R^2 < 0.4$ (Tables 5 and 6). Other underperformers include QGBRgs models at the 10$^{th}$, 20$^{th}$, 30$^{th}$, and 40$^{th}$ percentiles that are omitted from future consideration as predictive models. Based on the results in this section, a total of 16 ensemble models (e.g., 1-3, 5-6, 7-12, 17-21) are deemed worthy for use during the prediction phase. Lastly, the results in these tables reveal that observed and modeled mean values of the MALF and Mean annual low flow are significantly larger than the median (50%) values revealing the sample bias toward lower values and distribution that is not Gaussian. This finding supports the use of randomly shuffled split sets with alterative tunning criteria to increase the sampling distribution in line with the Central Limit Theorem. In doing so, the standard error of the prediction is expected to be reduced but there remains the bias-variance tradeoff issue to be resolved. Resolving this issue requires determining an optimal number of randomly shuffled split sets for use in the prediction phase: too few folds may result in high variance, while too many folds may result in high bias.

*3.1.2 Predictions*

The process of predicting naturalized hydrologic indices is undertaken across the Otago region. This process requires presenting independent catchment characteristics (i.e., those not used in the training and testing process) for the 317 gauged catchments, and catchment characteristics for the 18612 ungauged river reaches, to the 16 trained ensemble models. Statistical summaries of these two sets of independent catchment characteristics are presented in Tables 9 and 10. The reader can download these catchment characteristics as part of the complete New Zealand Freshwater Ecosystems geo-database by requesting access from the Department of Conservation (Department of Conservation, 2023). Differences in the statistical summaries presented in these tables are attributed to spatial sampling bias of the 317 gauged catchments draining into the Clutha, Taieri, Manuherekia Rivers and the Pacific Ocean; and spatial sampling bias of the 18612 ungauged river reaches randomly located across the entire Otago Region. Once the desired hydrologic index (target), e.g., Mean or MALF, is assigned then the relevant set of independent catchment characteristics (features) is presented to each ensemble model for simultaneous prediction of the chosen hydrologic index across the domain of interest, e.g., gauged catchments or ungauged river reaches.

*3.2 Simple stacking*

The result of simple stacking is presented for the 317 gauged catchments and 18,612 ungauged river reaches. Simple stacking involves computing 5-fold cross-validation statistics from the 80 model predictions (i.e., 16 models trained using 5 randomly shuffled subsets per model). These 5-fold cross-validation statistics are computed for each gauged site and each ungauged location across Otago. In this way, predictions at predefined percentiles (e.g., 5$^{th}$, 10$^{th}$, 25$^{th}$, 50$^{th}$, 75$^{th}$, 90th, 95$^{th}$ percentiles) and average value (expected or most likely) describe a discrete cumulative distribution function at each location. In this case, the largest prediction values are associated with the 95$^{th}$ percentile and are equal to or less than other predicted values at lessor percentiles, and smallest prediction values are associated with the 5th percentile that will be less than values at all other percentiles. In the interpretation of these prediction results, the average value is considered the most likely (expected) value when there are no outliers that skew the distribution. In cases where there are no outliers, the median and average value will be the same (or very similar). In cases where these values are skewed by outliers then the median value (representing 50% of the predicted values above and 50% of the predicted values below) is considered a more robust measure of the central tendency. For these reasons, both measures are presented for review and consideration when stacking the predictions as well as computing the traditional prediction intervals defined as the difference among the 5th and 95th percentiles and difference among the 25th and 75th percentiles (interquartile range).

*3.2.1 Probable naturalized hydrological indices at gauged catchments*

Given the large number of predictions at gauged catchments and ungauged river reaches, results in this section are sorted by the expected value for each hydrologic index and presented as Mean flow (Table 11) and MALF (Table 12) distributions of percentiles for the first and last 32 catchments (partial listing in Table 11 and complete listing in Appendix A). From these tables, the Makaroa River is determined to have the largest predicted natural flows with an expected value (most likely) for Mean flow of more than 63,095 l/s with the respective prediction interval and interquartile range of 52,026 l/s (from 28,168 to 80,195 l/a) and 11,790 l/s (from 58,408 to 70,199 l/s); and for MALF of more 16,900 with the respective prediction interval and interquartile range of 13,295 l/s (from 7,580 to 20,799 l/s) and 2,645 l/s (from 15,851 to 18,497 l/s). Inspection of these tables reveals that the expected values for the majority of predicted natural flow indices at the 317 gauged catchments are very similar to median values supporting the hypothesis that combining the predictions of many weak ensemble models (trained using a small number of catchment records) will (1) reduce the prediction bias at the expense of variance, and (2) reflect probability density functions that are normally distributed (expected value and median describe the same central tendency). That said, the smallest predicted natural flow indices differ from their median values for some catchments possibly related to greater uncertainty in describing the physical properties for the smallest catchments associated with (1) MALF predictions in the Dunedin & Coast, e.g., Jones Creek and Kaikorai Stream; and the North Otago Freshwater Management Unit, e.g., Aitchison Road Creek, Glen Creek, Oamaru North Creek and Welcome Creek; and (2) Mean flow predictions in the North Otago Freshwater Management Unit, e.g., Bow Alley Creek, Glen Creek, Hinahina Stream, Oamaru Airport Creek, Oamaru Creek, Oamaru North Creek, Peaks Road Creek, Waikoura Creek, and Welcome Creek. This difference in central tendencies as described using the expected and median values occurs in less than 10% of the gauged catchments but may become important should the regional council decide to provide consented abstractions in these catchments. In this case, the ensemble modelling could be refined using Learn Heuristics to define optimal catchment characteristics (number and type) for predictions with reduced uncertainty at these locations.

*3.2.2 Probable naturalized hydrological indices at ungauged stream reach segments*

Results of the stacked ensemble machine learning model predictions of naturalized hydrologic indices are presented at ungauged stream reaches across the Otago region (Fig. 6). To do so, catchment characteristics for each of the ungauged stream reach segments are presented to the trained ensemble machine learning models resulting in simultaneous predictions of Mean flows and MALF at 18612 ungauged river reach sites. Statistical stacking of the predicted hydrologic indices provided stochastic results of naturalized hydrologic indices at $5^{th}$, $25^{th}$, $50^{th}$, $75^{th}$, and $95^{th}$ percentiles for each ungauged river reach site. These predictions of hydrologic indices are summarized at each percentile in tabular format as a function of Strahler stream order (Table 13). Inspecting this Table 13 reveals a range of hydrologic indices at each quantile. For example, the range of Mean flow predictions for $7^{th}$ order streams at the $95^{th}$ percentile is 66.8 – 86.3 $m^3/s$ whereas the range of Mean flow predictions for $7^{th}$ order streams at the $5^{th}$ percentile is 42.6 – 60.7 $m^3/s$. Likewise, the range of MALF predictions for $7^{th}$ order streams at the $95^{th}$ percentile is 21.1 – 25.1 $m^3/s$, whereas the range MALF predictions for $7^{th}$ order streams at the $5^{th}$ percentile is 11.4 – 18.0 $m^3/s$. The difference between the $5^{th}$ and $95^{th}$ percentiles provide the prediction interval with insight into the magnitude of uncertainty. Extending this concept to the minimum and maximum values at a given percentile provides a spatial range of predictability for streams associated with a particular order. These relations also are presented as the series of maps showing the distribution of Mean flow (Fig. 8 a-e) and MALF (Fig. 9 a-e) across the Otago region. In these maps, the $5^{th}$ percentile portrays the driest condition (overly pessimistic) for which the likelihood is the 95% chance of being greater, whereas the $95^{th}$ percentile portrays the wettest condition (overly optimistic) for which there is the 5% chance of being wetter.

*3.2.3 Probable naturalized hydrological indices at gauged stream sites in the Taieri freshwater management unit*

The naturalized hydrologic indies are extracted from the regional prediction results at ungauged river reaches to demonstrate the broader utility of providing results for specified streamflow gauges. Currently, the Otago regional council is developing a traditional daily process-based hydrologic streamflow model of the Taieri FMU that will be calibrated using assuming current conditions. After this process, the anthropogenic constructs will be removed from the model to provide a deterministic simulation of naturalized Mean and MALF at the following streamflow gauge stations (N=11): Taieri at Outram, Taieri at Hindon, Taieri at Sutton, Taieri at Tiroiti, Taieri at Linn Burn, Kye Burn, Pig Burn, Sutton Creek, Deep Stream, Lee Stream, and Nenthorn. These deterministic process-based simulated results can then be compared to the stochastic ensemble model predictions of naturalized Mean and naturalized MALF summarized in Table 14, and in maps of Mean (Fig. 10) and MALF (Fig. 11). In comparing results, the deterministic process-based simulated values can be expected to fall somewhere between the $5^{th}$ and $95^{th}$ percentiles presented in the table.

Inspecting the table reveals that streamflow at the Taieri station at Outram is characterised as having the largest hydrologic indices and that station at Nenthorn the smallest hydrologic indices. In fact, the Mean flows at Taieri at Outram, Hindon, Sutton and Tiroiti have mean flows in the tens of thousands of l/s; Taieri at Linn Burn, Kye Burn, Pig Burn, and Sutton creek in the thousands; and Deep Stream, Lee Stream in the hundreds; and Nenthorn in the single or tenths l/s. A general rule of thumb observed here is that the MALF is roughly an order of magnitude less than the Mean flows. Exceptions to this rule are flows at Deep Stream, Lee Stream, and Nenthorn, whose Mean flows and MALF differ by a factor of about two. Of these streams, only streamflow at the Nenthorn gauge can be classified as ephemeral with a 5% chance to go dry in any given year, although other flows at 95$^{th}$ percentile indicate the likelihood for only slightly larger Mean flows and MALF of 1.56 l/s and 0.54 l/s. These results suggest the impracticality of supporting abstractions and limited ability to support aquatic ecology. Lastly, the prediction limits can be explored to understand the relative level of prediction uncertainty at each site. For example, the prediction intervals for Mean and MALF values at Outram is 2005 l/s and 2215 l/s and at Nenthorn is 1.56 l/s and 0.54 l/s.

*3.3. Limit setting*

The application of default limit setting guidelines (Hayes et al., 2021) are used to transform the predicted naturalized hydrologic indices at the 317 gauged catchments to their equivalent default minimum flows and default minimum allocation rates. The results of the default limit setting are presented for the largest 32 catchments (Table 15) and the smallest 32 catchments (Tables 16). For example, the largest default minimum flow and default allocation rate associated with the Makaroa River are more than 14,000 l/s and 5,000 l/s, respectively. These results are based on the average flows, so to compute their prediction interval and interquartile range requires that limit setting guidelines be applied across the range of percentiles for both the Mean and MALF set of predictions.

*3.4 Allocation status*

Results of the allocation status are presented for the 317 gauged catchments across the Otago region. To arrive at the default allocation status, the current allocation rate is subtracted from the default allocation rate resulting in the default allocation rate available. If this value is positive then the catchment status is deemed under-allocated with additional water available for future consenting, whereas those catchment status values that are negative are deemed over-allocated with a net deficit of catchment water available. Sustainability strategies required to manage catchment overallocation by the council policy team to assume some level of risk associated with the probability for this condition (Table 17). For example, at the 5$^{th}$ percentile there is a 5% chance that that the number of overallocated catchments will be 73 or greater and a 95% chance that the number of overallocated catchments will be 73 or less; at the 25$^{th}$ percentile there is a 25% chance that that the number of overallocated catchments will be 57 or greater and a 75% chance that the number of overallocated catchments will be 57 or less; at the 50$^{th}$ percentile there is a 50% chance that that the number of overallocated catchments will be 44 or greater and a 50% chance that the number of overallocated catchments will be 44 or less; at the 75$^{th}$ percentile there is a 75% chance that that the number of overallocated catchments will be 23 or greater and a 25% chance that the number of overallocated catchments will be 23 or less; at the 95$^{th}$ percentile there is a 95% chance that that the number of overallocated catchments will be 22 or greater and a 5% chance that the number of overallocated catchments will be 22 or less.

The allocation status is presented (in alphabetical order) as a function of percentiles, where 1 = overallocated, and 0 = underallocated (partial listing in Table 18 and complete listing in Appendix B).

Inspecting the partial list of overallocated catchments at the 95th percentile reflects those listed with a 95% chance (or less) of being overallocated (e.g., Arrow River, Bannock Burn, Basin Burn, Benger Burn, Butchers Creek (1), Cardrona River, Coal Creek (1), Coal Creek (2), Fraser River, Kakanui River, Lindis River, Low Burn (2), Manuherikia River, Pleasant River, Roaring Meg, Shingle Creek, Taieri River, Teviot River, Tinwald Burn, Waianakarua River, Water of Leith, and Welcome Creek) plus those catchments listed with a 75%, 50%, 25%, and 5% chance; the complete list of catchments at the 75th percentile includes those listed with a 75% chance (or less) of being overallocated (e.g., Luggate Creek) plus those catchments listed with a 50%, 25%, and 5% chance; the complete list of catchments at the 50th percentile includes those listed with a 50% chance (or less) of being overallocated (e.g., Albert Burn (1), Awamoa Creek, Awamoko Stream, Bendigo Creek, Bow Alley Creek, Camp Creek (1), Elbow Creek, Gentle Annie Creek, Hayes Creek, John Bull Creek, Pipeclay Gully Creek, Poison Creek, Pomahaka River, Quartz Reef Creek, Rastus Burn, Schoolhouse Creek, Scrubby Stream, Shag River, Tima Burn, Toms Creek, Waikouaiti River, Waitati River, and Waiwera River) plus those catchments listed with a 25%, and 5% chance; the complete list of catchments at the 25th percentile includes those listed with a 25% chance (or less) of being overallocated (e.g., Amisfield Burn, Burn Cottage Creek, Butchers Creek (2), Locharburn, Mt Pisa Creek, Roys Peak Creek, School Creek, Thomson Creek, Tokomairiro River, Trotters Creek, and Waitahuna River) plus those catchments listed with a 5% chance; and the complete list of catchments at the 5th percentile includes those listed with a 5% chance (or less) of being overallocated (e.g., Alpha Burn, Campbells Creek, Dead Horse Creek, Dinner Creek, Five Mile Creek (2), Franks Creek, Kaihiku Stream, Kingston Road Creek, Lake Dispute, Landon Creek, Long Gully Creek (1), Orokonui Creek, Puerua River, Seven Mile Creek, and Waikerikeri Creek). The probable location of over-allocated catchments in the Otago region are provided as a series of maps (Fig 12). For example, 73 over-allocated catchments at the 5th percentile (Fig. 12a), 57 over-allocated catchments at the 25th percentile (Fig. 12b), (46 over-allocation catchments at the expected value (Fig. 12c), 44 over-allocated catchments at the 50th percentile (Fig. 12d), 23 over-allocated catchments at the 75th percentile (Fig. 12e), and 22 over-allocated catchments at the 95th percentile (Fig. 12f).

## 4 Conclusions

Conclusions from the Otago regional catchment allocation study are as follows: (1) Training and testing of ensemble machine learning models resulted in 60 unbiased models of very good to excellent generalizability. (2) Limit setting of naturalized Mean and MALF predictions provides naturalized default minimum flows and allocation rates that when accounting for current consents resulted in quantifying the probable range of allocation status at 317 priority catchments. (3) Cross-validated flow statistics identified 46 as the most likely number (expected value) of over-allocated catchments across freshwater management units with a probable range of 22 to 77 over-allocated catchments. (4) Naturalized predictions are available for Mean flow and MALF at 18612 ungauged river reaches from which results are extracted at (or near) 11 streamflow gauge stations in the Taieri freshwater management unit. The proposed stacked ensemble modeling framework can be applied to inform sustainable stream management in other regional catchments across New Zealand and worldwide.

## Disclosure statement

No potential conflict of interest was reported by the authors.

## Data availability

Data will be made available on request.

**Funding**

This work was supported by the Otago Regional Council, Dunedin, New Zealand [grant number PO029742].

List of Appendices

Appendix A – Probable naturalized mean annual low flow (MALF) and daily mean flow (Mean) predicted at 317 gauged catchments in the Otago Region, New Zealand.

Appendix B - Naturalized allocation status predicted at 5$^{th}$, 25$^{th}$, 50$^{th}$, 75$^{th}$, 95$^{th}$ percentiles for 317 gauged catchments in the Otago Region, New Zealand. Over-allocated indicated as 1 and under-allocated indicated as 0.

Table 1. Hydrological Indices derived from observed mean daily flows.

| Index | Description | Calculation |
|---|---|---|
| Mean | Mean flow over all time | Mean of all daily flows |
| MALF | Mean of minimum 7- day flow in each year | Mean of minimum flow for each water year after having applied a running 7-day mean to the daily flows |

Table 2. Summary of physical catchment features explaining hydrologic variation across New Zealand.

| Feature | Description |
|---|---|
| Area | Log of catchment area ($m^2$) |
| Elevation | Average elevation in the upstream catchment (m) |
| Partilce size | Catchment average of particle size (mm) |
| Potential evapotranspiration (PET) | Annual potential evapotranspiration of catchment (mm) |
| Rainfall variation | Annual catchment rainfall coefficient of variation (mm) |
| Rain days | Catchment rain days greater than 10 mm/month (days/year) |
| Runoff volume | Percentage annual runoff volume from catchment area with slope > 30 degrees (%) |
| Slope | Average catchment slope (%) |

Table 3. Default limit setting guidelines expressed as a percentage of naturalized 7-day annual low flow (MALF) for maintaining flow regimes that present a low risk of more than minor effects on ecosystem health and wellbeing of Otago's streams and rivers, including their instream habitat, life-supporting capacity, and fisheries amenity (after Hayes et al., 2021).

| Limit | Surface water body with average Mean daily flow <= 5 m3/s | Surface water body with average Mean daily flow > 5 m3/s |
|---|---|---|
| Minimum flow | 90% of naturalised 7-day MALF | 80% of naturalised 7-day MALF |
| Allocation rate | 20% of naturalised 7-day MALF | 30% of naturalised 7-day MALF |

Table 4. Summary table of independent catchment characteristics and dependent hydrologic indices from the natural streamflow sites used in the base model training and testing phase across the Otago Region. PET = potential evapotranspiration (mm/unit time), Particle size = mm, Mean = mean of all daily flow, and MALF = Mean of minimum flow for each water year having applied a running 7day mean to the daily flows.

| | Log Area ($m^2$) | Elevation (m) | Partilce Size (mm) | PET (mm/unit time) | Rainfall Variaton (mm) | Rain Days (days/yr) | Runoff Volume (%) | Slope (%) | Mean (m3/s) | MALF (m3/s) |
|---|---|---|---|---|---|---|---|---|---|---|
| count | 100 | 100 | 100 | 100 | 100 | 100 | 100 | 100 | 100 | 100 |
| mean | 8.28 | 642 | 3.19 | 841 | 178 | 1.94 | 0.08 | 13.86 | 7.96 | 1.92 |
| std | 0.66 | 361 | 0.8 | 126 | 20.7 | 0.88 | 0.13 | 6.96 | 15.6 | 4.38 |
| min | 6.52 | 66.6 | 1.31 | 318 | 143 | 1 | 0 | 1.32 | 0.01 | 0 |
| 25% | 7.91 | 323 | 2.58 | 794 | 162 | 1.47 | 0 | 7.8 | 0.58 | 0.1 |
| 50% | 8.31 | 594 | 3.46 | 859 | 178 | 1.64 | 0.02 | 12.4 | 2.22 | 0.31 |
| 75% | 8.69 | 887 | 3.8 | 917 | 192 | 2.07 | 0.11 | 19.2 | 5.63 | 1.05 |
| max | 10.2 | 1362 | 4.82 | 1025 | 225 | 5.8 | 0.49 | 29.3 | 80.2 | 20.8 |

Table 5. Summary table of R-Squared values when predicting 7-day mean annual low flow (MALF) using two different randomly shuffled splits sets (random states 2 and 4) revealing that 11 of the 13 trained models generalized well with very good to extremely good coefficient of determinations. Similar findings are observed for other Ensemble models.

| Model | Hydrologic index | Hyperparmeter tuning | Model | R-Squared (Random state = 2) | Quality | R-Squared (Random state = 4) | Quality |
|---|---|---|---|---|---|---|---|
| 1 | MALF | no | Random forest regressor (RFR) | 0.99 | Extremely good | 0.91 | Extremely good |
| 2 | | | Gradient boosting regressor (GBR) | 0.98 | Extremely good | 0.91 | Extremely good |
| 3 | | | Extreme gradient boosting regressor (XGB) | 0.97 | Extremely good | 0.91 | Extremely good |
| 4 | | | Quantile gradient boosting regressor at 10th percentile (QGBR10) | -0.01 | No resolution | -0.31 | No resolution |
| 5 | | | Quantile gradient boosting regressor at 50th percentile (QGBR50) | 0.91 | Extremely good | 0.93 | Extremely good |
| 6 | | | Quantile gradient boosting regressor at 90th percentile (QGBR90) | 0.97 | Extremely good | 0.89 | Very good |
| 7 | | yes | RFR with random grid search (RFRgs) | 0.98 | Extremely good | 0.94 | Extremely good |
| 8 | | | RFR with random grid search and cross-validation (RFRgscv) | 0.88 | Very good | 0.89 | Very good |
| 9 | | | GBR with random grid search and cross-validation (GBRgscv) | 0.98 | Extremely good | 0.87 | Very good |
| 10 | | | XGB with random grid search and cross-validation (XGBgscv) | 0.99 | Extremely good | 0.92 | Extremely good |
| 11 | | | QGBR10 with random grid search (QGBR10gs) | -0.13 | No resolution | -0.16 | No resolution |
| 12 | | | QGBR50 with random grid search (QGBR50gs) | 0.94 | Extremely good | 0.91 | Extremely good |
| 13 | | | QGBR90 with random grid search (QGBR90gs) | 0.96 | Extremely good | 0.86 | Very good |

Table 6. Summary table of R-Squared values when predicting mean daily flows (Mean) using two different randomly shuffled splits sets (random states 2 and 4) revealing that 11 of the 13 trained models generalized well with very good to extremely good coefficient of determinations. Similar findings are observed for other Ensemble models.

| Model | Hydrologic index | Hyperparmeter tuning | Algorithm | R-Squared (Random state = 2) | Quality | R-Squared (Random state = 4) | Quality |
|---|---|---|---|---|---|---|---|
| 1 | Mean flow | no | Random forest regressor (RFR) | 0.87 | Very good | 0.91 | Extremely good |
| 2 | | | Gradient boosting regressor (GBR) | 0.82 | Extremely good | 0.93 | Extremely good |
| 3 | | | Extreme gradient boosting regressor (XGB) | 0.81 | Extremely good | 0.95 | Extremely good |
| 4 | | | Quantile gradient boosting regressor at 10th percentile (QGBR10) | -0.1 | No resolution | -0.11 | No resolution |
| 5 | | | Quantile gradient boosting regressor at 50th percentile (QGBR50) | 0.82 | Very good | 0.96 | Extremely good |
| 6 | | | Quantile gradient boosting regressor at 90th percentile (QGBR90) | 0.83 | Very good | 0.9 | Extremely good |
| 7 | | yes | RFR with random grid search (RFRgs) | 0.88 | Extremely good | 0.91 | Extremely good |
| 8 | | | RFR with random grid search and cross-validation (RFRgscv) | 0.83 | Very good | 0.85 | Very good |
| 9 | | | GBR with random grid search and cross-validation (GBRgscv) | 0.44 | Poor | 0.93 | Extremely good |
| 10 | | | XGB with random grid search and cross-validation (XGBgscv) | 0.99 | Extremely good | 0.92 | Extremely good |
| 11 | | | QGBR10 with random grid search (QGBR10gs) | -0.13 | No resolution | -0.16 | No resolution |
| 12 | | | QGBR50 with random grid search (QGBR50gs) | 0.82 | Very good | 0.96 | Extremely good |
| 13 | | | QGBR90 with random grid search (QGBR90gs) | 0.83 | Very good | 0.9 | Extremely good |

Table 7. Statistical comparison of 7-day mean annual low flow (MALF) observations (known) with MALF predictions (testing phase) using ensemble models with no hyperparameter tuning and with hyperparameter tuning. The randomly shuffled spilt set used in training and testing reflects assignment of random state numbers to 2 and 4. RFR = Random Forest Regressor, GBR = Gradient Boosting Regressor, XGB = Extreme Gradient Boosting regressor, QGBR = Quantile Gradient Boosting Regressor with numeral denoting the quantile; std = standard deviation, min = minimum, max = maximum, 25% = 25$^{th}$ percentile, 50% = 50$^{th}$ percentile, and 75% = 75$^{th}$ percentile, gs = grid search, rgcv = random grid search with cross-validation.

| Random State = 2 | | | No Hyperparameter Tuning | | | | |
|---|---|---|---|---|---|---|---|
| | MALF | RFR | GBR | XGB | QGBR10 | QGBR50 | QGBR90 |
| count | 20 | 20 | 20 | 20 | 20 | 20 | 20 |
| mean | 2.21 | 2.32 | 2.65 | 2.33 | 0.19 | 1.92 | 2.97 |
| std | 5.62 | 4.94 | 6.01 | 4.97 | 0.12 | 3.90 | 5.46 |
| min | 0.00 | 0.02 | 0.00 | 0.06 | 0.00 | -0.01 | 0.12 |
| 25% | 0.10 | 0.35 | 0.28 | 0.27 | 0.13 | 0.18 | 0.47 |
| 50% | 0.16 | 0.53 | 0.40 | 0.71 | 0.22 | 0.38 | 1.29 |
| 75% | 0.89 | 1.05 | 1.13 | 1.05 | 0.28 | 1.44 | 2.13 |
| max | 20.80 | 18.20 | 20.76 | 17.23 | 0.37 | 15.07 | 20.81 |
| Model | | 1 | 2 | 3 | 4 | 5 | 6 |

| Random State = 2 | | | | | | | Hyperparameter Tuning | | | | | | | | | |
|---|---|---|---|---|---|---|---|---|---|---|---|---|---|---|---|---|
| | MALF | RFRgs | RFRrgs | GBRgs | GBRrgcv | XGBgs | XGBrgs | QGBR10 | QGBR20 | QGBR30 | QGBR40 | QGBR50 | QGBR60 | QGBR70 | QGBR80 | QGBR90 |
| count | 20 | 20 | 20 | 20 | 20 | 20.00 | 20.00 | 20.00 | 20.00 | 20.00 | 20.00 | 20.00 | 20.00 | 20.00 | 20.00 | 20.00 |
| mean | 2.21 | 2.32 | 2.45 | 1.92 | 2.56 | 2.63 | 2.38 | 0.08 | 0.30 | 0.93 | 1.56 | 1.97 | 2.23 | 2.43 | 2.53 | 3.17 |
| std | 5.62 | 4.94 | 5.37 | 3.90 | 5.67 | 5.83 | 5.39 | 0.03 | 0.23 | 1.54 | 3.17 | 4.21 | 4.96 | 5.36 | 5.38 | 5.66 |
| min | 0.00 | 0.02 | 0.01 | -0.01 | 0.01 | 0.02 | -0.21 | 0.02 | -0.01 | -0.05 | 0.01 | 0.01 | 0.01 | 0.01 | 0.05 | 0.02 |
| 25% | 0.10 | 0.35 | 0.34 | 0.18 | 0.30 | 0.29 | 0.21 | 0.07 | 0.15 | 0.14 | 0.18 | 0.18 | 0.28 | 0.39 | 0.40 | 0.78 |
| 50% | 0.16 | 0.53 | 0.73 | 0.38 | 0.46 | 0.74 | 0.59 | 0.09 | 0.24 | 0.33 | 0.32 | 0.44 | 0.35 | 0.52 | 0.76 | 1.46 |
| 75% | 0.89 | 1.05 | 1.18 | 1.44 | 1.41 | 1.10 | 1.24 | 0.10 | 0.47 | 0.89 | 1.23 | 1.32 | 1.27 | 1.30 | 1.27 | 2.32 |
| max | 20.8 | 18.2 | 18.1 | 15.1 | 20.7 | 20.8 | 20.8 | 0.10 | 0.71 | 5.68 | 11.8 | 14.9 | 16.9 | 20.8 | 20.9 | 21.9 |
| Model | | 7 | 8 | 9 | 10 | 11 | 12 | 13 | 14 | 15 | 16 | 17 | 18 | 19 | 20 | 21 |

| Random State = 4 | | | No Hyperparameter Tuning | | | | |
|---|---|---|---|---|---|---|---|
| | MALF | RFR | GBR | XGB | QGBR10 | QGBR50 | QGBR90 |
| count | 20 | 20 | 20 | 20 | 20 | 20 | 20 |
| mean | 2.03 | 1.95 | 2.05 | 1.90 | 0.29 | 2.12 | 2.55 |
| std | 4.14 | 4.36 | 4.63 | 4.42 | 0.75 | 4.54 | 4.57 |
| min | 0.00 | 0.03 | 0.06 | 0.06 | 0.01 | -0.10 | 0.04 |
| 25% | 0.15 | 0.20 | 0.15 | 0.18 | 0.08 | 0.09 | 0.62 |
| 50% | 0.37 | 0.35 | 0.35 | 0.29 | 0.15 | 0.25 | 0.92 |
| 75% | 1.56 | 0.71 | 0.58 | 0.44 | 0.17 | 2.03 | 1.47 |
| max | 16.8 | 15.8 | 16.7 | 16.0 | 3.46 | 16.9 | 17.0 |
| Model | | 1 | 2 | 3 | 4 | 5 | 6 |

| Random State = 4 | | | | | | | Hyperparameter Tuning | | | | | | | | | |
|---|---|---|---|---|---|---|---|---|---|---|---|---|---|---|---|---|
| | MALF | RFRgs | RFRrgs | GBRgs | GBRrgcv | XGBgs | XGBrgscv | QGBR10 | QGBR20 | QGBR30 | QGBR40 | QGBR50 | QGBR60 | QGBR70 | QGBR80 | QGBR90 |
| count | 20 | 20 | 20 | 20 | 20 | 20 | 20 | 20 | 20 | 20 | 20 | 20 | 20 | 20 | 20 | 20 |
| mean | 2.03 | 1.95 | 1.95 | 2.12 | 2.09 | 2.00 | 2.05 | 0.14 | 0.70 | 0.45 | 0.99 | 1.91 | 1.92 | 2.26 | 2.35 | 2.93 |
| std | 4.14 | 4.36 | 4.64 | 4.54 | 4.62 | 4.64 | 4.60 | 0.16 | 1.93 | 0.52 | 1.73 | 4.32 | 4.74 | 4.88 | 4.86 | 4.32 |
| min | 0.00 | 0.03 | 0.00 | -0.10 | -0.03 | -0.03 | 0.04 | 0.00 | -1.23 | 0.01 | -0.05 | -0.10 | -0.49 | 0.15 | 0.24 | 0.71 |
| 25% | 0.15 | 0.20 | 0.07 | 0.09 | 0.06 | 0.16 | 0.24 | 0.05 | 0.12 | 0.12 | 0.13 | 0.12 | 0.01 | 0.15 | 0.71 | 1.36 |
| 50% | 0.37 | 0.35 | 0.27 | 0.25 | 0.36 | 0.28 | 0.40 | 0.09 | 0.16 | 0.25 | 0.39 | 0.29 | 0.07 | 0.20 | 0.71 | 1.69 |
| 75% | 1.56 | 0.71 | 0.78 | 2.03 | 0.97 | 0.55 | 0.65 | 0.15 | 0.23 | 0.57 | 0.76 | 0.98 | 0.50 | 0.66 | 0.71 | 1.94 |
| max | 16.8 | 15.8 | 16.8 | 16.9 | 16.7 | 16.7 | 16.7 | 0.56 | 6.28 | 1.83 | 6.05 | 14.8 | 16.8 | 17.1 | 17.4 | 17.3 |
| Model | | 7 | 8 | 9 | 10 | 11 | 12 | 13 | 14 | 15 | 16 | 17 | 18 | 19 | 20 | 21 |

Table 8. Statistical comparison of mean daily flow (Mean) observations (known) with Mean predictions (testing phase) using ensemble models with no hyperparameter tuning and with hyperparameter tuning. The randomly shuffled spilt set used in training and testing reflects assignment of random state numbers to 2 and 4. RFR = Random Forest Regressor, GBR = Gradient Boosting Regressor, XGB = Extreme Gradient Boosting regressor, QGBR = Quantile Gradient Boosting Regressor with numeral denoting the quantile; std = standard deviation, min = minimum, max = maximum, 25% = 25$^{th}$ percentile, 50% = 50$^{th}$ percentile, and 75% = 75$^{th}$ percentile, gs = grid search, rgcv = random grid search with cross-validation.

| Random State = 2 | No Hyperparameter Tuning | | | | | | |
|---|---|---|---|---|---|---|---|
| | Mean | RFR | GBR | XGB | QGBR10 | QGBR50 | QGBR90 |
| count | 20 | 20 | 20 | 20 | 20 | 20 | 20 |
| mean | 8.99 | 10.0 | 10.8 | 9.78 | 1.49 | 8.90 | 13.1 |
| std | 19.0 | 19.3 | 22.3 | 19.7 | 1.07 | 18.0 | 20.6 |
| min | 0.02 | 0.20 | -0.09 | 0.09 | 0.10 | 0.27 | 1.42 |
| 25% | 0.65 | 1.18 | 1.27 | 0.72 | 0.56 | 0.92 | 2.35 |
| 50% | 1.69 | 2.04 | 2.45 | 2.01 | 1.33 | 1.85 | 3.98 |
| 75% | 4.24 | 4.86 | 4.76 | 5.92 | 2.37 | 4.51 | 13.5 |
| max | 80.2 | 68.1 | 80.1 | 65.7 | 3.10 | 61.7 | 80.2 |
| Model | | 1 | 2 | 3 | 4 | 5 | 6 |

| Random State = 2 | Hyperparameter Tuning | | | | | | | | | | | | | | | |
|---|---|---|---|---|---|---|---|---|---|---|---|---|---|---|---|---|
| | Mean | RFRgs | RFRrgs | GBRgs | GBRrgcv | XGBgs | XGBrgs | QGBR10 | QGBR20 | QGBR30 | QGBR40 | QGBR50 | QGBR60 | QGBR70 | QGBR80 | QGBR90 |
| count | 20 | 20 | 20 | 20 | 20 | 20 | 20 | 20 | 20 | 20 | 20 | 20 | 20 | 20 | 20 | 20 |
| mean | 8.99 | 10.0 | 9.4 | 8.9 | 12.2 | 10.9 | 10.3 | 0.2 | 0.8 | 1.8 | 2.7 | 9.3 | 10.9 | 12.8 | 13.3 | 18.6 |
| std | 19.0 | 19.3 | 17.8 | 18.0 | 26.7 | 22.7 | 20.8 | 0.1 | 0.3 | 1.1 | 2.3 | 18.1 | 19.1 | 25.6 | 24.8 | 22.6 |
| min | 0.02 | 0.20 | 0.30 | 0.27 | -0.61 | 0.02 | -2.40 | 0.07 | 0.07 | -0.01 | -0.58 | -0.68 | 0.10 | 1.40 | 3.06 | 10.2 |
| 25% | 0.65 | 1.18 | 1.61 | 0.92 | 0.85 | 0.71 | 1.54 | 0.27 | 0.75 | 0.90 | 0.74 | 0.48 | 1.37 | 1.40 | 3.06 | 10.2 |
| 50% | 1.69 | 2.04 | 3.38 | 1.85 | 2.05 | 1.97 | 3.87 | 0.27 | 0.90 | 2.00 | 2.42 | 1.67 | 4.98 | 2.03 | 3.06 | 10.2 |
| 75% | 4.24 | 4.86 | 4.94 | 4.51 | 4.95 | 5.84 | 5.79 | 0.27 | 1.00 | 2.92 | 4.14 | 7.94 | 7.34 | 6.69 | 6.02 | 10.2 |
| max | 80.2 | 68.1 | 65.7 | 61.7 | 96.5 | 80.2 | 80.2 | 0.3 | 1.0 | 2.9 | 6.1 | 63.1 | 66.7 | 91.8 | 88.7 | 88.0 |
| Model | | 7 | 8 | 9 | 10 | 11 | 12 | 13 | 14 | 15 | 16 | 17 | 18 | 19 | 20 | 21 |

| Random State = 4 | No Hyperparameter Tuning | | | | | | |
|---|---|---|---|---|---|---|---|
| | Mean | RFR | GBR | XGB | QGBR10 | QGBR50 | QGBR90 |
| count | 20 | 20 | 20 | 20 | 20 | 20 | 20 |
| mean | 7.33 | 7.88 | 8.47 | 8.23 | 1.85 | 7.98 | 9.87 |
| std | 14.5 | 13.5 | 15.7 | 15.3 | 1.24 | 15.2 | 15.4 |
| min | 0.03 | 0.20 | 0.14 | 0.09 | 0.05 | -0.17 | 2.01 |
| 25% | 0.98 | 0.95 | 1.47 | 1.03 | 0.57 | 0.70 | 2.64 |
| 50% | 1.89 | 2.94 | 2.35 | 2.23 | 2.33 | 2.23 | 3.56 |
| 75% | 4.47 | 5.29 | 4.42 | 4.19 | 2.89 | 6.05 | 7.60 |
| max | 62.7 | 51.6 | 62.5 | 61.1 | 3.43 | 62.9 | 62.7 |
| Model | | 1 | 2 | 3 | 4 | 5 | 6 |

| Random State = 4 | Hyperparameter Tuning | | | | | | | | | | | | | | | |
|---|---|---|---|---|---|---|---|---|---|---|---|---|---|---|---|---|
| | Mean | RFRgs | RFRrgs | GBRgs | GBRrgcv | XGBgs | XGBrgscv | QGBR10 | QGBR20 | QGBR30 | QGBR40 | QGBR50 | QGBR60 | QGBR70 | QGBR80 | QGBR90 |
| count | 20 | 20 | 20 | 20 | 20 | 20 | 20 | 20 | 20 | 20 | 20 | 20 | 20 | 20 | 20 | 20 |
| mean | 7.33 | 7.88 | 7.80 | 7.98 | 8.35 | 8.57 | 8.71 | 1.54 | 2.69 | 5.39 | 6.13 | 7.89 | 8.39 | 8.80 | 9.42 | 11.04 |
| std | 14.5 | 13.5 | 13.4 | 15.2 | 15.3 | 15.9 | 15.6 | 1.07 | 2.66 | 9.22 | 11.4 | 15.5 | 15.6 | 15.7 | 15.4 | 14.8 |
| min | 0.03 | 0.20 | 0.19 | -0.17 | 0.29 | 0.09 | -0.01 | 0.05 | 0.03 | 0.13 | 0.05 | 0.16 | 0.13 | 0.30 | 1.15 | 1.71 |
| 25% | 0.98 | 0.95 | 0.88 | 0.70 | 1.87 | 1.07 | 2.08 | 0.64 | 0.60 | 0.68 | 0.71 | 0.82 | 1.17 | 1.49 | 2.55 | 4.90 |
| 50% | 1.89 | 2.94 | 2.84 | 2.23 | 2.33 | 2.23 | 2.99 | 1.62 | 2.02 | 2.16 | 2.34 | 2.03 | 2.09 | 2.34 | 3.43 | 5.45 |
| 75% | 4.47 | 5.29 | 5.23 | 6.05 | 4.52 | 4.51 | 4.68 | 2.41 | 3.63 | 4.17 | 4.70 | 4.87 | 6.06 | 6.80 | 6.71 | 8.72 |
| max | 62.7 | 51.6 | 51.3 | 62.9 | 61.8 | 62.7 | 62.6 | 3.53 | 9.09 | 32.4 | 44.5 | 62.9 | 62.7 | 62.7 | 62.7 | 62.7 |
| Model | | 7 | 8 | 9 | 10 | 11 | 12 | 13 | 14 | 15 | 16 | 17 | 18 | 19 | 20 | 21 |

Table 9. Summary table of independent catchment characteristics used to predict naturalized mean daily flow and naturalized 7-day mean annual low flow at 317 regulated priority catchments across the Otago Region.

| | Log Area ($m^2$) | Elevation (m) | Partilce Size (mm) | PET (mm/unit time) | Rainfall Variaton (mm) | Rain Days (days/yr) | Runoff Volume (%) | Slope (%) |
|---|---|---|---|---|---|---|---|---|
| count | 317 | 317 | 317 | 317 | 317 | 317 | 317 | 317 |
| mean | 7.12 | 477 | 3.17 | 952 | 172 | 1.83 | 0.08 | 14.3 |
| std | 0.74 | 345 | 1.02 | 115 | 20.5 | 0.76 | 0.13 | 6.57 |
| min | 5.49 | 11.1 | 0 | 400 | 141 | 0.68 | 0 | 0.23 |
| 25% | 6.61 | 147 | 2.7 | 880 | 155 | 1.4 | 0 | 9.75 |
| 50% | 7 | 453 | 3.55 | 960 | 171 | 1.7 | 0.02 | 14 |
| 75% | 7.54 | 733 | 3.89 | 1019 | 183 | 2.08 | 0.1 | 18.2 |
| max | 9.76 | 1386 | 5 | 1221 | 218 | 6.63 | 0.59 | 30.7 |

Table 10. Summary table of independent catchment characteristics used to predict naturalized mean daily flow and naturalized 7-day mean annual low flow at 18612 ungauged river reaches across the Otago Region.

| | Log Area ($m^2$) | Elevation (m) | Partilce Size (mm) | PET (mm/unit time) | Rainfall Variaton (mm) | Rain Days (days/yr) | Runoff Volume (%) | Slope (%) |
|---|---|---|---|---|---|---|---|---|
| count | 18612 | 18612 | 18606 | 18612 | 18612 | 18612 | 18612 | 18612 |
| mean | 7.5 | 833 | 3.39 | 772 | 176 | 17.4 | 0.17 | 2.62 |
| std | 0.87 | 438 | 0.84 | 191 | 20.5 | 9.71 | 0.23 | 1.62 |
| min | 5.7 | 8.94 | 0 | 102 | 141 | 0.14 | 0 | 0.66 |
| 25% | 6.9 | 472 | 3.12 | 677 | 160 | 8.79 | 0 | 1.49 |
| 50% | 7.31 | 841 | 3.75 | 820 | 176 | 15 | 0.02 | 1.96 |
| 75% | 7.94 | 1189 | 3.99 | 895 | 190 | 27.4 | 0.35 | 3.22 |
| max | 10.3 | 2105 | 5 | 1221 | 220 | 45.6 | 0.95 | 7.54 |

Table 11. Summary table of probable 5-fold cross-validated results when predicting the mean daily flow (Mean) and 7-day mean annual low flow (MALF) at various percentiles for the largest flows in the Otago Region (others not shown here). These values are sorted on the expected value of the Mean flows.

| | Hydrologic Index | MALF | MALF | MALF | MALF | MALF | MALF | Mean | Mean | Mean | Mean | Mean | Mean |
|---|---|---|---|---|---|---|---|---|---|---|---|---|---|
| ID | Cross-validation statistic<br>Catchment | 5th percentile (l/s) | 25th percentile (l/s) | 50th percentile (l/s) | Expected value (l/s) | 75h percentile (l/s) | 95h percentile (l/s) | 5th percentile (l/s) | 25th percentile (l/s) | 50th percentile (l/s) | Expected value (l/s) | 75h percentile (l/s) | 95h percentile (l/s) |
| 1 | Makarora River | 7503.8 | 15851.8 | 16831.9 | 16900.7 | 15851.8 | 20798.8 | 28168.9 | 58408.7 | 63395.7 | 63095.3 | 58408.7 | 80195.0 |
| 2 | Dart River | 9111.9 | 15030.6 | 17283.8 | 16904.1 | 15030.6 | 20438.2 | 32052.5 | 55606.2 | 62858.5 | 60889.0 | 55606.2 | 80318.6 |
| 3 | Greenstone River | 6538.6 | 10925.2 | 12821.0 | 13066.6 | 10925.2 | 20797.0 | 24272.3 | 39282.2 | 46726.9 | 51881.2 | 39282.2 | 80197.0 |
| 4 | Hunter River | 924.2 | 11238.0 | 13265.7 | 13293.4 | 11238.0 | 20778.4 | 17143.6 | 34720.9 | 43758.0 | 43209.7 | 34720.9 | 66410.9 |
| 5 | Rees River | 991.0 | 6292.8 | 12136.9 | 10637.7 | 6292.8 | 17410.1 | 6620.3 | 33138.4 | 42574.0 | 38729.3 | 33138.4 | 56535.0 |
| 6 | Dingle Burn | 1043.3 | 7815.9 | 9562.4 | 9932.0 | 7815.9 | 20776.3 | 9457.5 | 25976.8 | 30823.8 | 32457.6 | 25976.8 | 65212.0 |
| 7 | Shotover River | 2177.9 | 8900.3 | 10117.1 | 10742.5 | 8900.3 | 16534.8 | 21700.9 | 28410.6 | 31668.7 | 32255.4 | 28410.6 | 53892.7 |
| 8 | Matukituki River | 1291.3 | 6168.8 | 8840.1 | 8074.9 | 6168.8 | 16247.9 | 5589.3 | 25281.9 | 30223.2 | 28365.8 | 25281.9 | 53386.3 |
| 9 | Big Hopwood Burn | 1167.6 | 5212.1 | 7412.2 | 8204.9 | 5212.1 | 20783.9 | 6480.8 | 15330.2 | 21488.4 | 25829.8 | 15330.2 | 77816.5 |
| 10 | Taieri River | 871.7 | 2053.3 | 3292.0 | 3321.4 | 2053.3 | 6130.5 | 12456.9 | 19849.8 | 25222.5 | 23428.2 | 19849.8 | 31559.2 |
| 11 | Lochy River | 924.2 | 4233.2 | 4952.8 | 6674.1 | 4233.2 | 17272.9 | 6556.7 | 12588.7 | 15780.4 | 20686.8 | 12588.7 | 80171.1 |
| 12 | Pomahaka River | 741.1 | 1616.5 | 2285.4 | 2395.0 | 1616.5 | 4062.2 | 5770.2 | 18513.0 | 21588.7 | 20189.9 | 18513.0 | 26900.4 |
| 13 | Timaru River | 1043.3 | 4335.9 | 5160.7 | 6040.5 | 4335.9 | 13197.5 | 4885.3 | 10753.5 | 14319.3 | 15972.0 | 10753.5 | 31800.7 |
| 14 | Minaret Burn | 1057.6 | 2746.1 | 3608.1 | 5223.4 | 2746.1 | 14706.3 | 3371.3 | 8018.2 | 11877.4 | 14984.3 | 8018.2 | 44520.6 |
| 15 | Manuherikia River | 178.8 | 960.7 | 1218.0 | 1304.7 | 960.7 | 2729.8 | 8919.2 | 13463.9 | 14497.9 | 14652.2 | 13463.9 | 19824.8 |
| 16 | Nevis River | 1705.1 | 2994.3 | 3559.3 | 3579.4 | 2994.3 | 5124.3 | 8183.7 | 10306.3 | 11809.0 | 11990.4 | 10306.3 | 17622.3 |
| 17 | Waipati River | 111.6 | 756.3 | 2600.1 | 2641.4 | 756.3 | 9947.7 | 1472.0 | 2487.3 | 7989.4 | 8535.9 | 2487.3 | 25879.6 |
| 18 | Catlins River | 345.3 | 754.8 | 979.5 | 1010.9 | 754.8 | 2342.6 | 4866.8 | 7143.5 | 8298.8 | 8435.4 | 7143.5 | 15382.8 |
| 19 | Tautuku River | 151.4 | 686.5 | 2643.8 | 2612.4 | 686.5 | 9947.7 | 1395.4 | 2296.2 | 7528.0 | 8161.4 | 2296.2 | 25703.9 |
| 20 | Tahakopa River | 418.3 | 739.3 | 913.2 | 1449.3 | 739.3 | 7633.5 | 3226.5 | 4996.6 | 5995.3 | 6664.8 | 4996.6 | 18659.2 |
| 21 | Waitahuna River | 69.0 | 361.1 | 529.5 | 596.1 | 361.1 | 1509.2 | 2847.6 | 4618.4 | 6194.5 | 6161.4 | 4618.4 | 10022.6 |
| 22 | Tokomairiro River | 69.0 | 289.5 | 397.4 | 508.2 | 289.5 | 1481.5 | 2329.4 | 4706.2 | 5874.9 | 6027.9 | 4706.2 | 9013.4 |
| 23 | Lindis River | 144.9 | 434.3 | 636.6 | 720.2 | 434.3 | 1624.9 | 3303.0 | 4973.5 | 5518.9 | 5874.0 | 4973.5 | 9097.4 |
| 24 | Von River | 630.3 | 878.3 | 1111.0 | 1260.6 | 878.3 | 3046.4 | 3376.3 | 4277.2 | 5170.5 | 5405.0 | 4277.2 | 10714.4 |
| 25 | Kakanui River | 0.1 | 310.5 | 418.6 | 485.8 | 310.5 | 1400.4 | 2285.8 | 3728.2 | 4344.6 | 4638.9 | 3728.2 | 9862.4 |
| 26 | Shag River | 54.2 | 342.8 | 593.5 | 645.0 | 342.8 | 1447.2 | 2106.7 | 3164.0 | 3808.3 | 4056.3 | 3164.0 | 7514.7 |
| 27 | Waikouaiti River | 77.9 | 228.3 | 315.5 | 438.4 | 228.3 | 1717.6 | 1735.1 | 2387.9 | 3592.6 | 3546.1 | 2387.9 | 8607.6 |
| 28 | Arrow River | 248.3 | 619.6 | 740.9 | 716.4 | 619.6 | 1057.8 | 1893.5 | 2808.4 | 3140.2 | 3453.4 | 2808.4 | 6574.2 |
| 29 | Teviot River | 171.6 | 466.2 | 638.3 | 702.5 | 466.2 | 1566.3 | 1502.2 | 2563.1 | 3066.7 | 3394.5 | 2563.1 | 6801.2 |
| 30 | Buckler Burn | 398.8 | 803.6 | 1046.2 | 1123.9 | 803.6 | 2978.2 | 1167.2 | 2458.5 | 3129.3 | 3358.7 | 2458.5 | 7193.7 |
| 31 | Staircase Creek | 364.7 | 794.2 | 1076.8 | 1244.7 | 794.2 | 3759.8 | 1086.5 | 1799.5 | 2807.2 | 3044.1 | 1799.5 | 9369.5 |
| 32 | Cardrona River | 0.1 | 361.6 | 425.8 | 497.0 | 361.6 | 950.5 | 1076.5 | 2207.8 | 2550.0 | 3000.8 | 2207.8 | 5826.8 |
| … | | | | | | | | | | | | | |

Table 12. Summary table of probable 5-fold cross-validated results when predicting the mean daily flow (Mean) and 7-day mean annual low flow (MALF) at various percentiles for the smallest (N=32) flows in the Otago Region (others not shown here). These values are sorted on the expected value of the Mean flows.

| ID | Hydrologic Index<br>Cross-validation statistic<br>Catchment | MALF<br>5th percentile (l/s) | MALF<br>25th percentile (l/s) | MALF<br>50th percentile (l/s) | MALF<br>Expected value (l/s) | MALF<br>75h percentile (l/s) | MALF<br>95h percentile (l/s) | Mean<br>5th percentile (l/s) | Mean<br>25th percentile (l/s) | Mean<br>50th percentile (l/s) | Mean<br>Expected value (l/s) | Mean<br>75h percentile (l/s) | Mean<br>95h percentile (l/s) |
|---|---|---|---|---|---|---|---|---|---|---|---|---|---|
| … | | | | | | | | | | | | | |
| 286 | Hilderthorpe | 0.1 | 0.1 | 86.2 | 80.0 | 0.1 | 591.6 | 0.1 | 0.1 | 261.6 | 172.0 | 0.1 | 2258.7 |
| 287 | Hall Road Creek | 0.1 | 0.1 | 11.5 | 0.1 | 0.1 | 104.2 | 0.1 | 64.8 | 159.8 | 166.6 | 64.8 | 585.9 |
| 288 | Pillans Stream | 0.1 | 0.1 | 13.0 | 36.8 | 0.1 | 569.2 | 0.1 | 38.8 | 126.6 | 147.1 | 38.8 | 1058.1 |
| 289 | Isas Creek | 0.1 | 0.1 | 13.4 | 37.1 | 0.1 | 569.2 | 0.1 | 38.8 | 127.6 | 146.1 | 38.8 | 1058.1 |
| 290 | Waitangi Stream | 0.1 | 0.1 | 13.4 | 37.1 | 0.1 | 569.2 | 0.1 | 38.8 | 127.6 | 146.1 | 38.8 | 1058.1 |
| 291 | Finnies Creek | 0.1 | 45.9 | 84.9 | 137.1 | 45.9 | 699.2 | 0.1 | 29.1 | 116.4 | 134.8 | 29.1 | 651.7 |
| 292 | Post Office Creek | 0.1 | 0.1 | 66.6 | 50.0 | 0.1 | 360.8 | 0.1 | 0.1 | 131.1 | 133.4 | 0.1 | 1751.0 |
| 293 | Orore Creek | 0.1 | 20.6 | 27.5 | 62.2 | 20.6 | 502.3 | 0.1 | 24.6 | 96.7 | 124.7 | 24.6 | 2307.7 |
| 294 | Allangrange (S) | 0.1 | 0.1 | 11.6 | 4.1 | 0.1 | 79.7 | 0.1 | 41.6 | 131.1 | 119.3 | 41.6 | 426.7 |
| 295 | Pannetts Road Creek | 0.1 | 0.1 | 9.1 | 0.1 | 0.1 | 59.5 | 0.1 | 39.7 | 117.3 | 105.3 | 39.7 | 366.2 |
| 296 | King Road Creek | 0.1 | 0.1 | 46.0 | 5.2 | 0.1 | 401.1 | 0.1 | 0.1 | 46.6 | 89.0 | 0.1 | 2013.9 |
| 297 | Clydevale Creek | 0.1 | 0.1 | 5.9 | 0.1 | 0.1 | 59.5 | 0.1 | 18.3 | 118.9 | 86.1 | 18.3 | 614.4 |
| 298 | Oven Hill Creek | 0.1 | 0.9 | 20.2 | 17.8 | 0.9 | 160.1 | 0.1 | 16.2 | 108.6 | 82.2 | 16.2 | 624.2 |
| 299 | Three Brothers Gully | 0.1 | 0.1 | 33.8 | 34.1 | 0.1 | 315.5 | 0.1 | 0.1 | 71.1 | 79.0 | 0.1 | 732.7 |
| 300 | Kakaho Creek | 0.1 | 0.1 | 8.9 | 3.3 | 0.1 | 405.1 | 0.1 | 0.1 | 52.9 | 75.3 | 0.1 | 1558.9 |
| 301 | Kaik Road Creek | 0.1 | 0.1 | 91.8 | 79.4 | 0.1 | 591.6 | 0.1 | 0.1 | 301.0 | 72.1 | 0.1 | 4103.7 |
| 302 | Landon Creek | 0.1 | 26.8 | 85.4 | 143.9 | 26.8 | 951.4 | 0.1 | 0.1 | 107.1 | 62.3 | 0.1 | 2416.9 |
| 303 | Shagree Creek | 0.1 | 1.8 | 16.1 | 17.1 | 1.8 | 95.6 | 0.1 | 13.0 | 73.1 | 58.3 | 13.0 | 245.3 |
| 304 | Reids Stream | 0.1 | 0.1 | 11.2 | 5.4 | 0.1 | 93.0 | 0.1 | 5.8 | 77.1 | 51.1 | 5.8 | 329.7 |
| 305 | Jones Creek | 0.1 | 0.1 | 37.4 | 0.1 | 0.1 | 292.2 | 0.1 | 0.1 | 223.4 | 42.9 | 0.1 | 1184.8 |
| 306 | Awamoko Stream | 0.1 | 5.9 | 44.5 | 41.2 | 5.9 | 386.2 | 0.1 | 0.1 | 157.4 | 28.8 | 0.1 | 618.6 |
| 307 | Kaikorai Stream | 0.1 | 52.6 | 99.8 | 131.0 | 52.6 | 482.1 | 0.1 | 0.1 | 205.5 | 8.6 | 0.1 | 1139.1 |
| 308 | Aitchison Road Cree | 0.1 | 0.1 | 10.0 | 0.1 | 0.1 | 464.3 | 0.1 | 0.1 | 0.1 | 0.1 | 0.1 | 459.9 |
| 309 | Bow Alley Creek | 0.1 | 16.3 | 27.6 | 38.8 | 16.3 | 209.4 | 0.1 | 0.1 | 43.6 | 0.1 | 0.1 | 746.4 |
| 310 | Glen Creek | 0.1 | 0.1 | 22.0 | 0.1 | 0.1 | 512.9 | 0.1 | 0.1 | 38.4 | 0.1 | 0.1 | 2246.4 |
| 311 | Hinahina Stream | 0.1 | 0.1 | 8.9 | 16.2 | 0.1 | 211.9 | 0.1 | 0.1 | 42.2 | 0.1 | 0.1 | 153.8 |
| 312 | Oamaru Airport Cre | 0.1 | 0.1 | 95.5 | 15.8 | 0.1 | 574.7 | 0.1 | 0.1 | 266.7 | 0.1 | 0.1 | 2117.0 |
| 313 | Oamaru Creek | 0.1 | 0.1 | 48.0 | 72.5 | 0.1 | 636.0 | 0.1 | 0.1 | 35.0 | 0.1 | 0.1 | 1510.8 |
| 314 | Oamaru North Creek | 0.1 | 0.1 | 22.0 | 0.1 | 0.1 | 512.9 | 0.1 | 0.1 | 37.1 | 0.1 | 0.1 | 2246.4 |
| 315 | Peaks Road Creek | 0.1 | 0.1 | 30.3 | 24.9 | 0.1 | 529.8 | 0.1 | 0.1 | 53.3 | 0.1 | 0.1 | 2219.7 |
| 316 | Waikoura Creek | 0.1 | 0.1 | 43.9 | 48.6 | 0.1 | 491.8 | 0.1 | 0.1 | 36.1 | 0.1 | 0.1 | 1440.6 |
| 317 | Welcome Creek | 0.1 | 0.1 | 26.4 | 0.1 | 0.1 | 512.9 | 0.1 | 0.1 | 39.0 | 0.1 | 0.1 | 2205.4 |

Table 13. Summary table of naturalized flow indices predicted at 18,612 ungauged river reaches spanning 7 Strahler stream orders across the Otago Region, New Zealand.

| Hydrologic index | Mean | Mean | Mean | Mean | Mean | MALF | MALF | MALF | MALF | MALF |
|---|---|---|---|---|---|---|---|---|---|---|
| Strahler stream order | 5th percentile (m3/s) | 25th percentile (m3/s) | 50h percentile (m3/s) | 75th percentile (m3/s) | 95th percentile (m3/s) | 5th percentile (m3/s) | 25th percentile (m3/s) | 50h percentile (m3/s) | 75th percentile (m3/s) | 95th percentile (m3/s) |
| 1 | 0.0-0.1 | 0.0-0.4 | 0.0-0.7 | 0.1-1.6 | 0.1-6.7 | 0.0-0.0 | 0.0-0.1 | 0.0-0.2 | 0.02-0.35 | 0.07-0.69 |
| 2 | 0.1-0.5 | 0.4-0.9 | 0.7-1.3 | 1.6-1.8 | 6.7-8.5 | 0.0-0.02 | 0.1-0.2 | 0.2-0.3 | 0.35-0.50 | 0.69-0.92 |
| 3 | 0.5-1.0 | 0.9-1.5 | 1.3-2.2 | 1.8-2.8 | 8.5-9.2 | 0.02-0.12 | 0.2-0.3 | 0.3-0.5 | 0.50-0.72 | 0.92-1.34 |
| 4 | 1.0-1.6 | 1.5-2.7 | 2.2-3.4 | 2.8-4.2 | 9.2-10.7 | 0.12-0.29 | 0.3-0.6 | 0.5-0.8 | 0.72-1.19 | 1.34-3.29 |
| 5 | 1.6-7.9 | 2.7-16.2 | 3.4-18.3 | 4.2-19.5 | 10.7-19.6 | 0.29-1.02 | 0.6-1.8 | 0.8-4.0 | 1.19-5.79 | 3.29-11.9 |
| 6 | 7.9-46.2 | 16.2-59.1 | 18.3-64 | 19.5-67.7 | 19.6-66.8 | 1.02-11.4 | 1.8-14.8 | 4.0-16.9 | 5.79-20.5 | 11.9-21.1 |
| 7 | 46.2-60.7 | 59.1-79.8 | 64-82.5 | 67.7-85.6 | 66.8-86.3 | 11.4-18.0 | 14.8-20.2 | 16.9-20.8 | 20.5-22.5 | 21.1-25.1 |

Table 14. Summary of naturalized flow indices predicted at (or near) streamflow gauging stations in the Taieri freshwater management unit: Taieri at Outram, Taieri at Hindon, Taieri at Sutton, Taieri at Tiroiti, Taieri at Linn Burn, Kye Burn, Pig Burn, Sutton Creek, Deep Stream, Lee Stream, and Nenthorn.

| | | | | Mean, l/s | | | | | MALF, l/s | | | | |
|---|---|---|---|---|---|---|---|---|---|---|---|---|---|
| N | Station | Easting | Northing | 5th percentile | 25th percentile | 50th percentile | 75th percentile | 95th percentile | 5th percentile | 25th percentile | 50th percentile | 75th percentile | 95th percentile |
| 1 | Taieri at Outram | 1385894 | 4918991 | 28188 | 29880 | 29974 | 30193 | 30193 | 3200 | 4254 | 4379 | 5121 | 5415 |
| 2 | Taieri at Hindon | 1393183 | 4934433 | 25918 | 27526 | 29431 | 30002 | 30002 | 2647 | 3220 | 4183 | 5047 | 5383 |
| 3 | Taieri at Sutton | 1376225 | 4949116 | 17884 | 18193 | 18473 | 19095 | 19095 | 2285 | 2418 | 2525 | 3232 | 4914 |
| 4 | Taieri at Tiroiti | 1385916 | 4984851 | 16628 | 17616 | 18293 | 18835 | 18835 | 1899 | 2176 | 2351 | 3246 | 4914 |
| 5 | Taieri at Linn Burn | 1351010 | 4958393 | 2665 | 2920 | 3384 | 3482 | 3482 | 148 | 258 | 423 | 570 | 1001 |
| 6 | Kye Burn | 1384708 | 4996733 | 2199 | 2714 | 2907 | 2923 | 2923 | 91.5 | 173 | 220 | 284 | 642 |
| 7 | Pig Burn | 1374122 | 4983925 | 1246 | 1483 | 2649 | 3255 | 3255 | 133 | 271 | 579 | 799 | 1100 |
| 8 | Sutton Creek | 1373363 | 4946708 | 1300 | 1635 | 2239 | 2715 | 2715 | 151 | 184 | 203 | 307 | 541 |
| 9 | Deep Stream | 1370377 | 4935501 | 275 | 768 | 856 | 1562 | 1562 | 127 | 310 | 391 | 564 | 826 |
| 10 | Lee Stream | 1377138 | 4924570 | 275 | 760 | 856 | 1562 | 1562 | 39.8 | 224 | 354 | 402 | 729 |
| 11 | Nenthorn | 1385683 | 4948654 | 0.00 | 0.19 | 0.41 | 1.56 | 1.56 | 0.00 | 0.003 | 0.06 | 0.13 | 0.54 |

Table 15. Summary table of the largest 5-fold cross-validated default catchment minimum flows and catchment default allocation rates (catchments records 1-32). These catchment records are sorted (maximum to minimum) on expected value for minimum flows.

| N | Default: Cross-validation Statistic: Catchment | Minimum Flow 5th percentile (l/s) | Minimum Flow 25th percentile (l/s) | Minimum Flow 50th percentile (l/s) | Minimum Flow Expected value (l/s) | Minimum Flow 75th percentile (l/s) | Minimum Flow 95th percentile (l/s) | Allocation Rate 5th percentile (l/s) | Allocation Rate 25th percentile (l/s) | Allocation Rate 50th percentile (l/s) | Allocation Rate Expected value (l/s) | Allocation Rate 75th percentile (l/s) | Allocation Rate 95th percentile (l/s) |
|---|---|---|---|---|---|---|---|---|---|---|---|---|---|
| 1 | Dart River | 7289.5 | 12024.5 | 13827.1 | 13523.3 | 16350.6 | 16350.6 | 2733.6 | 4509.2 | 5185.1 | 5071.2 | 6131.5 | 6131.5 |
| 2 | Makarora River | 6003.0 | 12681.4 | 13465.6 | 13520.6 | 16639.0 | 16639.0 | 2251.1 | 4755.5 | 5049.6 | 5070.2 | 6239.6 | 6239.6 |
| 3 | Hunter River | 739.4 | 8990.4 | 10612.6 | 10634.7 | 16622.7 | 16622.7 | 277.3 | 3371.4 | 3979.7 | 3988.0 | 6233.5 | 6233.5 |
| 4 | Greenstone River | 5230.9 | 8740.2 | 10256.8 | 10453.3 | 16637.6 | 16637.6 | 1961.6 | 3277.6 | 3846.3 | 3920.0 | 6239.1 | 6239.1 |
| 5 | Shotover River | 1742.3 | 7120.2 | 8093.7 | 8594.0 | 13227.8 | 13227.8 | 653.4 | 2670.1 | 3035.1 | 3222.8 | 4960.4 | 4960.4 |
| 6 | Rees River | 792.8 | 5034.3 | 9709.5 | 8510.2 | 13928.1 | 13928.1 | 297.3 | 1887.8 | 3641.1 | 3191.3 | 5223.0 | 5223.0 |
| 7 | Dingle Burn | 834.7 | 6252.7 | 7649.9 | 7945.6 | 16621.1 | 16621.1 | 313.0 | 2344.8 | 2868.7 | 2979.6 | 6232.9 | 6232.9 |
| 8 | Big Hopwood Burn | 934.1 | 4169.7 | 5929.8 | 6563.9 | 16627.1 | 16627.1 | 350.3 | 1563.6 | 2223.7 | 2461.5 | 6235.2 | 6235.2 |
| 9 | Matukituki River | 1033.0 | 4935.1 | 7072.1 | 6459.9 | 12998.3 | 12998.3 | 387.4 | 1850.7 | 2652.0 | 2422.5 | 4874.4 | 4874.4 |
| 10 | Lochy River | 739.4 | 3386.6 | 3962.2 | 5339.3 | 13818.3 | 13818.3 | 277.3 | 1270.0 | 1485.8 | 2002.2 | 5181.9 | 5181.9 |
| 11 | Timaru River | 939.0 | 3468.7 | 4128.6 | 4832.4 | 10558.0 | 10558.0 | 208.7 | 1300.8 | 1548.2 | 1812.2 | 3959.2 | 3959.2 |
| 12 | Minaret Burn | 951.8 | 2196.9 | 2886.4 | 4178.7 | 11765.1 | 11765.1 | 211.5 | 823.8 | 1082.4 | 1567.0 | 4411.9 | 4411.9 |
| 13 | Nevis River | 1364.1 | 2395.4 | 2847.4 | 2863.5 | 4099.4 | 4099.4 | 511.5 | 898.3 | 1067.8 | 1073.8 | 1537.3 | 1537.3 |
| 14 | Taieri River | 697.3 | 1642.6 | 2633.6 | 2657.1 | 4904.4 | 4904.4 | 261.5 | 616.0 | 987.6 | 996.4 | 1839.1 | 1839.1 |
| 15 | Waipati River | 100.4 | 680.6 | 2080.0 | 2113.1 | 7958.2 | 7958.2 | 22.3 | 151.3 | 780.0 | 792.4 | 2984.3 | 2984.3 |
| 16 | Tautuku River | 136.3 | 617.8 | 2115.1 | 2089.9 | 7958.2 | 7958.2 | 30.3 | 137.3 | 793.2 | 783.7 | 2984.3 | 2984.3 |
| 17 | Pomahaka River | 592.9 | 1293.2 | 1828.3 | 1916.0 | 3249.8 | 3249.8 | 222.3 | 484.9 | 685.6 | 718.5 | 1218.7 | 1218.7 |
| 18 | Tahakopa River | 376.5 | 665.3 | 730.6 | 1159.4 | 6106.8 | 6106.8 | 83.7 | 147.9 | 274.0 | 434.8 | 2290.0 | 2290.0 |
| 19 | Staircase Creek | 328.2 | 714.8 | 969.1 | 1120.2 | 3007.9 | 3383.8 | 72.9 | 158.8 | 215.4 | 248.9 | 752.0 | 1127.9 |
| 20 | Roaring Meg | 229.5 | 639.1 | 855.3 | 1069.0 | 2962.1 | 3332.4 | 51.0 | 142.0 | 190.1 | 237.6 | 740.5 | 1110.8 |
| 21 | Manuherikia River | 143.1 | 768.5 | 974.4 | 1043.8 | 2183.8 | 2183.8 | 53.7 | 288.2 | 365.4 | 391.4 | 818.9 | 818.9 |
| 22 | Black Gorge Creek | 278.8 | 682.4 | 979.4 | 1016.9 | 2372.9 | 2669.5 | 62.0 | 151.6 | 217.7 | 226.0 | 593.2 | 889.8 |
| 23 | Buckler Burn | 358.9 | 723.3 | 941.6 | 1011.6 | 2382.5 | 2680.4 | 79.8 | 160.7 | 209.2 | 224.8 | 595.6 | 893.5 |
| 24 | Von River | 567.3 | 790.5 | 888.8 | 1008.5 | 2437.1 | 2437.1 | 126.1 | 175.7 | 333.3 | 378.2 | 913.9 | 913.9 |
| 25 | McKinlays Creek | 324.4 | 716.3 | 920.5 | 951.7 | 2121.4 | 2386.6 | 72.1 | 159.2 | 204.6 | 211.5 | 530.4 | 795.5 |
| 26 | Alpha Burn | 121.9 | 599.4 | 892.8 | 911.4 | 2144.4 | 2144.4 | 27.1 | 133.2 | 198.4 | 202.5 | 476.5 | 476.5 |
| 27 | Twelve Mile Creek | 351.6 | 678.4 | 861.3 | 899.0 | 2225.5 | 2503.7 | 78.1 | 150.7 | 191.4 | 199.8 | 556.4 | 834.6 |
| 28 | Afton Burn | 437.2 | 706.9 | 843.3 | 898.7 | 1615.5 | 1817.4 | 97.1 | 157.1 | 187.4 | 199.7 | 403.9 | 605.8 |
| 29 | Estuary Burn | 287.3 | 641.4 | 806.0 | 859.9 | 2180.3 | 2180.3 | 63.9 | 142.5 | 179.1 | 191.1 | 484.5 | 484.5 |
| 30 | Twenty Four Mile | 296.3 | 600.0 | 808.4 | 842.2 | 2276.1 | 2560.6 | 65.8 | 133.3 | 179.6 | 187.2 | 569.0 | 853.5 |
| 31 | Little Hopwood Bu | 0.1 | 646.0 | 835.0 | 828.2 | 1536.4 | 1536.4 | 0.0 | 143.5 | 185.6 | 184.1 | 341.4 | 341.4 |
| 32 | Carsons Creek | 44.5 | 550.2 | 845.3 | 813.4 | 1456.2 | 1638.2 | 9.9 | 122.3 | 187.8 | 180.8 | 364.0 | 546.1 |
| … | | | | | | | | | | | | | |

Table 16. Summary table of the smallest 5-fold cross-validated default catchment minimum flows and catchment default allocation rates (catchments records 286-317). These catchment records are sorted (maximum to minimum) on expected value for minimum flows.

| N | Default: Cross-validation Statistic: Catchment | Minimum Flow 5th percentile (l/s) | Minimum Flow 25th percentile (l/s) | Minimum Flow 50th percentile (l/s) | Minimum Flow Expected value (l/s) | Minimum Flow 75th percentile (l/s) | Minimum Flow 95th percentile (l/s) | Allocation Rate 5th percentile (l/s) | Allocation Rate 25th percentile (l/s) | Allocation Rate 50th percentile (l/s) | Allocation Rate Expected value (l/s) | Allocation Rate 75th percentile (l/s) | Allocation Rate 95th percentile (l/s) |
|---|---|---|---|---|---|---|---|---|---|---|---|---|---|
| … | | | | | | | | | | | | | |
| 286 | Kurinui Creek | 0.1 | 0.1 | 95.6 | 46.2 | 321.3 | 321.3 | 0.0 | 0.0 | 21.3 | 10.3 | 71.4 | 71.4 |
| 287 | Post Office Creek | 0.1 | 0.1 | 60.0 | 45.0 | 324.7 | 324.7 | 0.0 | 0.0 | 13.3 | 10.0 | 72.2 | 72.2 |
| 288 | Waikoura Creek | 0.1 | 0.1 | 39.5 | 43.8 | 442.6 | 442.6 | 0.0 | 0.0 | 8.8 | 9.7 | 98.4 | 98.4 |
| 289 | Pink Gate Creek | 0.1 | 2.9 | 31.4 | 43.4 | 401.0 | 401.0 | 0.0 | 0.6 | 7.0 | 9.6 | 89.1 | 89.1 |
| 290 | Pleasant River | 0.1 | 0.1 | 58.8 | 41.5 | 219.7 | 219.7 | 0.0 | 0.0 | 13.1 | 9.2 | 48.8 | 48.8 |
| 291 | Pringle Road Creek | 0.1 | 1.5 | 27.4 | 38.3 | 307.8 | 307.8 | 0.0 | 0.3 | 6.1 | 8.5 | 68.4 | 68.4 |
| 292 | Awamoko Stream | 0.1 | 5.3 | 40.0 | 37.1 | 347.6 | 347.6 | 0.0 | 1.2 | 8.9 | 8.2 | 77.2 | 77.2 |
| 293 | Bow Alley Creek | 0.1 | 14.7 | 24.9 | 34.9 | 188.5 | 188.5 | 0.0 | 3.3 | 5.5 | 7.8 | 41.9 | 41.9 |
| 294 | Isas Creek | 0.1 | 0.1 | 12.1 | 33.4 | 512.2 | 512.2 | 0.0 | 0.0 | 2.7 | 7.4 | 113.8 | 113.8 |
| 295 | Waitangi Stream | 0.1 | 0.1 | 12.1 | 33.4 | 512.2 | 512.2 | 0.0 | 0.0 | 2.7 | 7.4 | 113.8 | 113.8 |
| 296 | Pillans Stream | 0.1 | 0.1 | 11.7 | 33.2 | 512.2 | 512.2 | 0.0 | 0.0 | 2.6 | 7.4 | 113.8 | 113.8 |
| 297 | Three Brothers Gully Creek | 0.1 | 0.1 | 30.4 | 30.7 | 284.0 | 284.0 | 0.0 | 0.0 | 6.8 | 6.8 | 63.1 | 63.1 |
| 298 | Peaks Road Creek | 0.1 | 0.1 | 27.3 | 22.4 | 476.8 | 476.8 | 0.0 | 0.0 | 6.1 | 5.0 | 106.0 | 106.0 |
| 299 | Oven Hill Creek | 0.1 | 0.9 | 18.2 | 16.0 | 144.1 | 144.1 | 0.0 | 0.2 | 4.0 | 3.6 | 32.0 | 32.0 |
| 300 | Shagree Creek | 0.1 | 1.6 | 14.5 | 15.4 | 86.0 | 86.0 | 0.0 | 0.4 | 3.2 | 3.4 | 19.1 | 19.1 |
| 301 | Hinahina Stream | 0.1 | 0.1 | 8.0 | 14.6 | 190.7 | 190.7 | 0.0 | 0.0 | 1.8 | 3.2 | 42.4 | 42.4 |
| 302 | Oamaru Airport Creek | 0.1 | 0.1 | 86.0 | 14.2 | 517.3 | 517.3 | 0.0 | 0.0 | 19.1 | 3.2 | 114.9 | 114.9 |
| 303 | Coutts Gully Stream | 0.1 | 6.0 | 23.3 | 10.6 | 246.6 | 246.6 | 0.0 | 1.3 | 5.2 | 2.4 | 54.8 | 54.8 |
| 304 | Reids Stream | 0.1 | 0.1 | 10.1 | 4.9 | 83.7 | 83.7 | 0.0 | 0.0 | 2.2 | 1.1 | 18.6 | 18.6 |
| 305 | King Road Creek | 0.1 | 0.1 | 41.4 | 4.7 | 361.0 | 361.0 | 0.0 | 0.0 | 9.2 | 1.0 | 80.2 | 80.2 |
| 306 | Allangrange (S) | 0.1 | 0.1 | 10.4 | 3.7 | 71.7 | 71.7 | 0.0 | 0.0 | 2.3 | 0.8 | 15.9 | 15.9 |
| 307 | Kakaho Creek | 0.1 | 0.1 | 8.0 | 2.9 | 364.6 | 364.6 | 0.0 | 0.0 | 1.8 | 0.7 | 81.0 | 81.0 |
| 308 | Allangrange (N) | 0.1 | 2.4 | 16.2 | 1.4 | 80.4 | 80.4 | 0.0 | 0.5 | 3.6 | 0.3 | 17.9 | 17.9 |
| 309 | Aitchison Road Creek | 0.1 | 0.1 | 9.0 | 0.1 | 417.9 | 417.9 | 0.0 | 0.0 | 2.0 | 0.0 | 92.9 | 92.9 |
| 310 | Clydevale Creek | 0.1 | 0.1 | 5.3 | 0.1 | 53.6 | 53.6 | 0.0 | 0.0 | 1.2 | 0.0 | 11.9 | 11.9 |
| 311 | Glen Creek | 0.1 | 0.1 | 19.8 | 0.1 | 461.6 | 461.6 | 0.0 | 0.0 | 4.4 | 0.0 | 102.6 | 102.6 |
| 312 | Hall Road Creek | 0.1 | 0.1 | 10.4 | 0.1 | 93.8 | 93.8 | 0.0 | 0.0 | 2.3 | 0.0 | 20.8 | 20.8 |
| 313 | Jones Creek | 0.1 | 0.1 | 33.6 | 0.1 | 263.0 | 263.0 | 0.0 | 0.0 | 7.5 | 0.0 | 58.4 | 58.4 |
| 314 | Oamaru North Creek | 0.1 | 0.1 | 19.8 | 0.1 | 461.6 | 461.6 | 0.0 | 0.0 | 4.4 | 0.0 | 102.6 | 102.6 |
| 315 | Otara Stream | 0.1 | 0.1 | 30.0 | 0.1 | 175.1 | 175.1 | 0.0 | 0.0 | 6.7 | 0.0 | 38.9 | 38.9 |
| 316 | Pannetts Road Creek | 0.1 | 0.1 | 8.2 | 0.1 | 53.6 | 53.6 | 0.0 | 0.0 | 1.8 | 0.0 | 11.9 | 11.9 |
| 317 | Welcome Creek | 0.1 | 0.1 | 23.7 | 0.1 | 461.6 | 461.6 | 0.0 | 0.0 | 5.3 | 0.0 | 102.6 | 102.6 |

Table 17. Summary table of probable status for 317 gauged catchments across the Otago region.

| | Cross Validated Flow Statistic | | | | | |
|---|---|---|---|---|---|---|
| Catchment status | 5th percentile | 25th percentile | Expected value | 50th percentile | 75th percentile | 95th percentile |
| Over allocated | 73 | 57 | 46 | 44 | 23 | 22 |
| Under allocated | 244 | 260 | 271 | 273 | 294 | 295 |
| Total | 317 | 317 | 317 | 317 | 317 | 317 |

Table 18. Summary table of probable allocation status for the first 75 of 317 gauged catchments across the Otago Region. Catchment allocation status: 1 = over-allocated and, 0 = under-allocated.

| N | Management Unit | Rhoe | Default: Cross-validation Statistic: Catchment | Allocation status 5th percentile | Allocation status 25th percentile | Allocation status Expected value | Allocation status 50th percentile | Allocation status 75th percentile | Allocation status 95th percentile | Easting | Northing | Area km2 | Stream Order |
|---|---|---|---|---|---|---|---|---|---|---|---|---|---|
| 1 | Clutha-Mata Au | Dunstan | Albert Burn (1) | 1 | 1 | 1 | 1 | 0 | 0 | 1310781.36 | 5027425.06 | 11.52 | 3 |
| 2 | Clutha-Mata Au | Upper Lakes | Alpha Burn | 1 | 0 | 0 | 0 | 0 | 0 | 1284418.29 | 5045673.51 | 17.31 | 3 |
| 3 | Clutha-Mata Au | Dunstan | Amisfield Burn | 1 | 1 | 0 | 0 | 0 | 0 | 1306055.07 | 5016661.04 | 29.16 | 3 |
| 4 | Clutha-Mata Au | Dunstan | Arrow River | 1 | 1 | 1 | 1 | 1 | 1 | 1275397.30 | 5008048.75 | 242.63 | 5 |
| 5 | North Otago | North Otago | Awamoa Creek | 1 | 1 | 1 | 1 | 0 | 0 | 1437653.44 | 4999144.40 | 23.42 | 3 |
| 6 | North Otago | North Otago | Awamoko Stream | 1 | 1 | 1 | 1 | 0 | 0 | 1432317.50 | 5026792.80 | 110.75 | 5 |
| 7 | Clutha-Mata Au | Dunstan | Bannock Burn | 1 | 1 | 1 | 1 | 1 | 1 | 1298892.93 | 5000161.60 | 91.78 | 4 |
| 8 | Clutha-Mata Au | Dunstan | Basin Burn | 1 | 1 | 1 | 1 | 1 | 1 | 1308548.91 | 5020686.30 | 24.24 | 4 |
| 9 | Clutha-Mata Au | Dunstan | Bendigo Creek | 1 | 1 | 1 | 1 | 0 | 0 | 1308874.46 | 5019769.45 | 51.01 | 4 |
| 10 | Clutha-Mata Au | Roxburgh | Benger Burn | 1 | 1 | 1 | 1 | 1 | 1 | 1317989.01 | 4939072.63 | 134.79 | 5 |
| 11 | North Otago | North Otago | Bow Alley Creek | 1 | 1 | 1 | 1 | 0 | 0 | 1432536.12 | 4988789.89 | 18.34 | 3 |
| 12 | Clutha-Mata Au | Dunstan | Burn Cottage Creek | 1 | 1 | 0 | 0 | 0 | 0 | 1301107.67 | 5007147.73 | 11.25 | 3 |
| 13 | Clutha-Mata Au | Roxburgh | Butchers Creek (1) | 1 | 1 | 1 | 1 | 1 | 1 | 1315000.88 | 4978593.16 | 35.24 | 4 |
| 14 | Clutha-Mata Au | Roxburgh | Butchers Creek (2) | 1 | 1 | 0 | 0 | 0 | 0 | 1311000.74 | 4954547.44 | 13.20 | 3 |
| 15 | Clutha-Mata Au | Dunstan | Camp Creek (1) | 1 | 1 | 1 | 0 | 0 | 0 | 1281515.96 | 5006940.00 | 15.28 | 3 |
| 16 | Clutha-Mata Au | Dunstan | Campbells Creek | 1 | 0 | 0 | 0 | 0 | 0 | 1302014.35 | 5004805.78 | 10.58 | 3 |
| 17 | Clutha-Mata Au | Dunstan | Cardrona River | 1 | 1 | 1 | 1 | 1 | 1 | 1298778.35 | 5044980.10 | 345.03 | 5 |
| 18 | Clutha-Mata Au | Roxburgh | Coal Creek (1) | 1 | 1 | 1 | 1 | 1 | 1 | 1310597.00 | 4968371.53 | 48.61 | 4 |
| 19 | Clutha-Mata Au | Roxburgh | Coal Creek (2) | 1 | 1 | 1 | 1 | 1 | 1 | 1311789.35 | 4956571.25 | 22.44 | 4 |
| 20 | Clutha-Mata Au | Dunstan | Dead Horse Creek | 1 | 0 | 0 | 0 | 0 | 0 | 1305649.37 | 5038963.13 | 12.88 | 3 |
| 21 | Clutha-Mata Au | Upper Lakes | Dinner Creek | 1 | 0 | 0 | 0 | 0 | 0 | 1301757.16 | 5067444.27 | 4.45 | 2 |
| 22 | Clutha-Mata Au | Roxburgh | Elbow Creek | 1 | 1 | 1 | 1 | 0 | 0 | 1311855.28 | 4961110.07 | 10.88 | 3 |
| 23 | Clutha-Mata Au | Upper Lakes | Five Mile Creek (2) | 1 | 0 | 0 | 0 | 0 | 0 | 1253705.59 | 5001321.52 | 5.88 | 2 |
| 24 | Clutha-Mata Au | Dunstan | Franks Creek | 1 | 0 | 0 | 0 | 0 | 0 | 1282046.30 | 5006858.98 | 3.71 | 3 |
| 25 | Clutha-Mata Au | Roxburgh | Fraser River | 1 | 1 | 1 | 1 | 1 | 1 | 1314454.77 | 4983038.14 | 314.98 | 5 |
| 26 | Clutha-Mata Au | Dunstan | Gentle Annie Creek | 1 | 1 | 1 | 1 | 0 | 0 | 1288291.47 | 5005966.32 | 21.57 | 4 |
| 27 | Clutha-Mata Au | Dunstan | Hayes Creek | 1 | 1 | 1 | 1 | 0 | 0 | 1269272.75 | 5008035.97 | 58.67 | 5 |
| 28 | Clutha-Mata Au | Dunstan | John Bull Creek | 1 | 1 | 1 | 1 | 0 | 0 | 1303845.19 | 5011094.39 | 22.95 | 3 |
| 29 | Clutha-Mata Au | Lower Clutha | Kaihiku Stream | 1 | 0 | 0 | 0 | 0 | 0 | 1342341.83 | 4880125.99 | 157.73 | 5 |
| 30 | North Otago | North Otago | Kakanui River | 1 | 1 | 1 | 1 | 1 | 1 | 1434922.60 | 4993945.20 | 893.71 | 5 |
| 31 | Clutha-Mata Au | Dunstan | Kingston Road Creek | 1 | 0 | 0 | 0 | 0 | 0 | 1265002.11 | 5004365.36 | 22.71 | 3 |
| 32 | Clutha-Mata Au | Upper Lakes | Lake Dispute | 1 | 0 | 0 | 0 | 0 | 0 | 1249609.85 | 5000274.80 | 5.48 | 3 |
| 33 | North Otago | North Otago | Landon Creek | 1 | 0 | 0 | 0 | 0 | 0 | 1444016.56 | 5008012.78 | 11.95 | 3 |
| 34 | Clutha-Mata Au | Dunstan | Lindis River | 1 | 1 | 1 | 1 | 1 | 1 | 1310932.54 | 5024201.24 | 1038.44 | 6 |
| 35 | Clutha-Mata Au | Dunstan | Locharburn | 1 | 1 | 0 | 0 | 0 | 0 | 1307993.66 | 5021999.98 | 7.89 | 3 |
| 36 | Clutha-Mata Au | Dunstan | Long Gully Creek (1) | 1 | 0 | 0 | 0 | 0 | 0 | 1295946.98 | 5002626.46 | 23.09 | 4 |
| 37 | Clutha-Mata Au | Dunstan | Low Burn (2) | 1 | 1 | 1 | 1 | 1 | 1 | 1301528.95 | 5009958.24 | 51.41 | 4 |
| 38 | Clutha-Mata Au | Dunstan | Luggate Creek | 1 | 1 | 1 | 1 | 1 | 0 | 1305573.88 | 5038927.11 | 127.66 | 5 |
| 39 | Clutha-Mata Au | Manuherekia | Manuherikia River | 1 | 1 | 1 | 1 | 1 | 1 | 1317092.53 | 4981863.30 | 3033.60 | 7 |
| 40 | Clutha-Mata Au | Dunstan | Mt Pisa Creek | 1 | 1 | 0 | 0 | 0 | 0 | 1305452.83 | 5017891.11 | 2.95 | 3 |
| 41 | Dunedin & Coast | Dunedin & Coast | Orokonui Creek | 1 | 0 | 0 | 0 | 0 | 0 | 1412288.03 | 4930426.72 | 4.28 | 2 |
| 42 | Clutha-Mata Au | Dunstan | Pipeclay Gully Creek | 1 | 1 | 1 | 0 | 0 | 0 | 1297292.62 | 5001395.67 | 9.80 | 3 |
| 43 | North Otago | North Otago | Pleasant River | 1 | 1 | 1 | 1 | 1 | 1 | 1422602.85 | 4951250.07 | 127.98 | 5 |
| 44 | Clutha-Mata Au | Dunstan | Poison Creek | 1 | 1 | 1 | 1 | 0 | 0 | 1310622.37 | 5032748.34 | 7.46 | 3 |
| 45 | Clutha-Mata Au | Lower Clutha | Pomahaka River | 1 | 1 | 1 | 1 | 0 | 0 | 1334780.81 | 4883515.69 | 1952.29 | 6 |
| 46 | Catlins | Catlins | Puerua River | 1 | 0 | 0 | 0 | 0 | 0 | 1352394.22 | 4860880.34 | 205.53 | 4 |
| 47 | Clutha-Mata Au | Dunstan | Quartz Reef Creek | 1 | 1 | 1 | 1 | 0 | 0 | 1303067.00 | 5009345.08 | 32.09 | 3 |
| 48 | Clutha-Mata Au | Dunstan | Rastus Burn | 1 | 1 | 1 | 1 | 0 | 0 | 1269669.87 | 5007946.77 | 14.27 | 3 |
| 49 | Clutha-Mata Au | Dunstan | Roaring Meg | 1 | 1 | 1 | 1 | 1 | 1 | 1290284.41 | 5009429.00 | 128.69 | 4 |
| 50 | Clutha-Mata Au | Upper Lakes | Roys Peak Creek | 1 | 1 | 0 | 0 | 0 | 0 | 1289826.36 | 5045848.03 | 3.73 | 2 |
| 51 | Clutha-Mata Au | Dunstan | School Creek | 1 | 1 | 0 | 0 | 0 | 0 | 1309309.21 | 5018809.21 | 5.99 | 4 |
| 52 | Clutha-Mata Au | Dunstan | Schoolhouse Creek | 1 | 1 | 1 | 1 | 0 | 0 | 1311150.25 | 5025644.28 | 6.67 | 3 |
| 53 | Clutha-Mata Au | Dunstan | Scrubby Stream | 1 | 1 | 1 | 1 | 0 | 0 | 1293595.87 | 5006364.85 | 5.13 | 2 |
| 54 | Clutha-Mata Au | Upper Lakes | Seven Mile Creek | 1 | 0 | 0 | 0 | 0 | 0 | 1252764.91 | 5001009.19 | 3.17 | 2 |
| 55 | North Otago | North Otago | Shag River | 1 | 1 | 1 | 1 | 0 | 0 | 1429514.73 | 4961507.05 | 543.19 | 5 |
| 56 | Clutha-Mata Au | Roxburgh | Shingle Creek | 1 | 1 | 1 | 1 | 1 | 1 | 1311473.05 | 4963596.15 | 34.76 | 3 |
| 57 | Taieri | Taieri | Taieri River | 1 | 1 | 1 | 1 | 1 | 1 | 1383545.37 | 4896144.25 | 5704.78 | 7 |
| 58 | Clutha-Mata Au | Roxburgh | Teviot River | 1 | 1 | 1 | 1 | 1 | 1 | 1312582.37 | 4950505.94 | 329.77 | 5 |
| 59 | Dunedin & Coast | Dunedin & Coast | Thomson Creek | 1 | 1 | 0 | 0 | 0 | 0 | 1413992.92 | 4923131.33 | 7.25 | 3 |
| 60 | Clutha-Mata Au | Roxburgh | Tima Burn | 1 | 1 | 1 | 1 | 0 | 0 | 1319265.70 | 4938052.03 | 44.17 | 4 |
| 61 | Clutha-Mata Au | Dunstan | Tinwald Burn | 1 | 1 | 1 | 1 | 1 | 1 | 1307577.31 | 5018854.29 | 18.20 | 3 |
| 62 | Dunedin & Coast | Dunedin & Coast | Tokomairiro River | 1 | 1 | 0 | 0 | 0 | 0 | 1372228.88 | 4877426.27 | 395.70 | 6 |
| 63 | Clutha-Mata Au | Dunstan | Toms Creek | 1 | 1 | 1 | 1 | 0 | 0 | 1279059.79 | 5007790.14 | 6.52 | 3 |
| 64 | North Otago | North Otago | Trotters Creek | 1 | 1 | 0 | 0 | 0 | 0 | 1431478.13 | 4970921.71 | 32.54 | 3 |
| 65 | North Otago | North Otago | Waianakarua River | 1 | 1 | 1 | 1 | 1 | 1 | 1431429.20 | 4986733.44 | 260.72 | 5 |
| 66 | Clutha-Mata Au | Roxburgh | Waikerikeri Creek | 1 | 0 | 0 | 0 | 0 | 0 | 1312665.42 | 4987333.91 | 39.51 | 4 |
| 67 | North Otago | North Otago | Waikouaiti River | 1 | 1 | 1 | 1 | 0 | 0 | 1417705.34 | 4943253.10 | 426.36 | 5 |
| 68 | Clutha-Mata Au | Lower Clutha | Waitahuna River | 1 | 1 | 0 | 0 | 0 | 0 | 1336279.50 | 4883279.83 | 406.46 | 5 |
| 69 | Dunedin & Coast | Dunedin & Coast | Waitati River | 1 | 1 | 1 | 1 | 0 | 0 | 1411623.24 | 4930659.39 | 46.26 | 4 |
| 70 | Clutha-Mata Au | Lower Clutha | Waiwera River | 1 | 1 | 1 | 1 | 0 | 0 | 1334989.63 | 4883143.36 | 208.94 | 5 |
| 71 | Dunedin & Coast | Dunedin & Coast | Water of Leith | 1 | 1 | 1 | 1 | 1 | 1 | 1407853.53 | 4917333.17 | 48.68 | 4 |
| 72 | Clutha-Mata Au | Upper Lakes | Waterfall Creek (1) | 1 | 0 | 0 | 0 | 0 | 0 | 1290687.20 | 5044457.18 | 8.94 | 2 |
| 73 | North Otago | North Otago | Welcome Creek | 1 | 1 | 1 | 1 | 1 | 1 | 1449887.53 | 5022882.86 | 19.79 | 2 |

## List of Figures

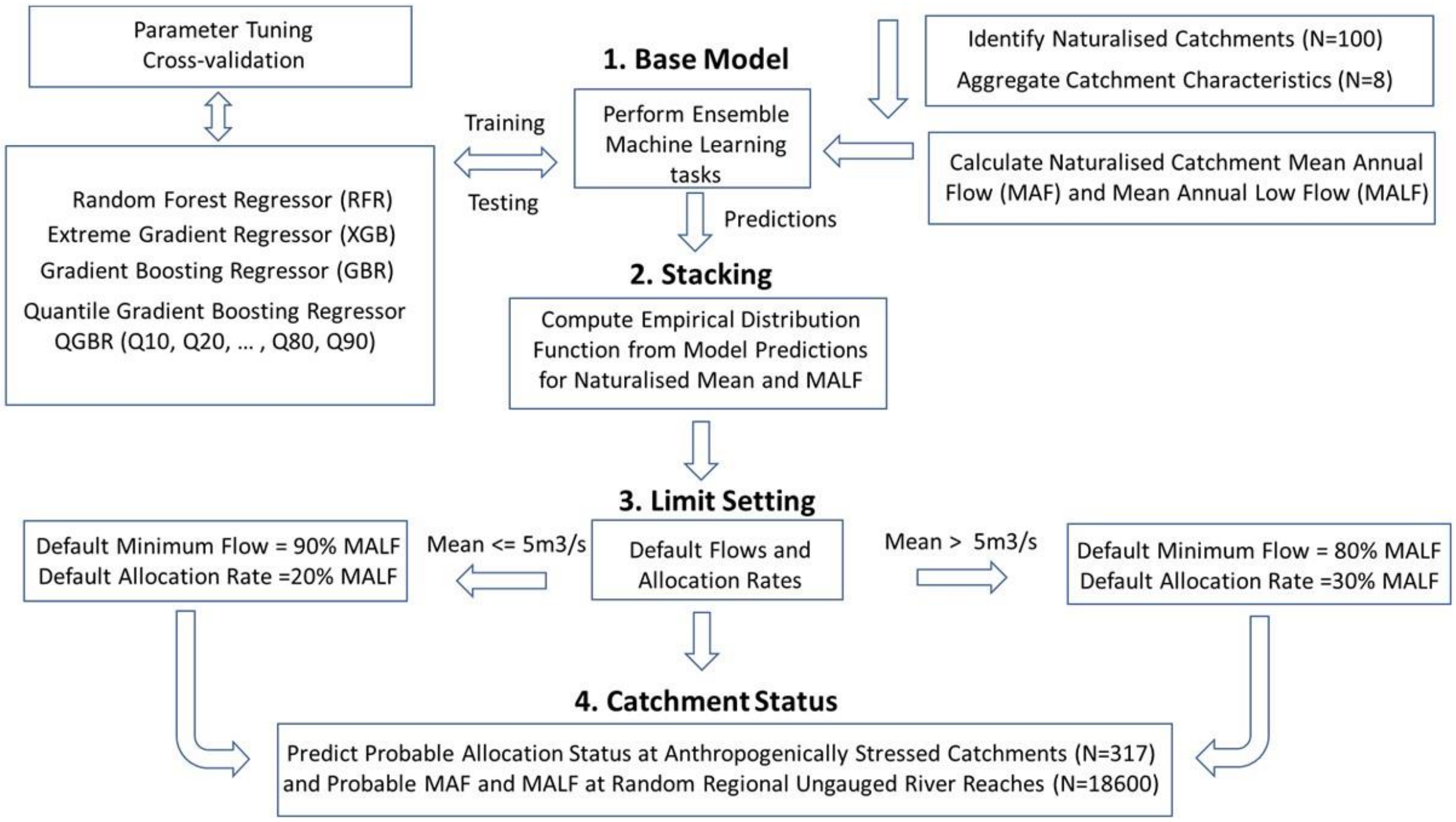

Fig 1. Flowchart illustrating the stacked ensemble machine learning model used to predict probable naturalized hydrology and allocation status across gauged catchments and ungauged reaches in the Otago Region, New Zealand.

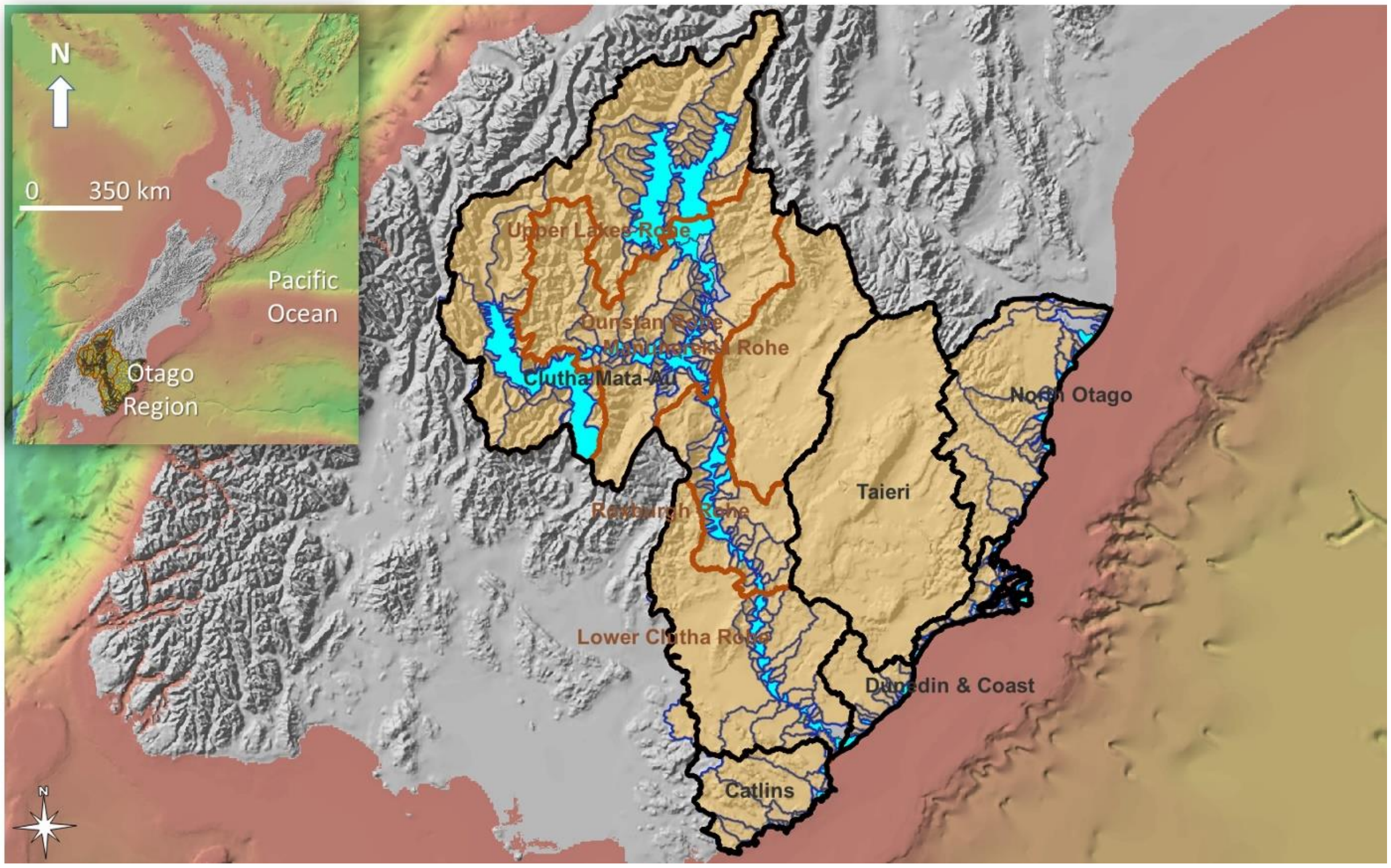

Fig. 2. Location map showing water management regions across the Otago Region, New Zealand. The region has 5 Freshwater Management Units (outlined and labeled in black) that include the Clutha (Mata-Au), Catlins, Dunedin & Coast, North Otago and Taieri. The Clutha comprises 5 smaller indigenous (iwi) management units (outlined and labeled in brown) called Rohe that include the Dunstan, Lower Clutha, Manuherekia, Roxburgh, and Upper Lakes.

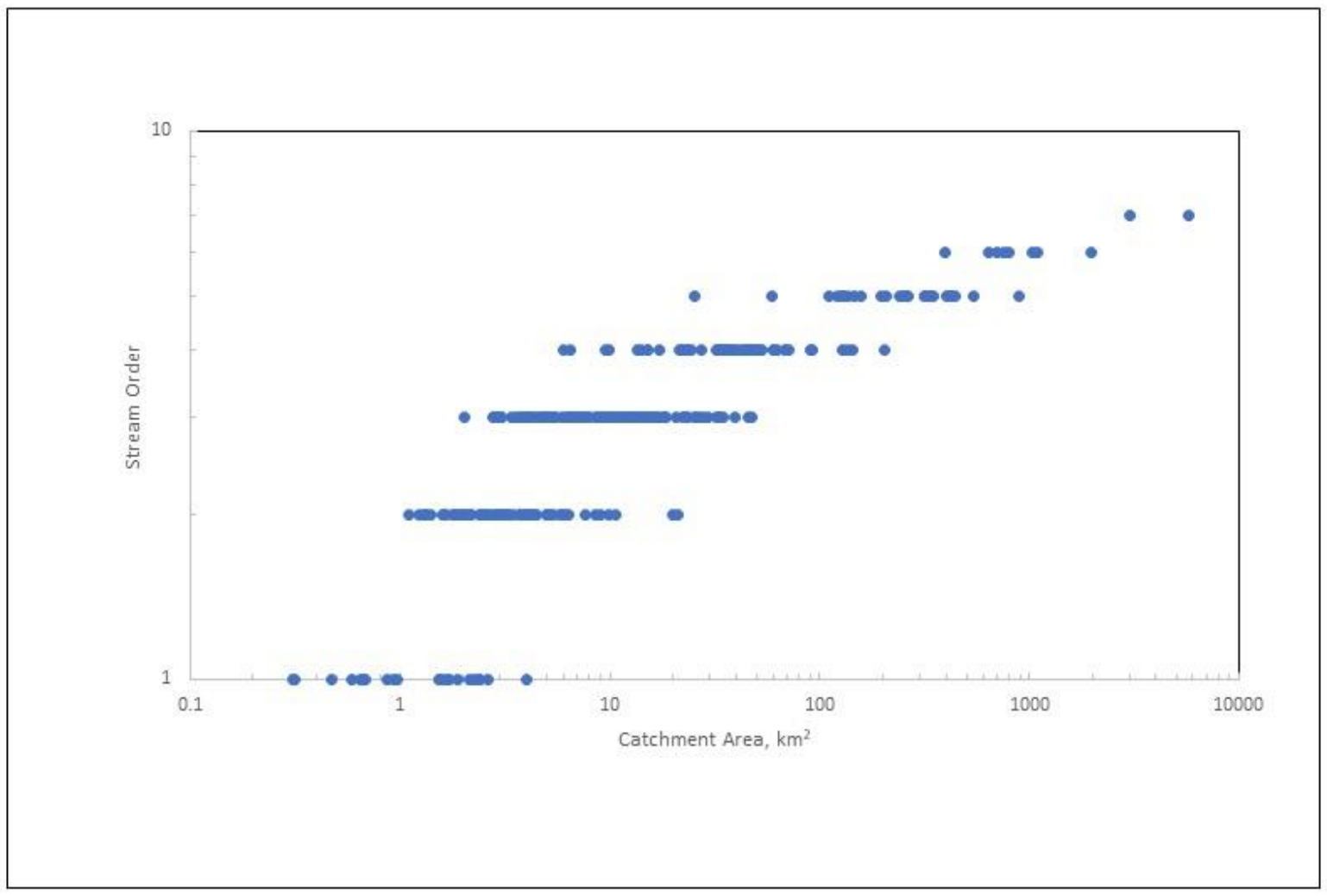

Fig. 3. Plot showing the distribution of natural streamflow gauging stations (blue dots) with respect to the Strahler stream order and catchment area in the Otago Region, New Zealand.

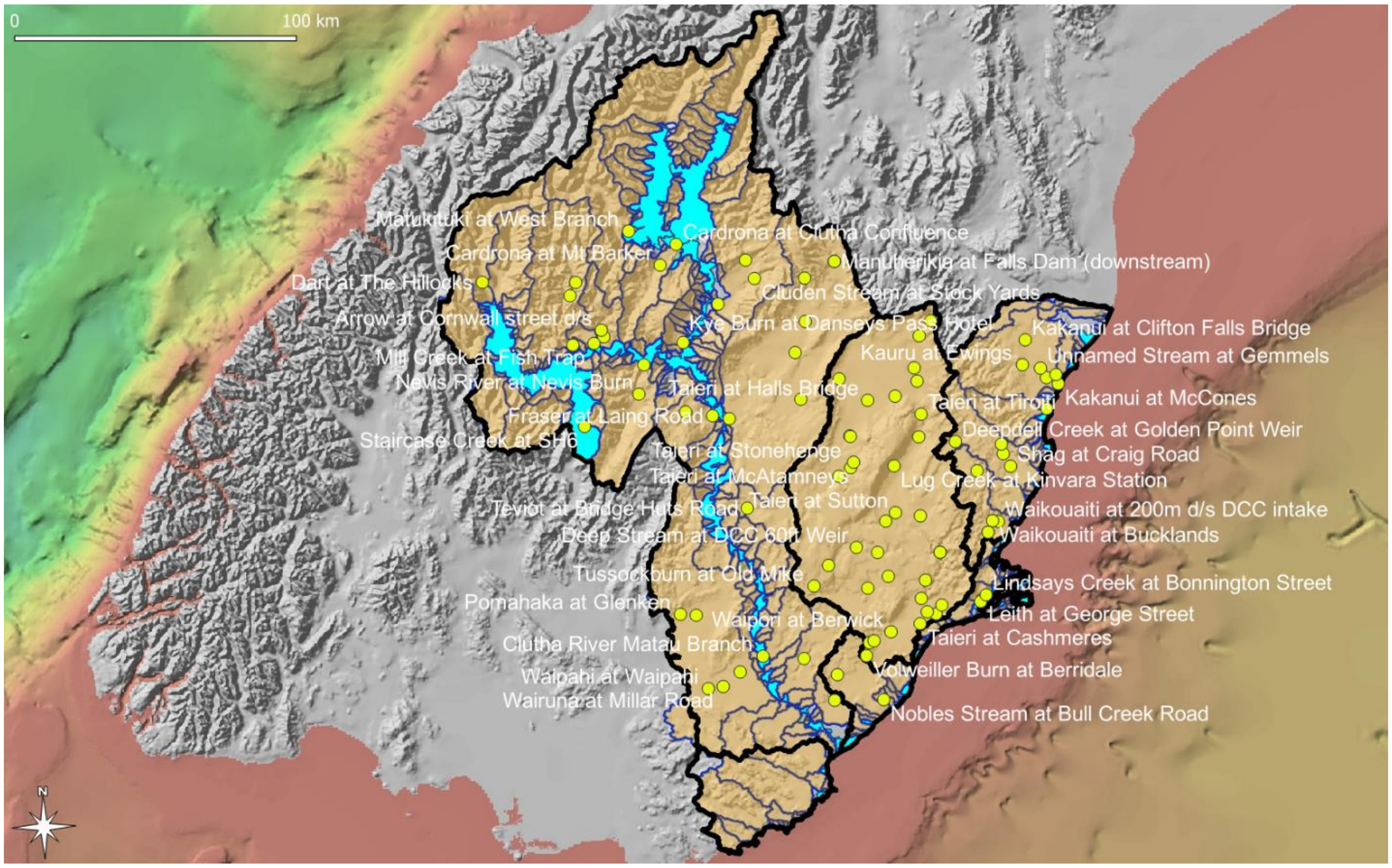

Fig. 4. Location map showing names (white text) of 100 gauging stations (yellow dots) that recorded natural flows across Otago, New Zealand.

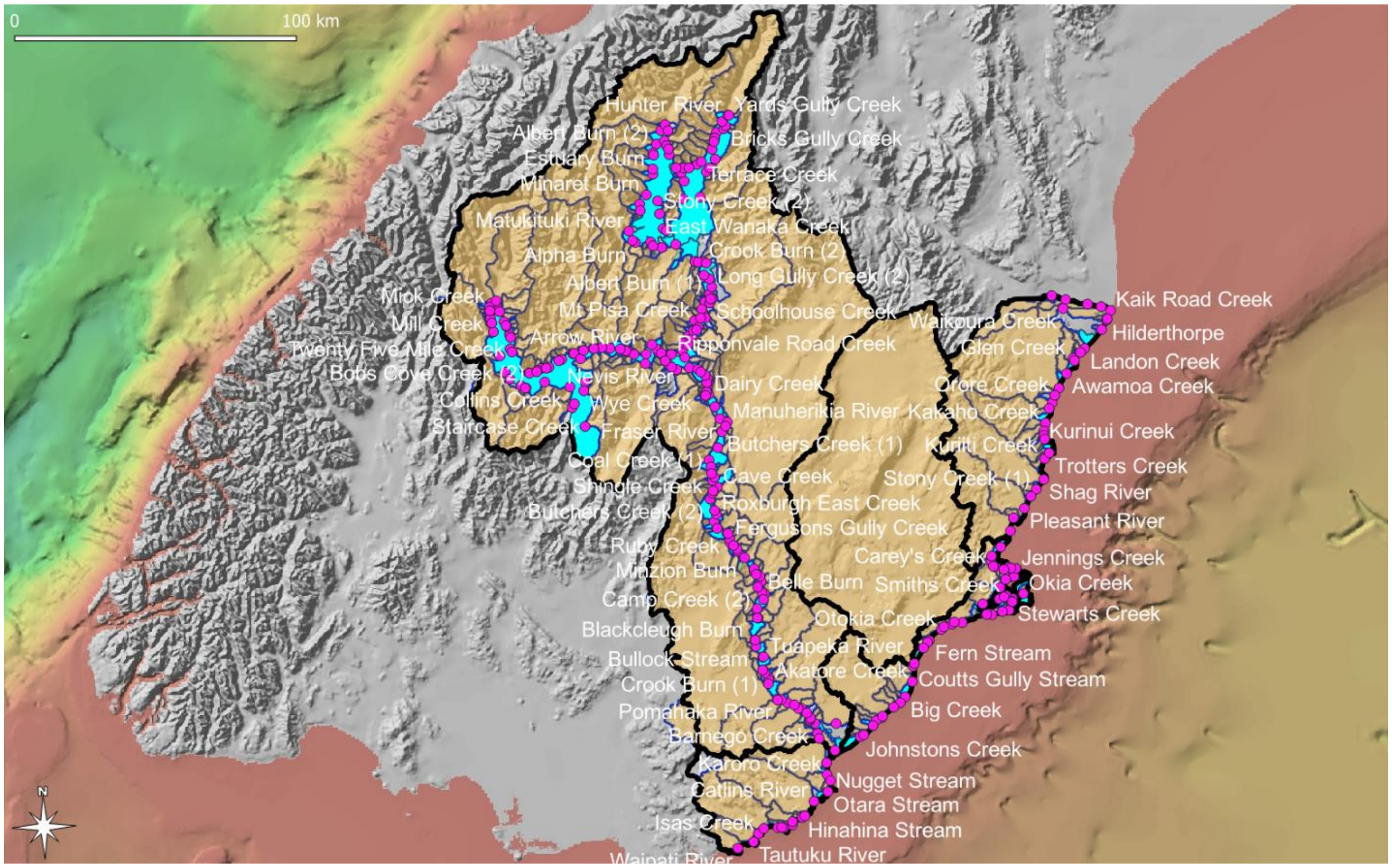

Fig. 5. Location map of 317 regulated gauged catchments where the naturalized annual mean flow and 7-day mean annual low flow, and catchment allocation status are predicted at 5$^{th}$, 25$^{th}$, 50$^{th}$, 75$^{th}$, and 95$^{th}$ percentiles across the Otago Region, New Zealand. The black outlines are the catchment boundaries and purple dots are the streamflow gauge stations.

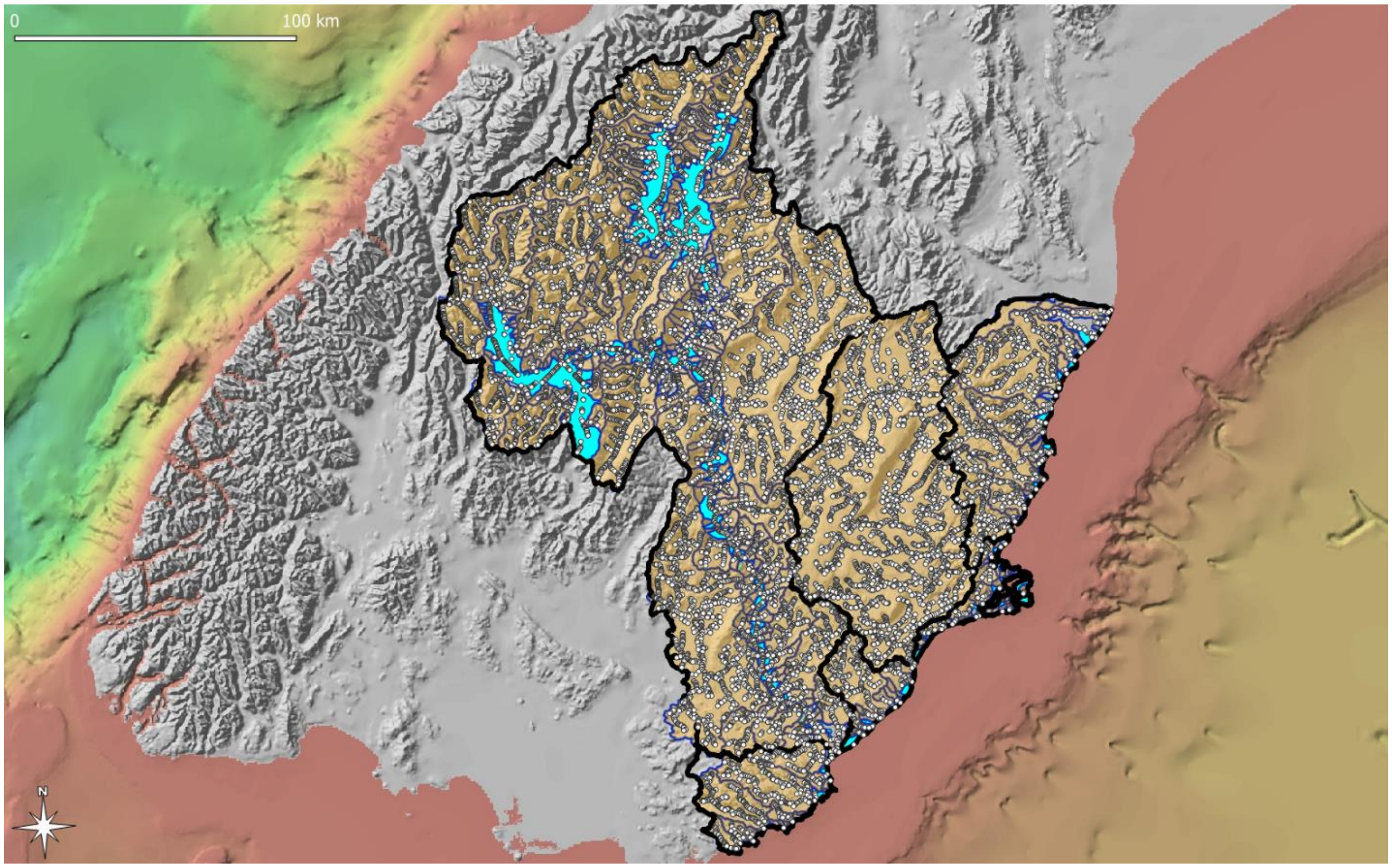

Fig. 6. Location map of 18,612 ungauged river reaches where the naturalized mean daily flow and 7-day mean annual low flow are predicted at 5th, 25th, 50th, 75th, and 95th percentiles across the Otago Region, New Zealand.

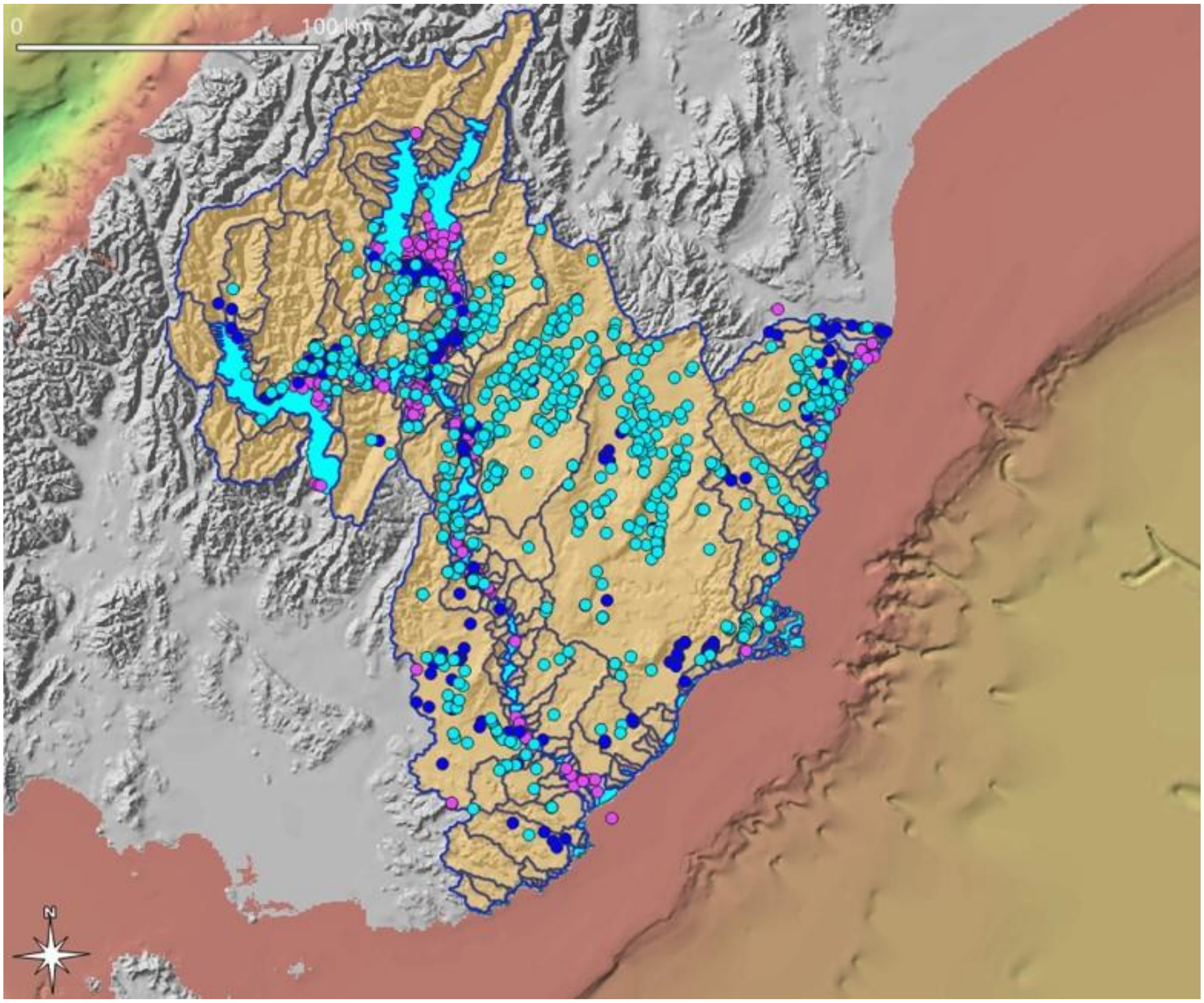

Fig. 7. Location map of surface water (light blue), groundwater (dark blue), and undifferentiated consented abstraction points (also called takes) across the Otago Region, New Zealand. The groundwater and undifferentiated (purple) takes are assumed to be hydrologically connected to the adjacent stream.

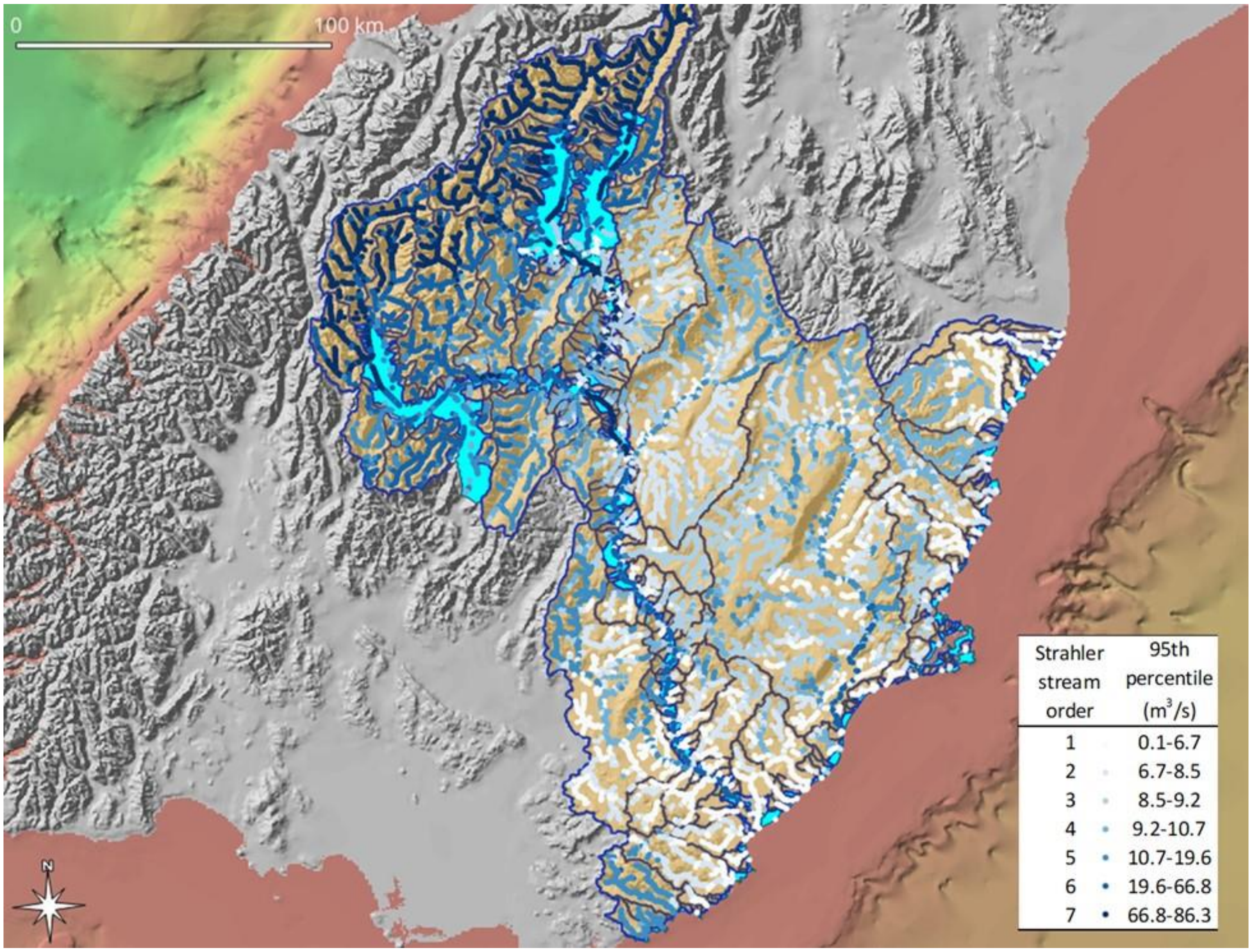

8(a)

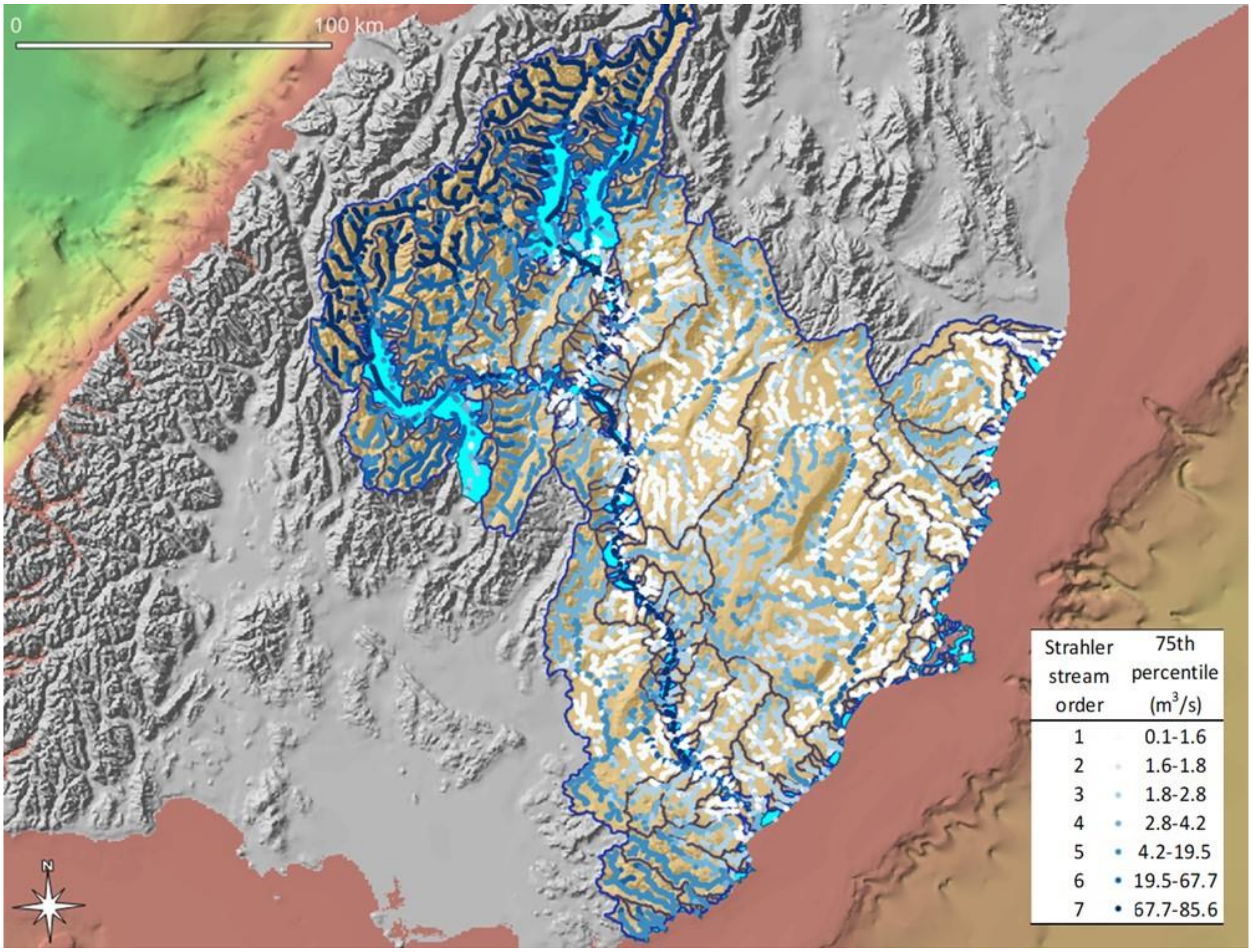

8(b)

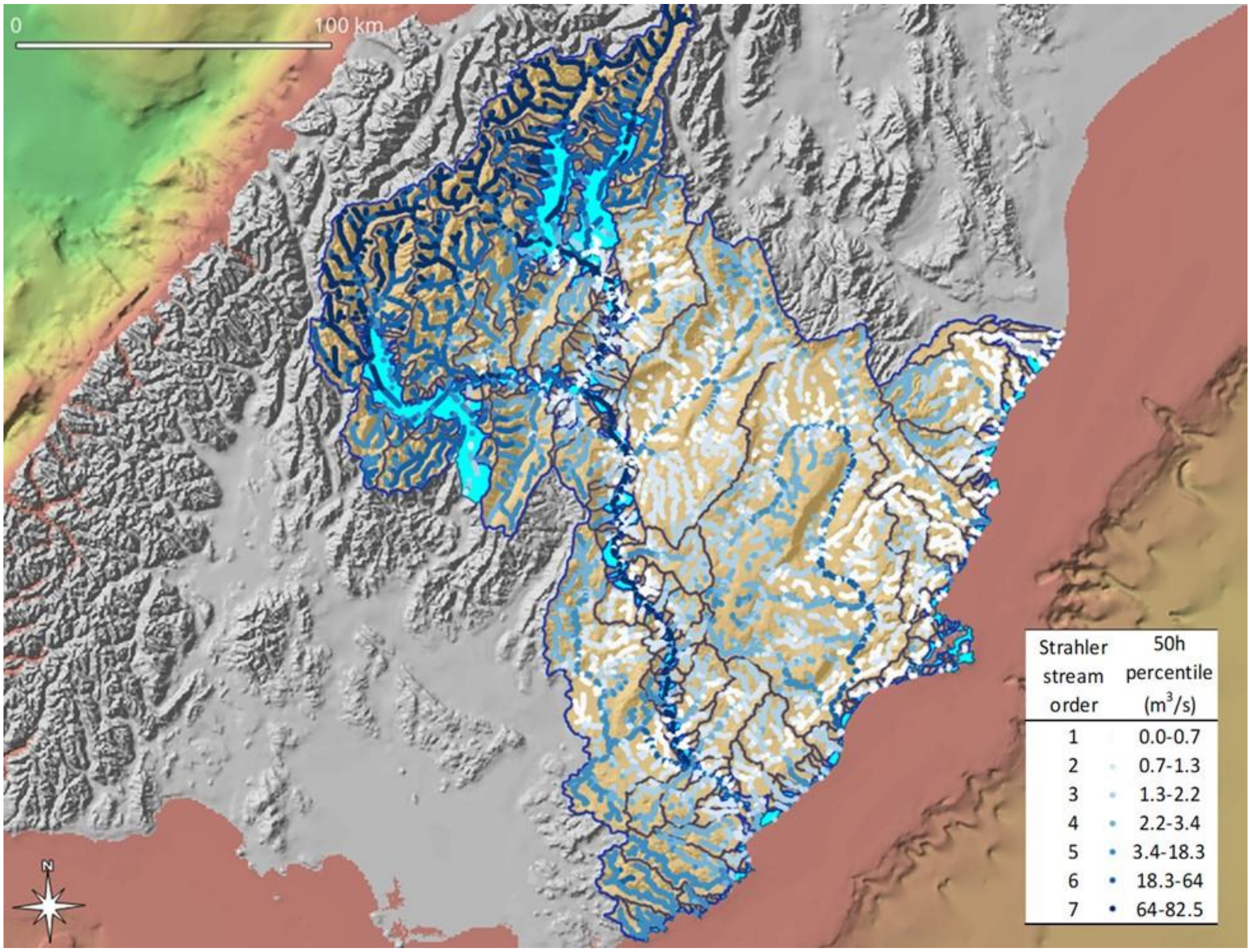

8(c)

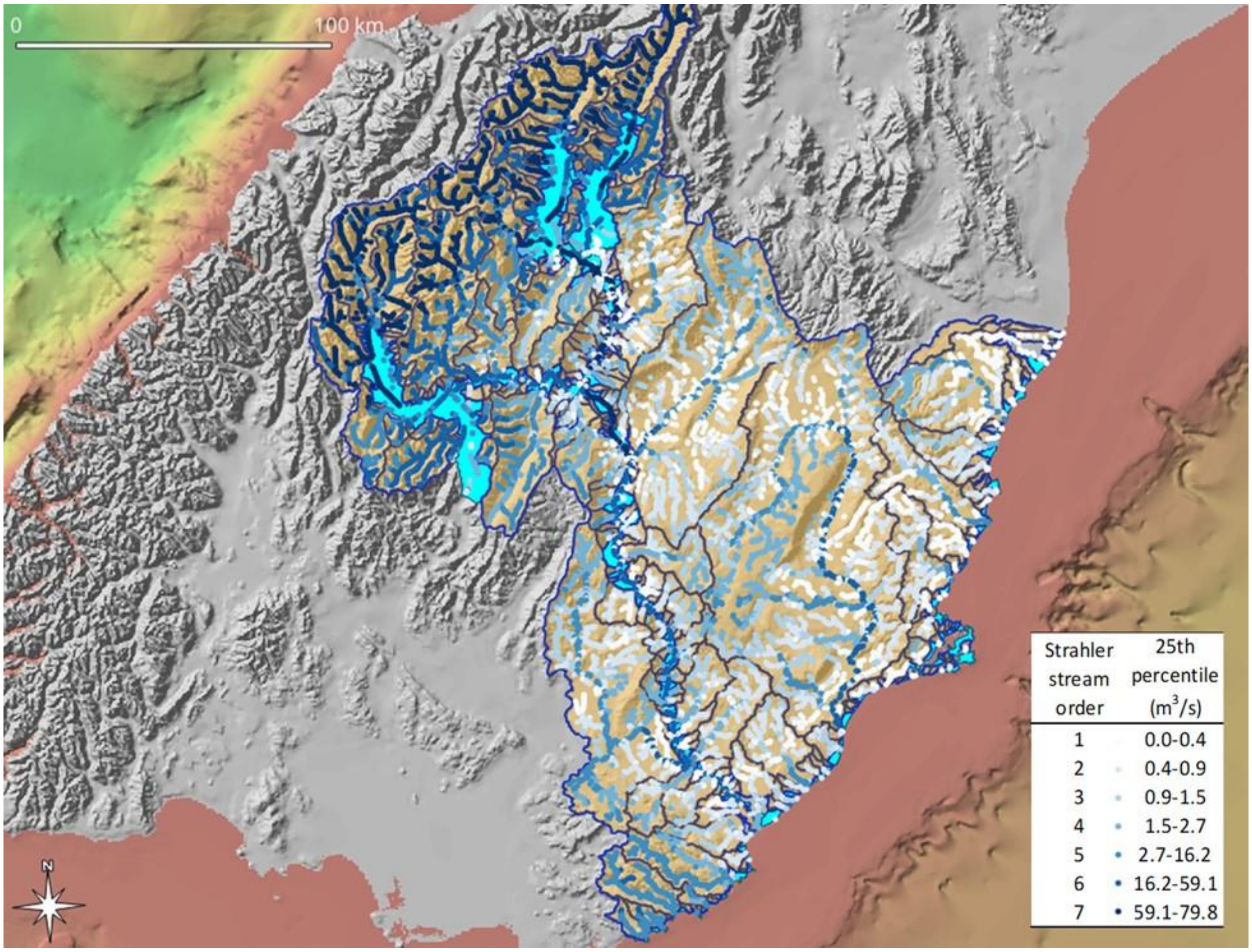

8(d)

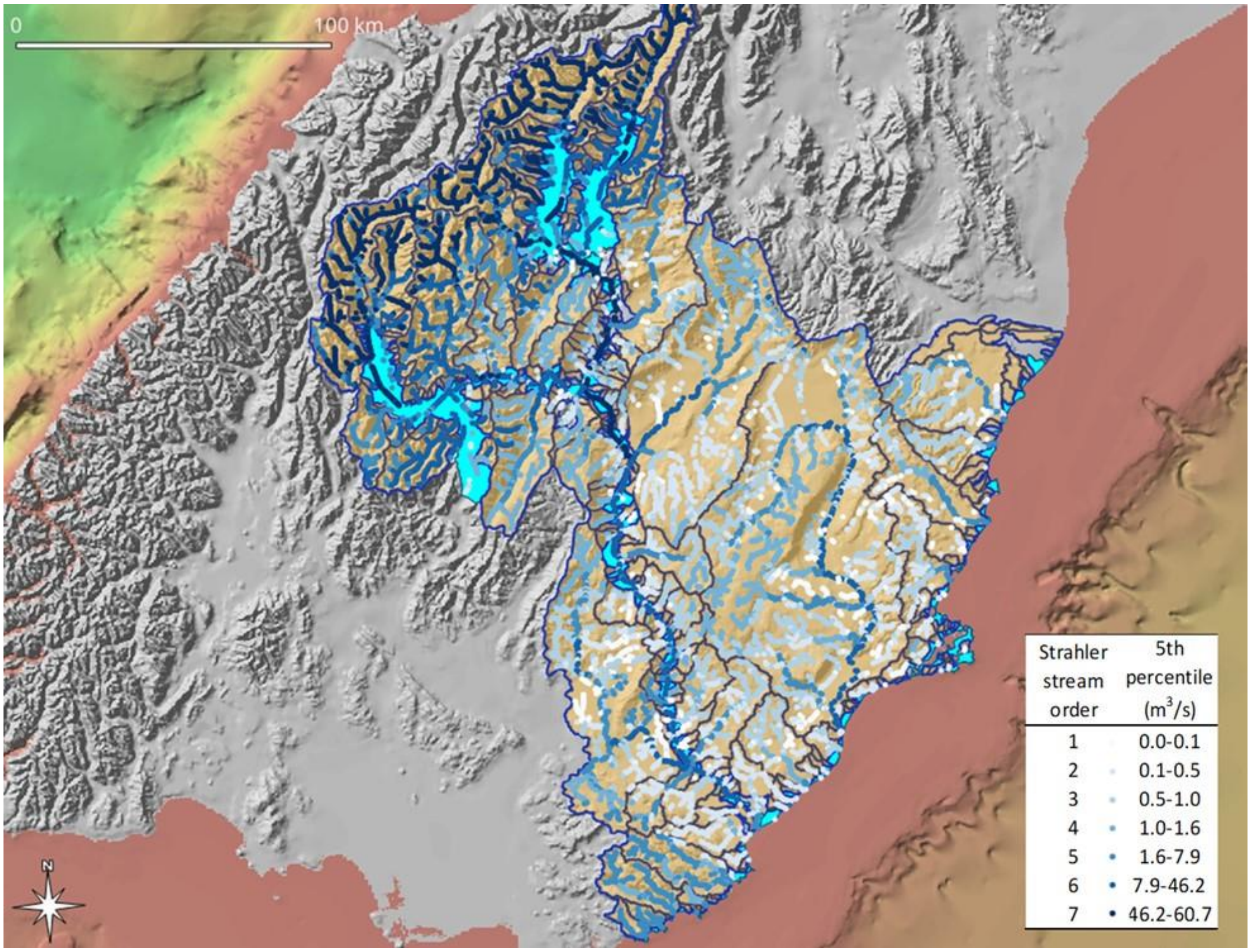

8(e)

Fig. 8. Probable naturalized Mean daily flow predictions at 18,612 ungauged sites across the Otago Region: (a) 95$^{th}$ percentile, (b) 75$^{th}$ percentile, (c) 50th percentile, (d) 25th percentile, and (e) 5$^{th}$ percentile.

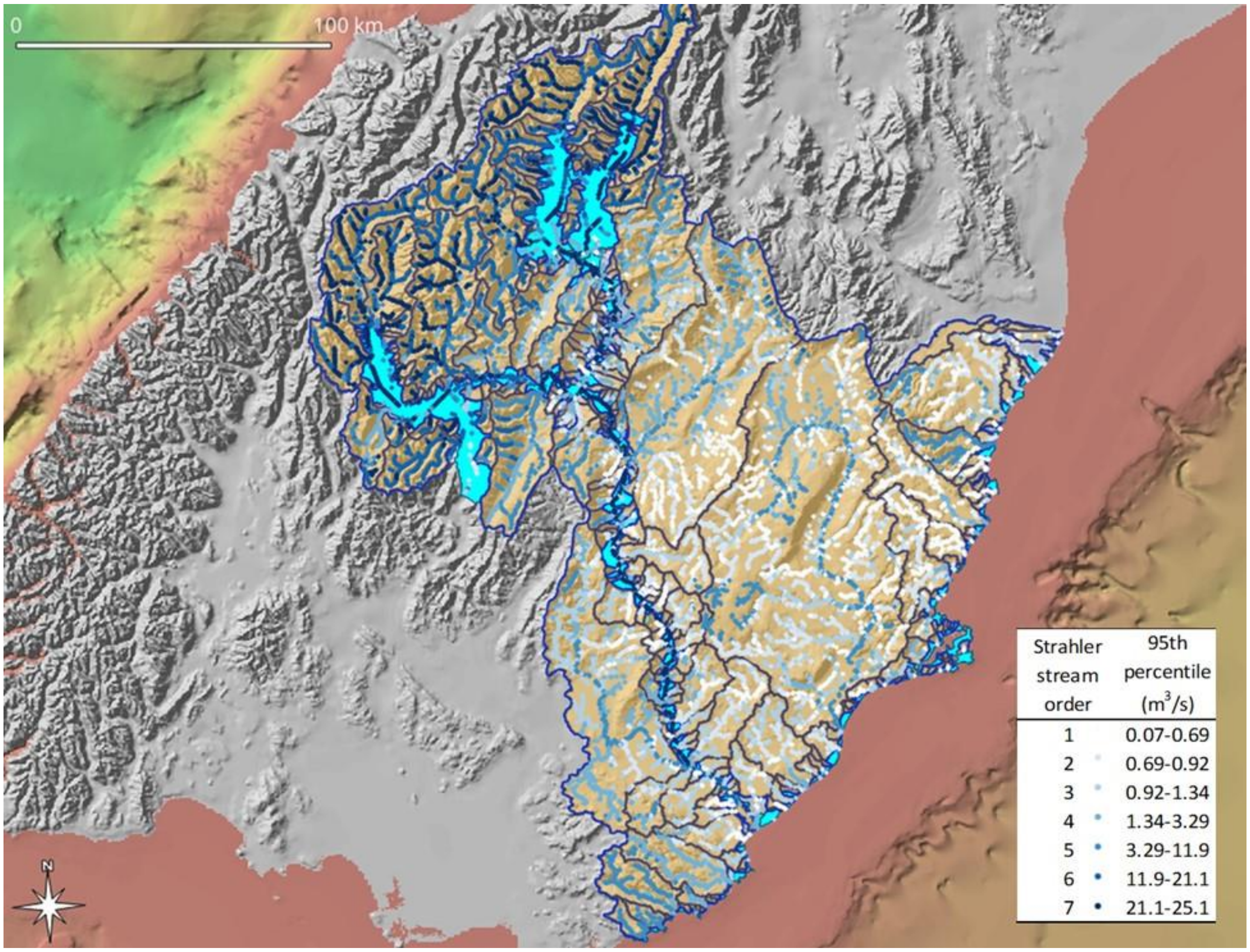

9(a)

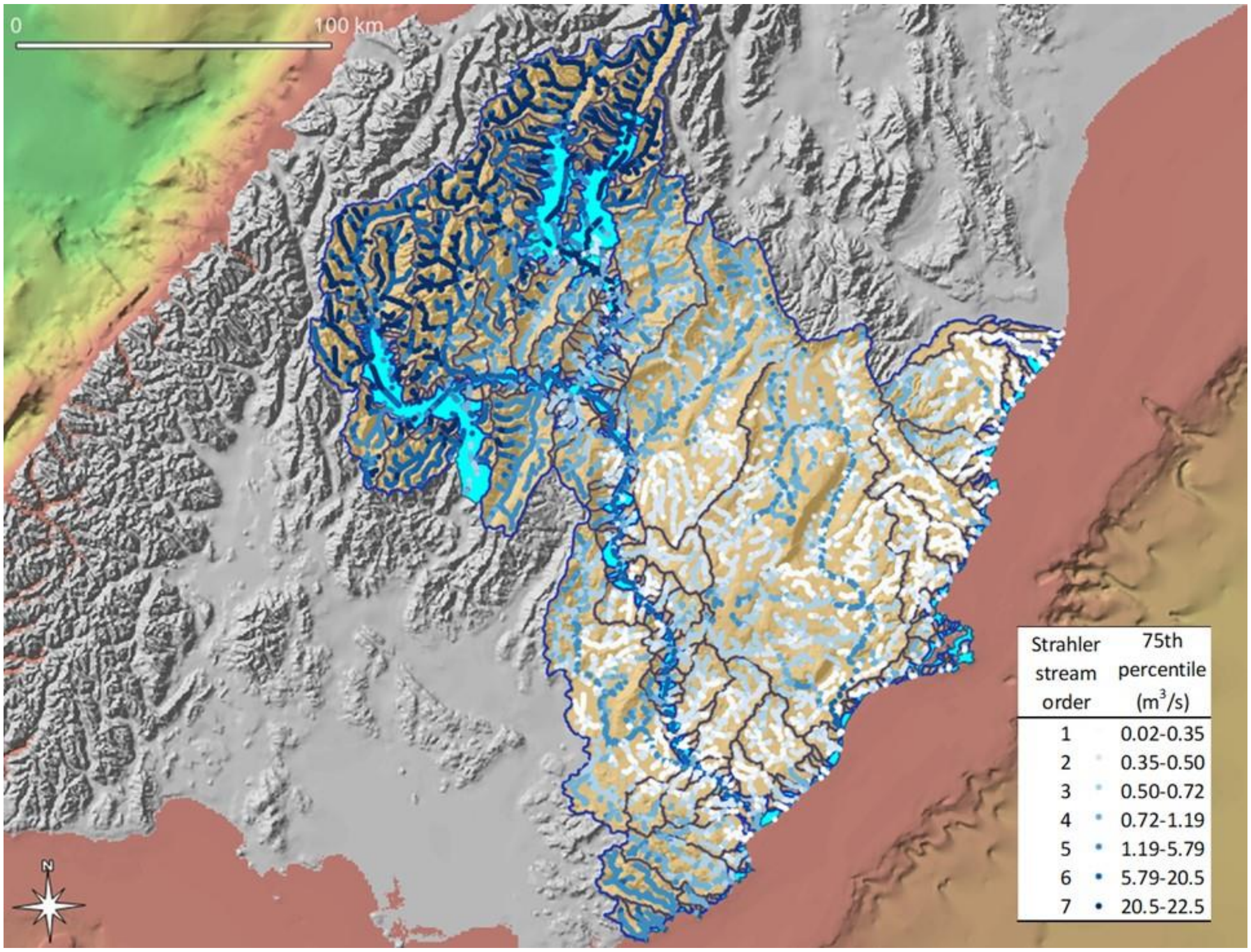

9(b)

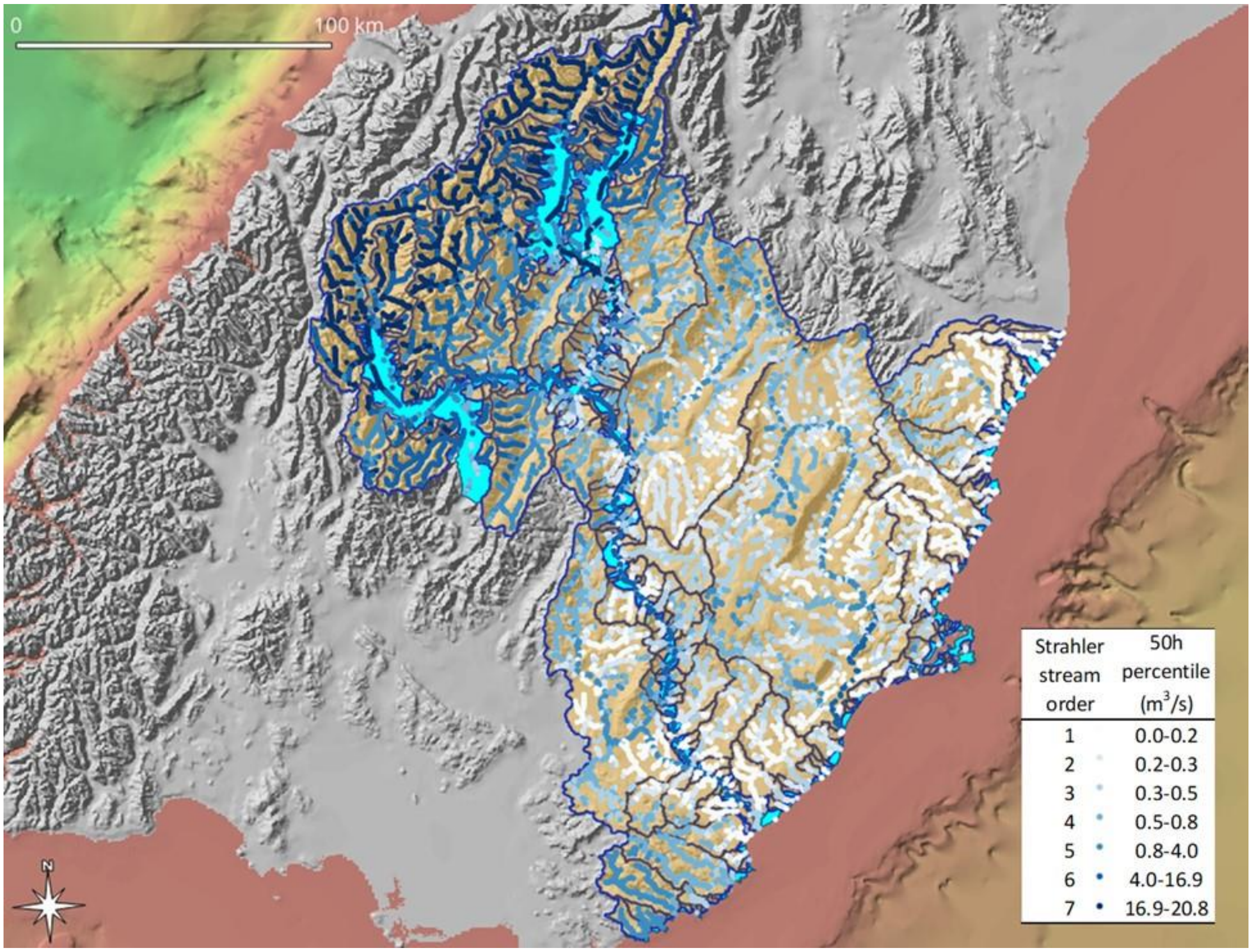

9(c)

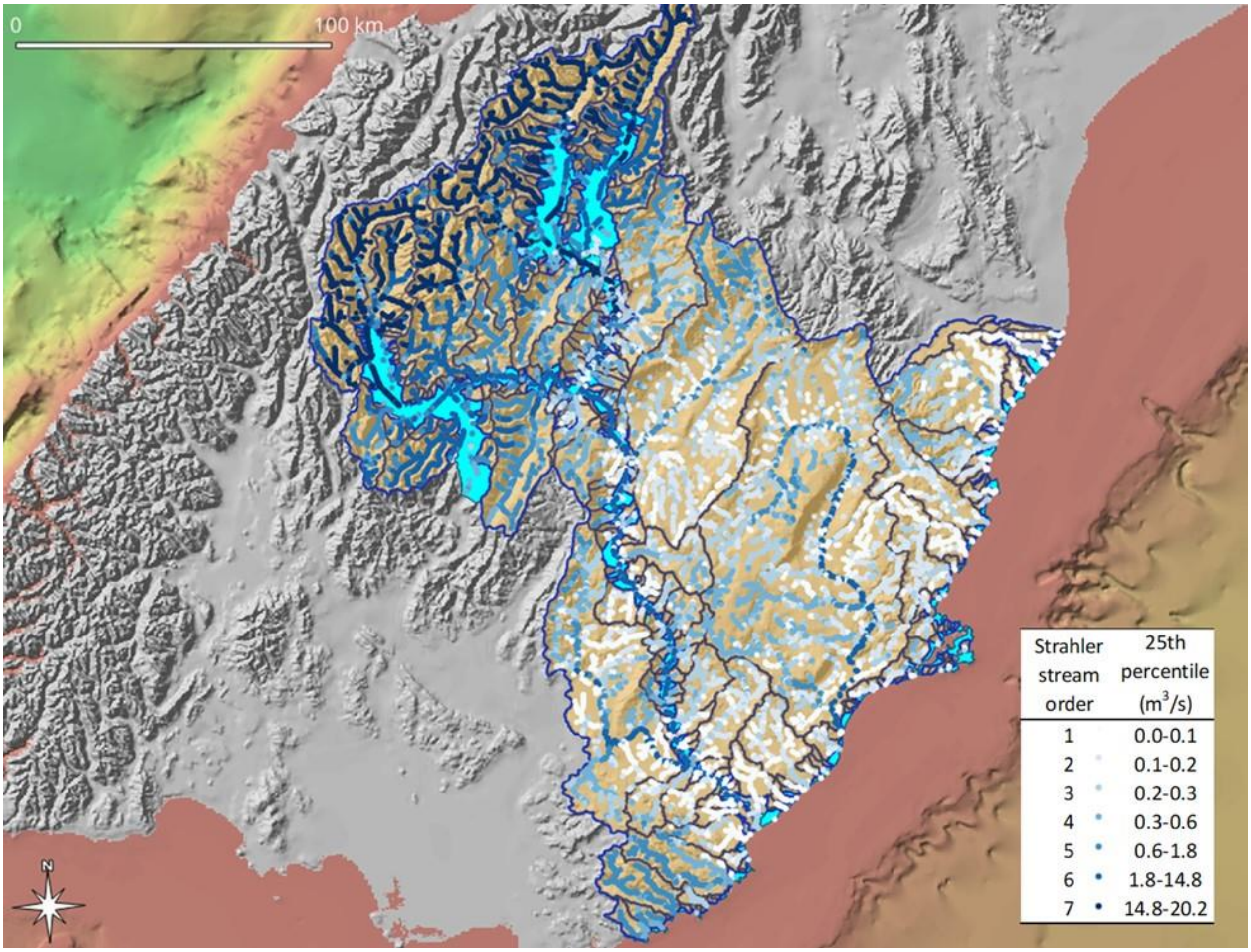

9(d)

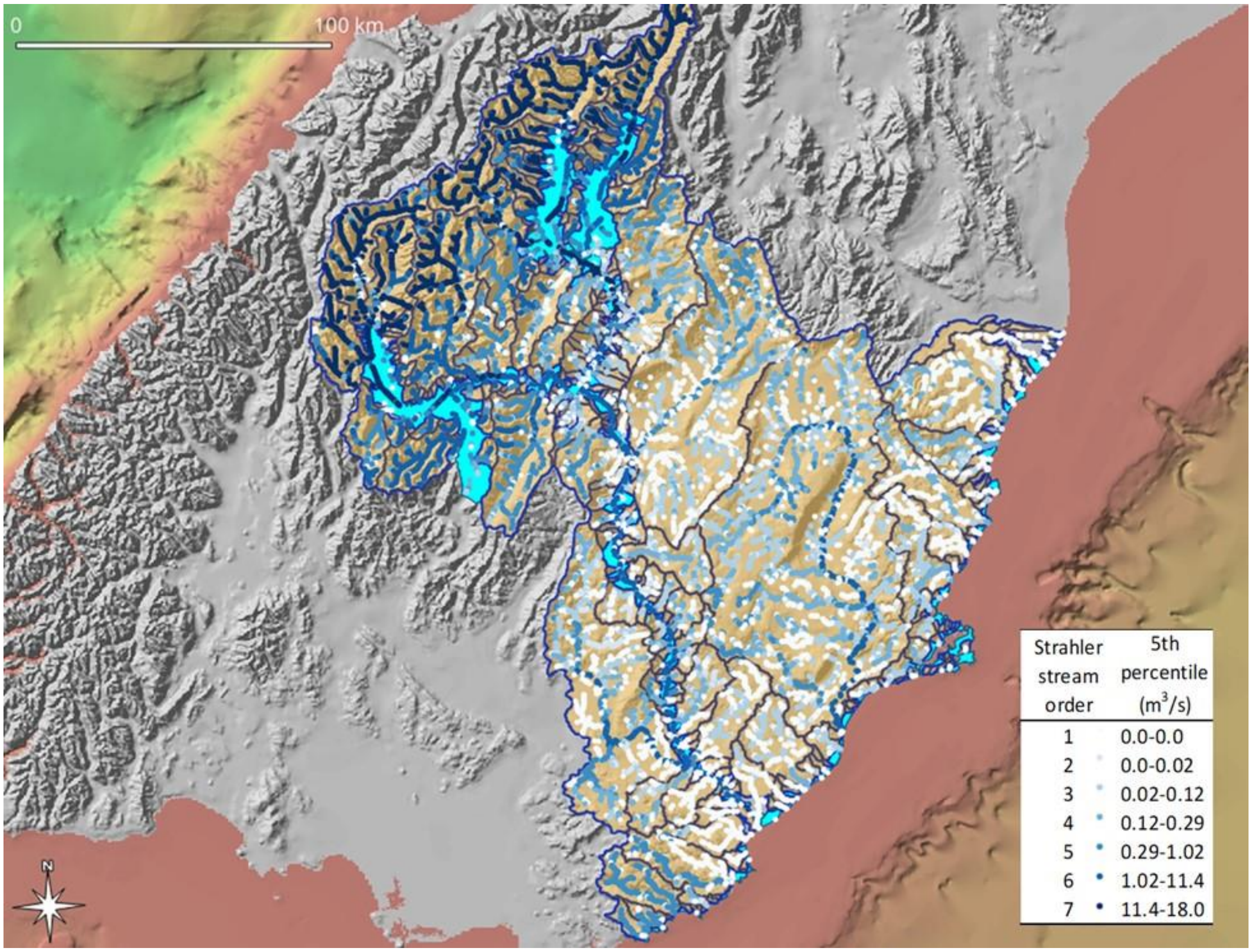

9(e)

Fig. 9. Probable naturalized 7-day mean annual low flow (MALF) predictions at 18,612 ungauged sites across the Otago Region: (a) 95th percentile, (b) 75th percentile, (c) 50th percentile, (d) 25th percentile, and (e) 5th percentile.

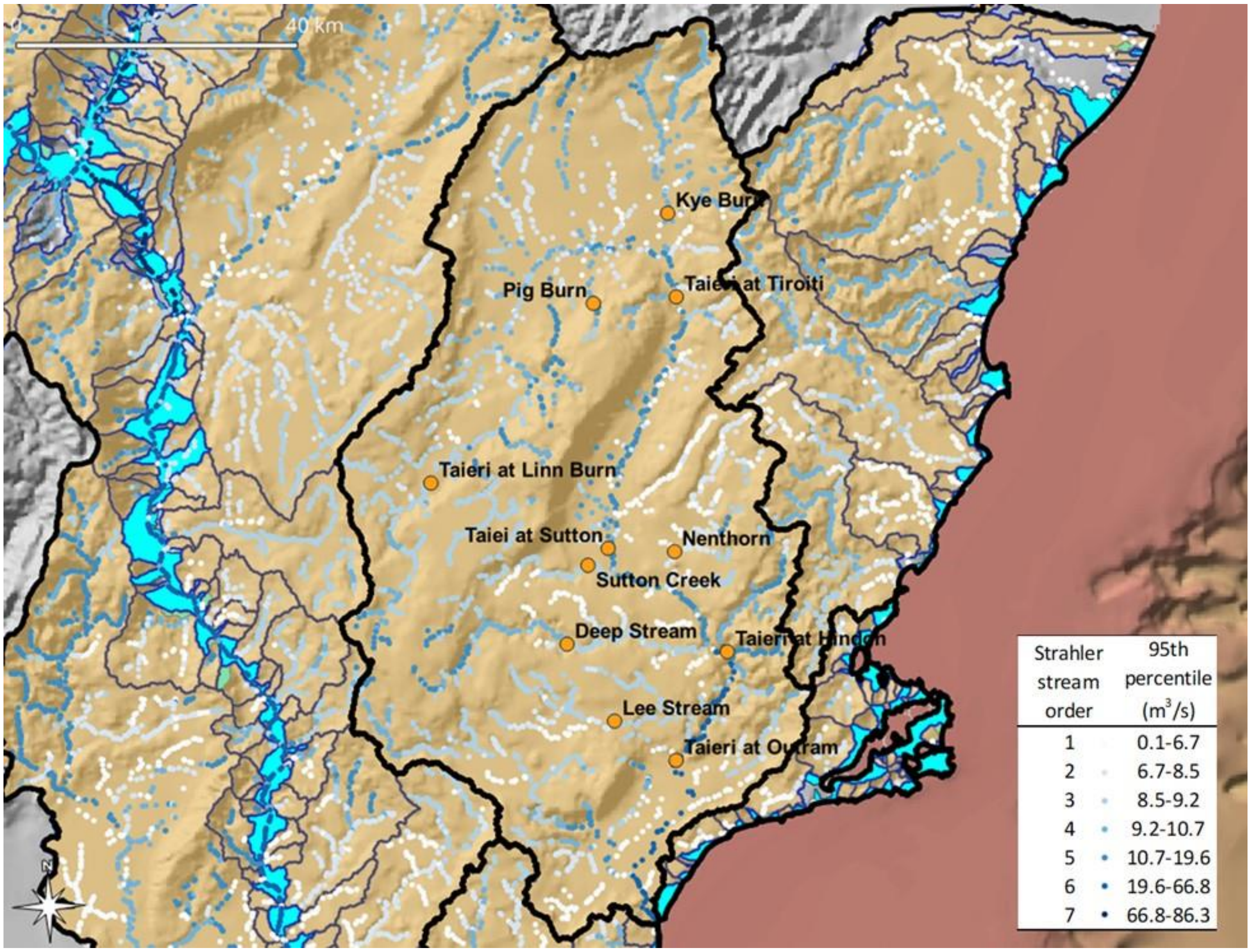

10(a)

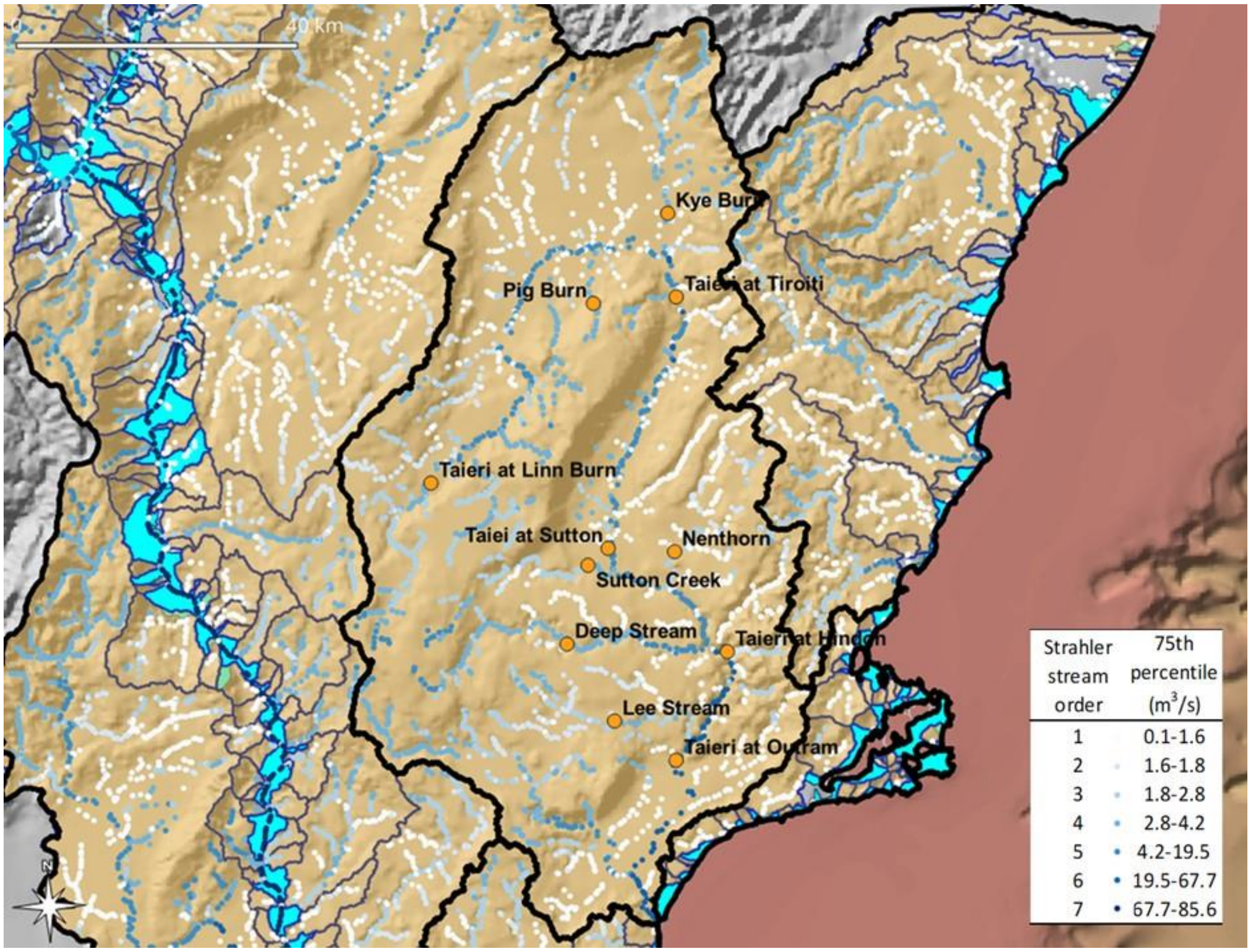

10(b)

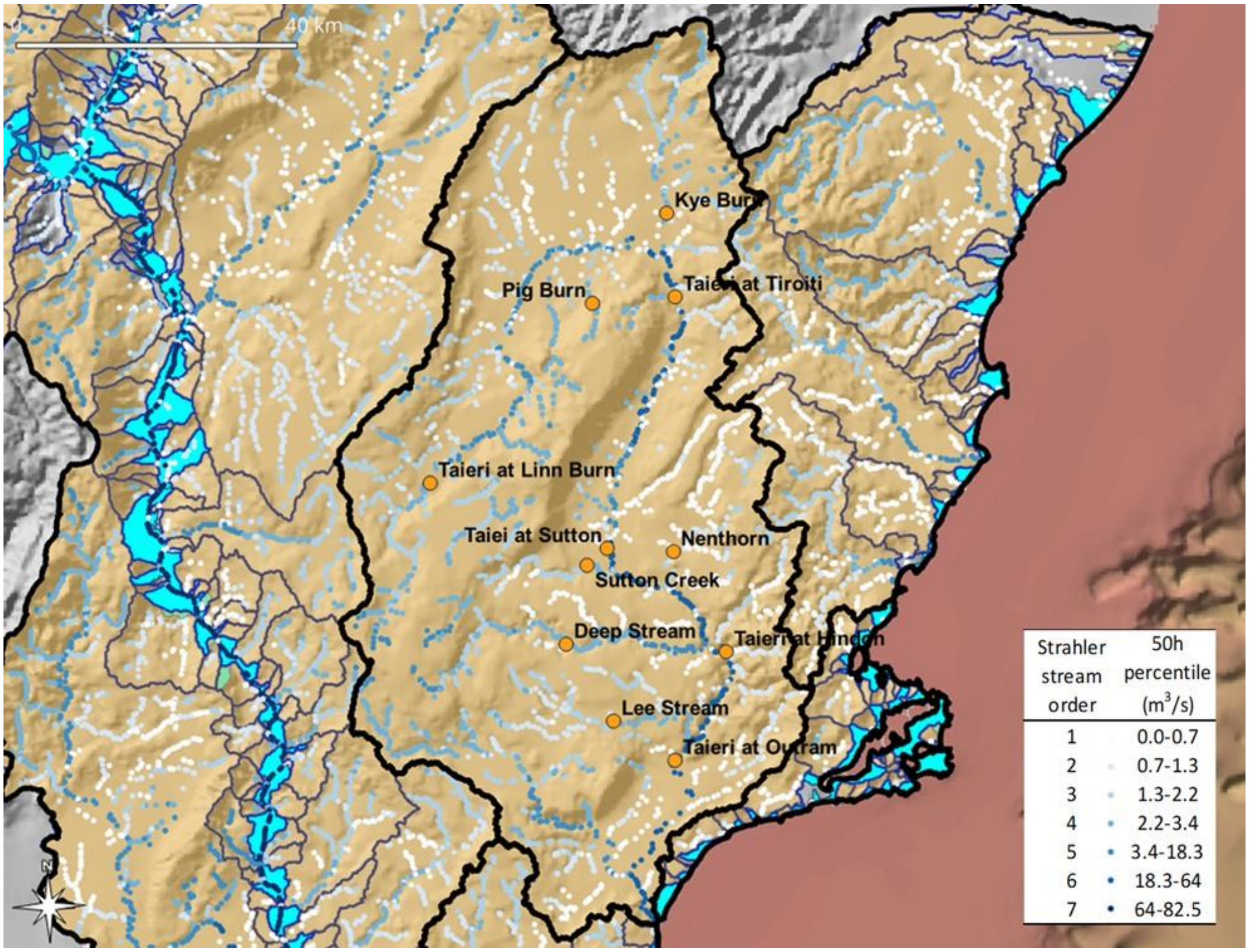

10(c)

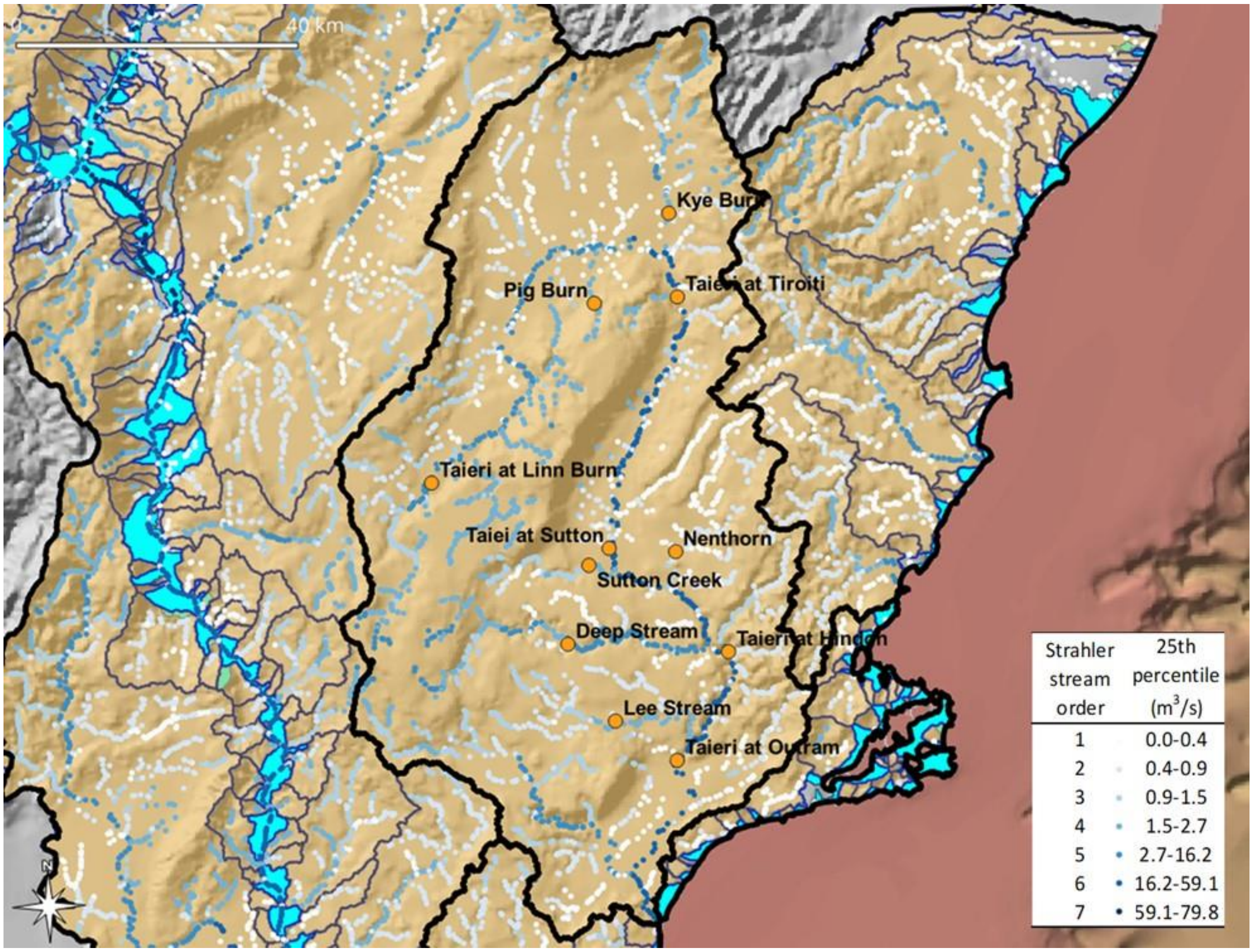

10(d)

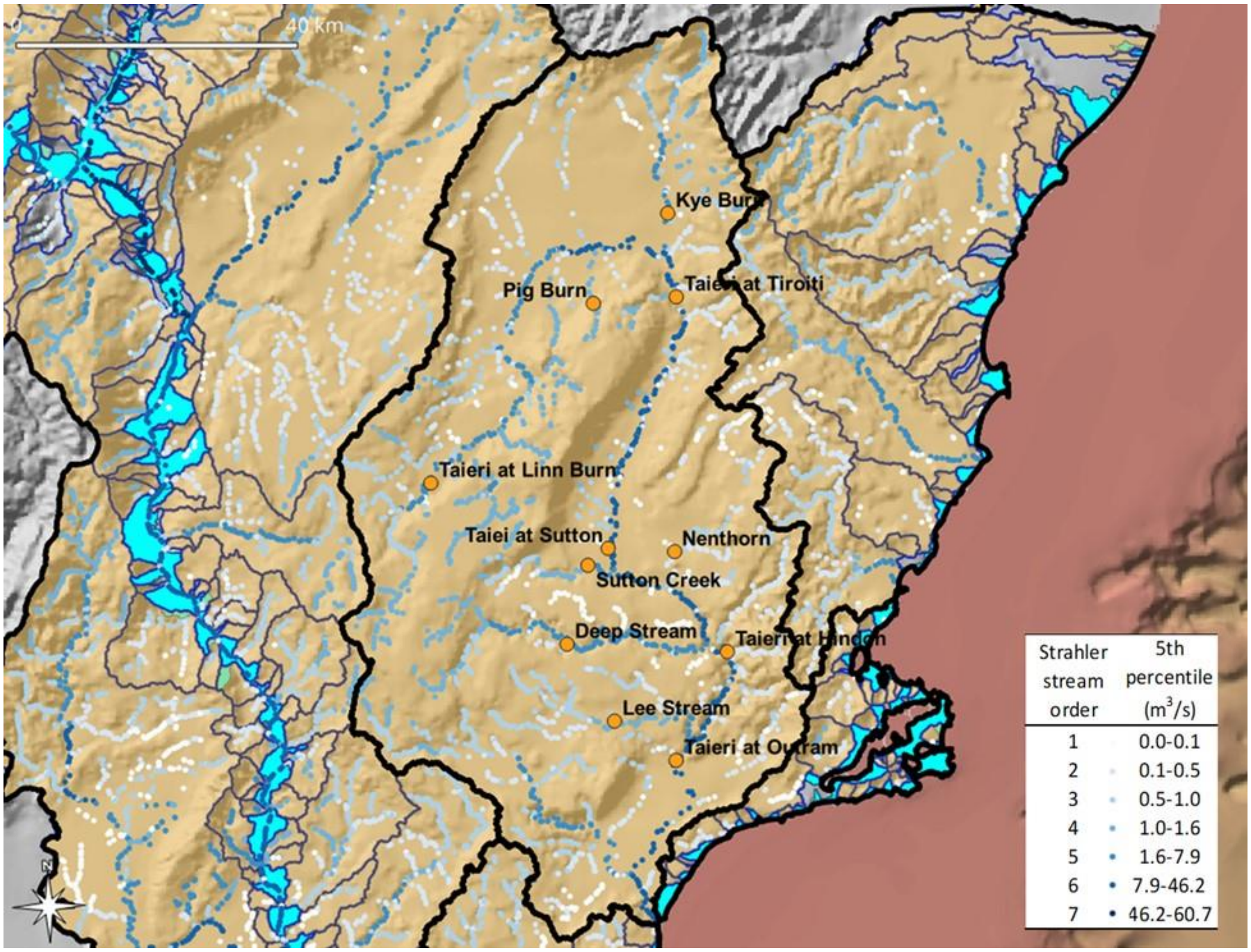

10(e)

Fig. 10. Probable naturalized mean daily flow (Mean) predictions across the Taieri freshwater management unit: (a) 95$^{th}$ percentile, (b) 75th percentile, (c) 50th percentile, (d) 25th percentile, and (e) 5$^{th}$ percentile. The natural flow statistics are extracted at 11 Taieri surface water stations: Taieri at Outram, Taieri at Hindon, Taieri at Sutton, Taieri at Tiroiti, Taieri at Linn Burn, Kye Burn, Pig Burn, Sutton Creek, Deep Stream, Lee Stream, and Nenthorn (see table 14).

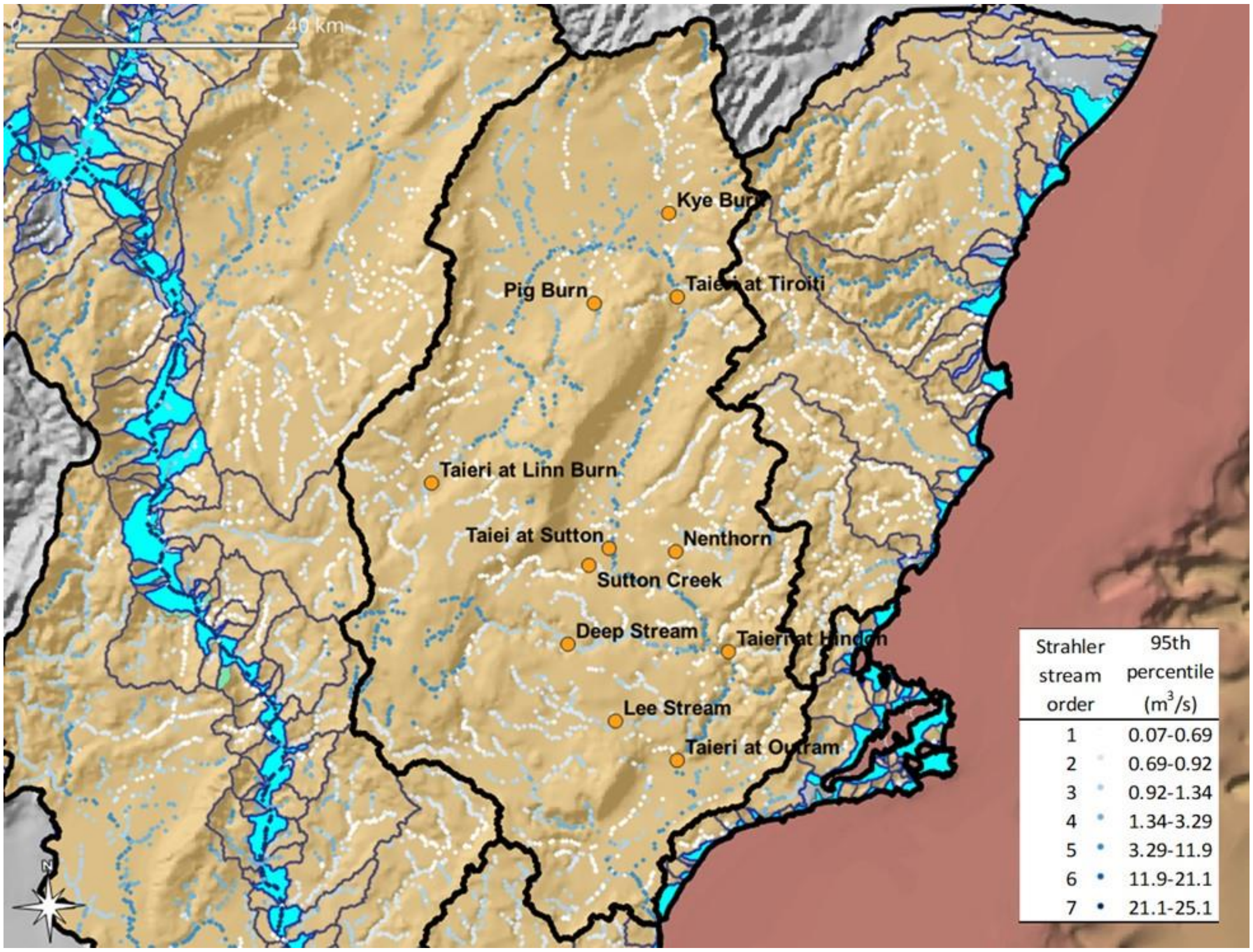

11(a)

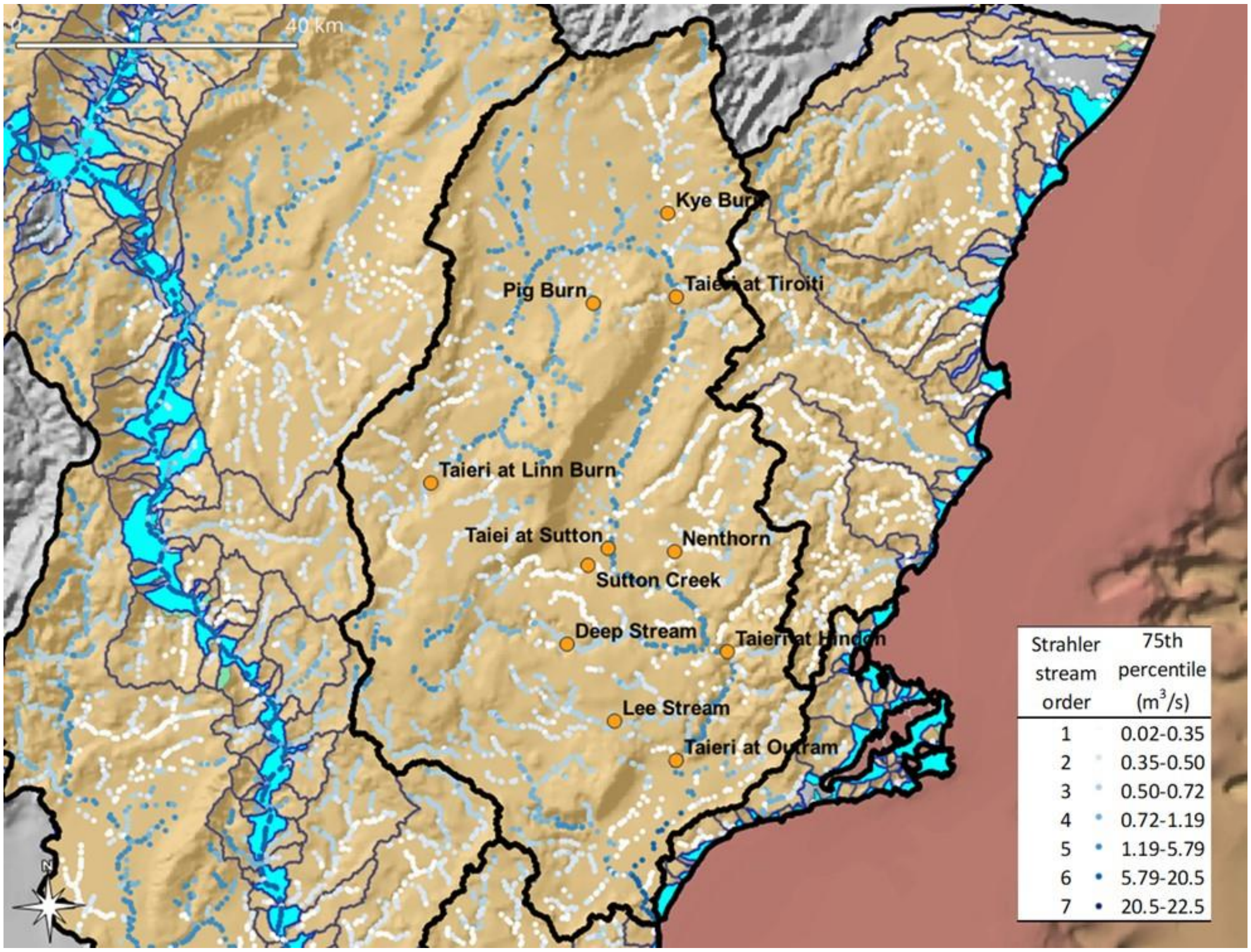

11(b)

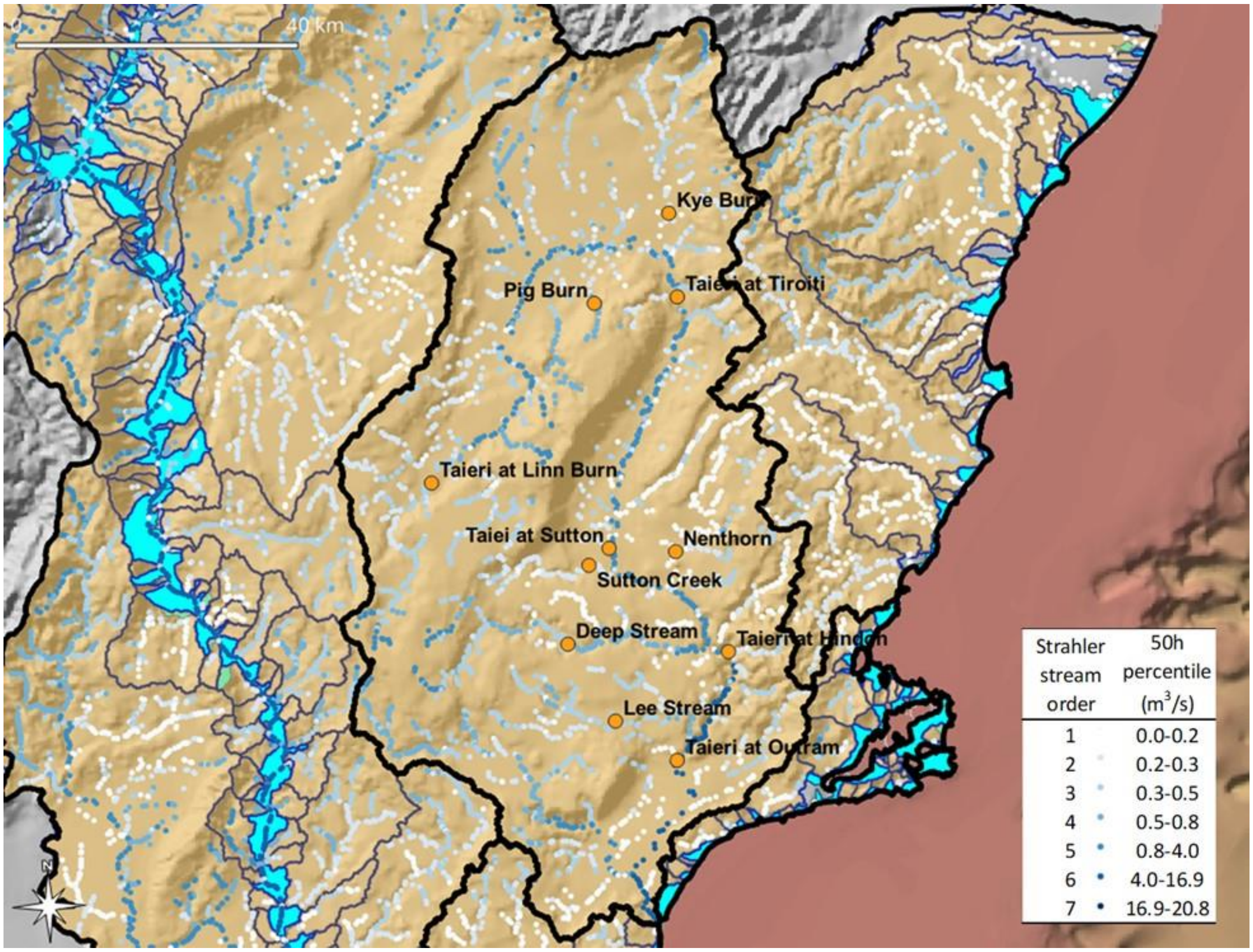

11(c)

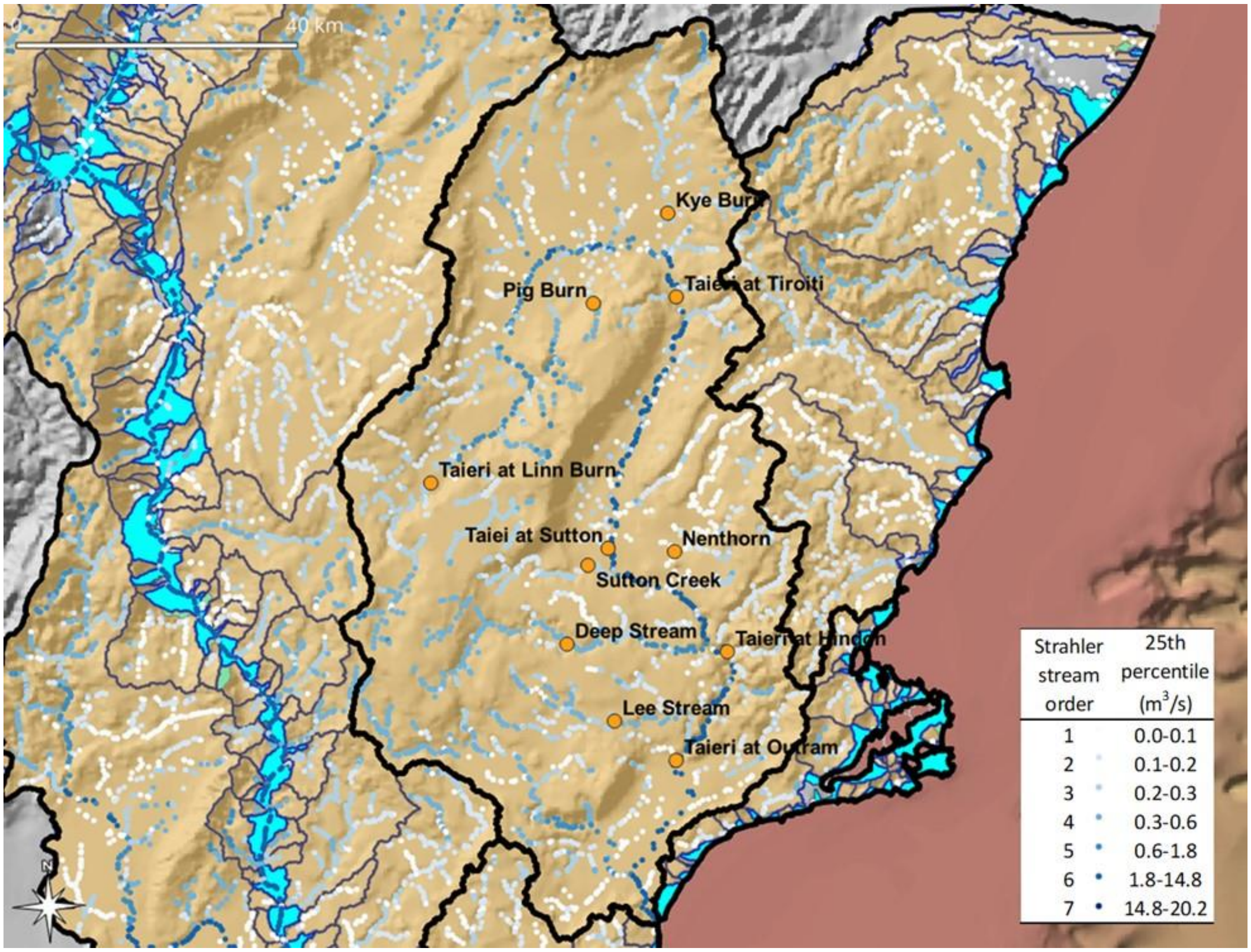

11(d)

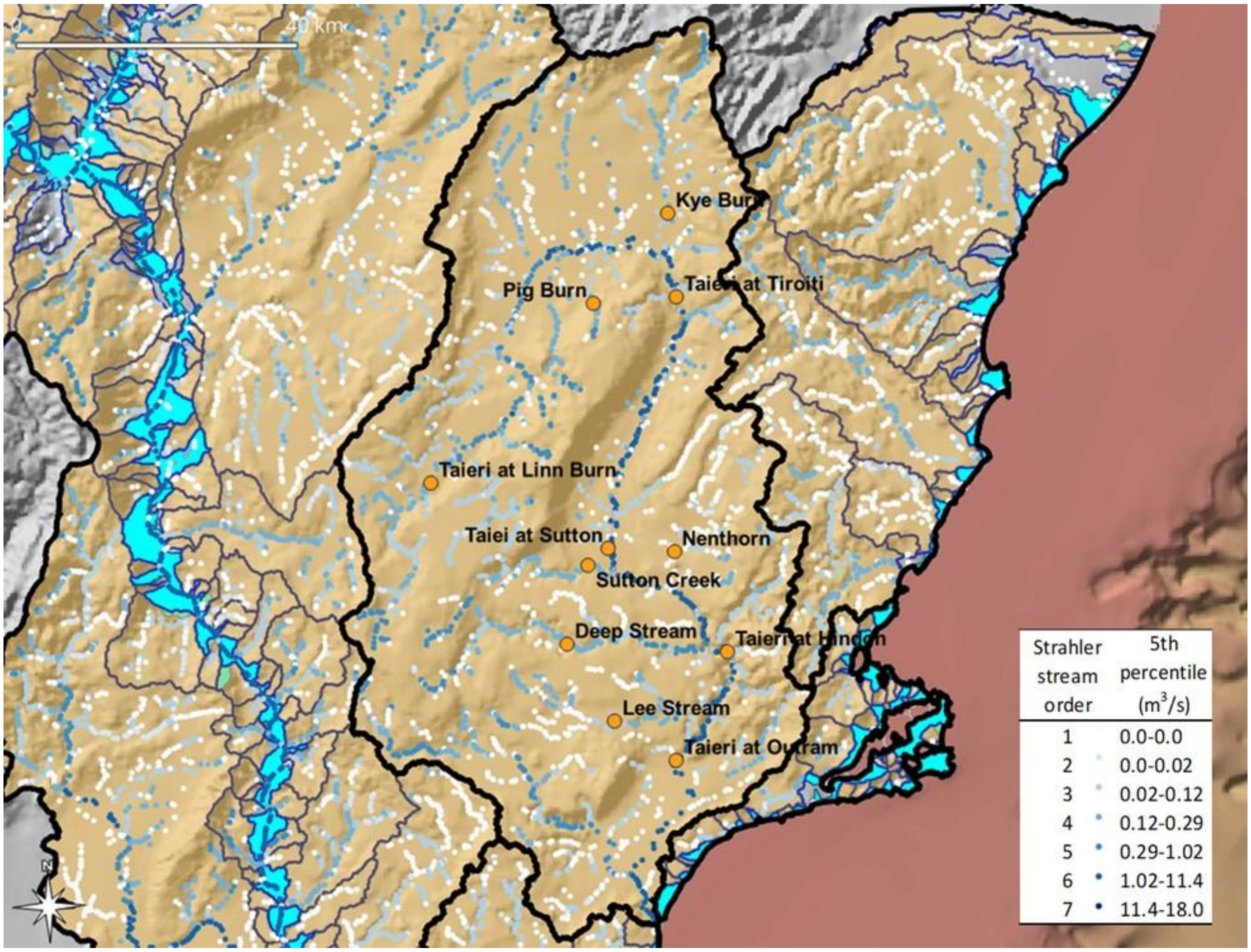

11(e)

Fig. 11. Probable naturalized 7-day mean annual low flow (MALF) predictions across the Taieri freshwater management unit: (a) 25$^{th}$ percentile, (b) 50th percentile, (c) 75th percentile. The natural flow statistics are extracted at 11 Taieri surface water stations: Taieri at Outram, Taieri at Hindon, Taieri at Sutton, Taieri at Tiroiti, Taieri at Linn Burn, Kye Burn, Pig Burn, Sutton Creek, Deep Stream, Lee Stream, and Nenthorn (see table 14).

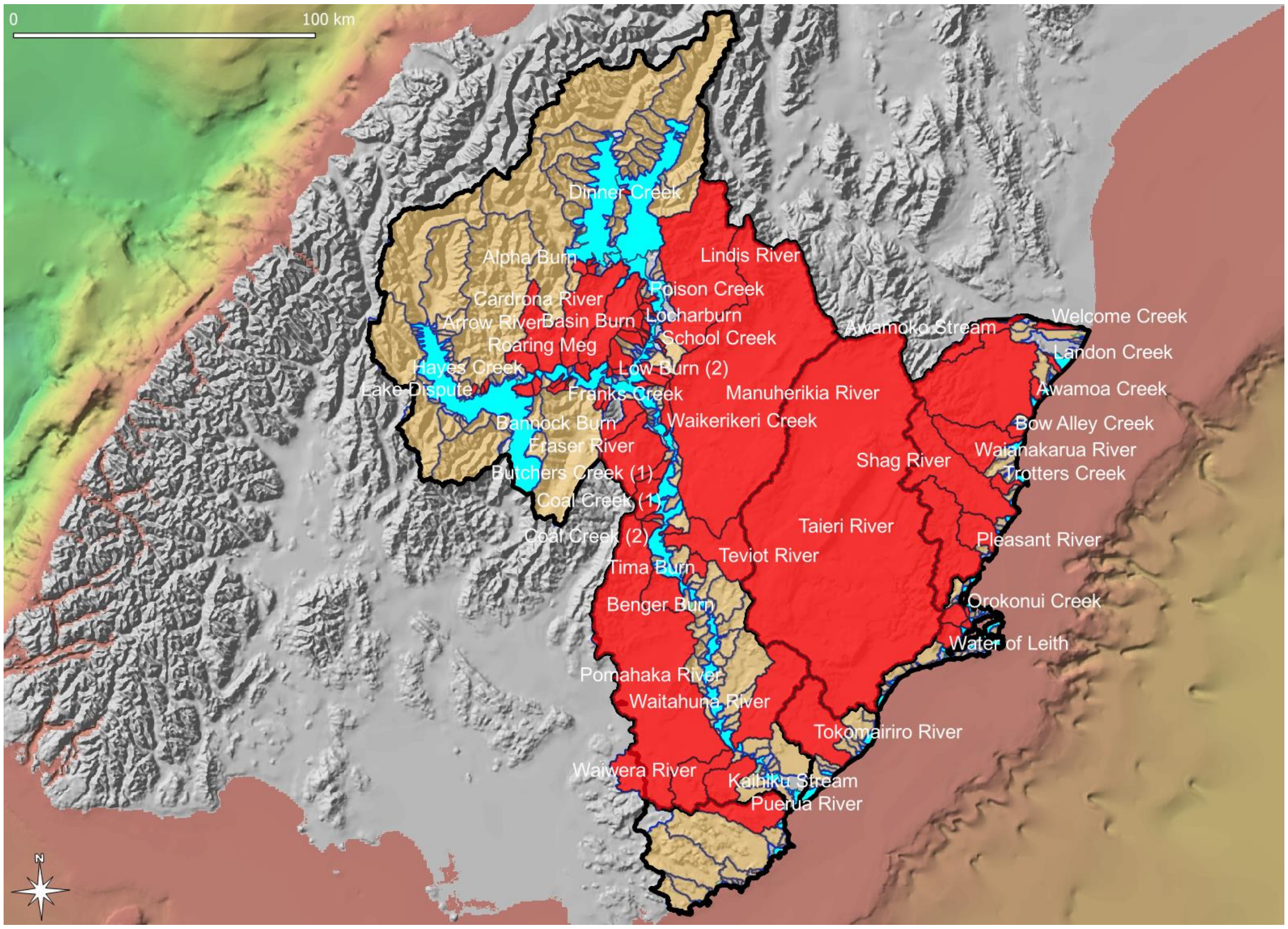

12(a)

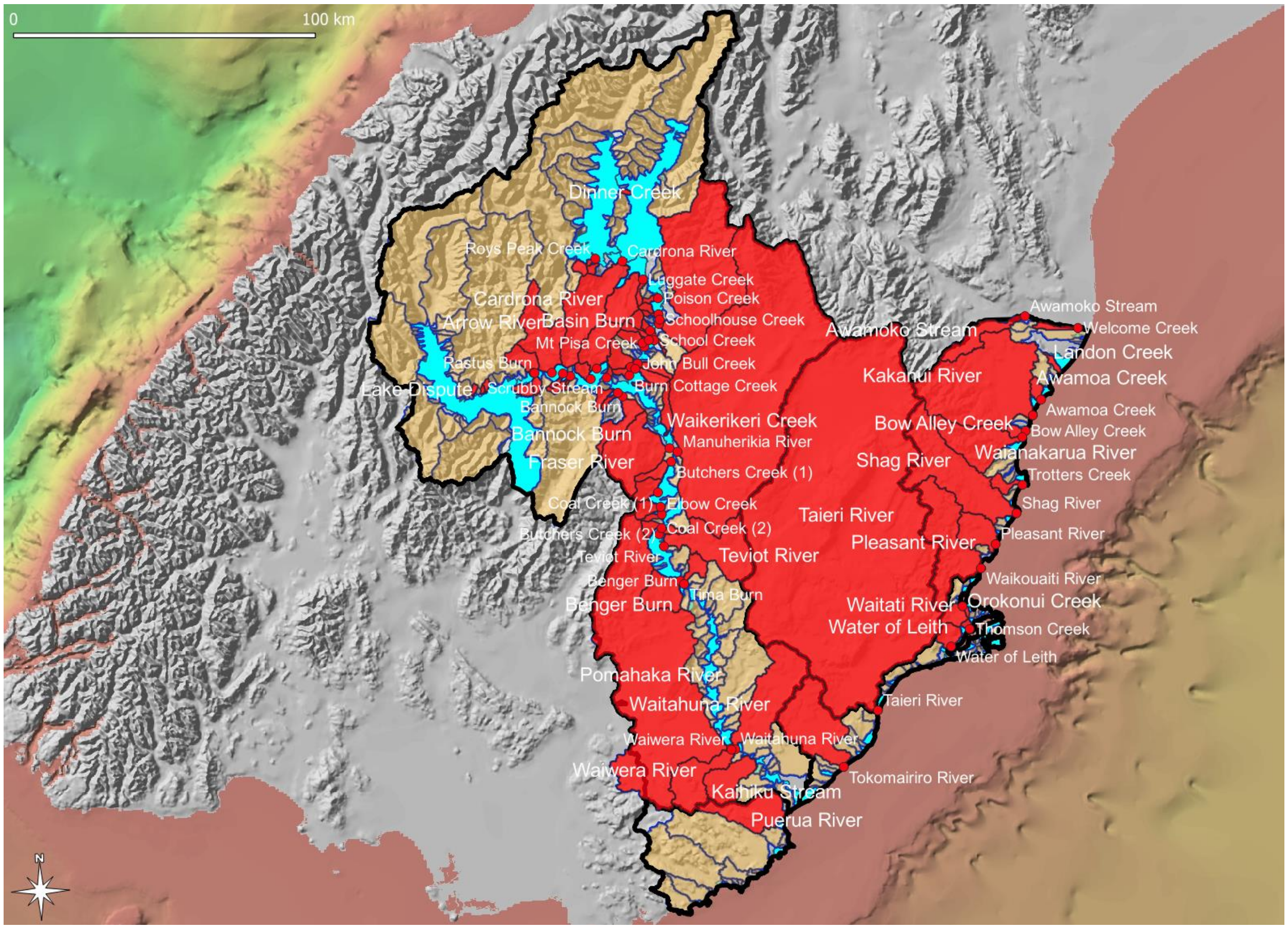

12(b)

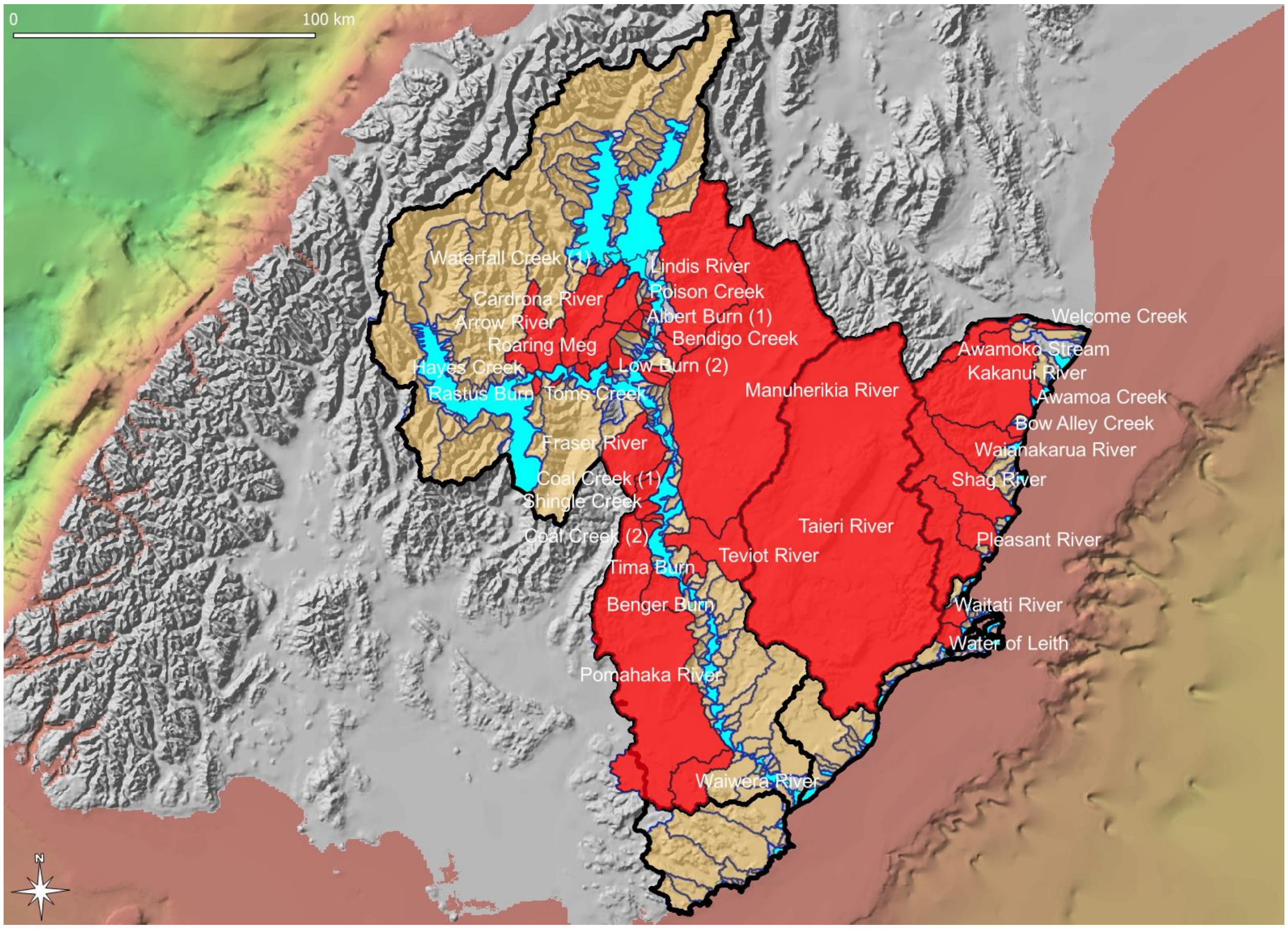
12(c)

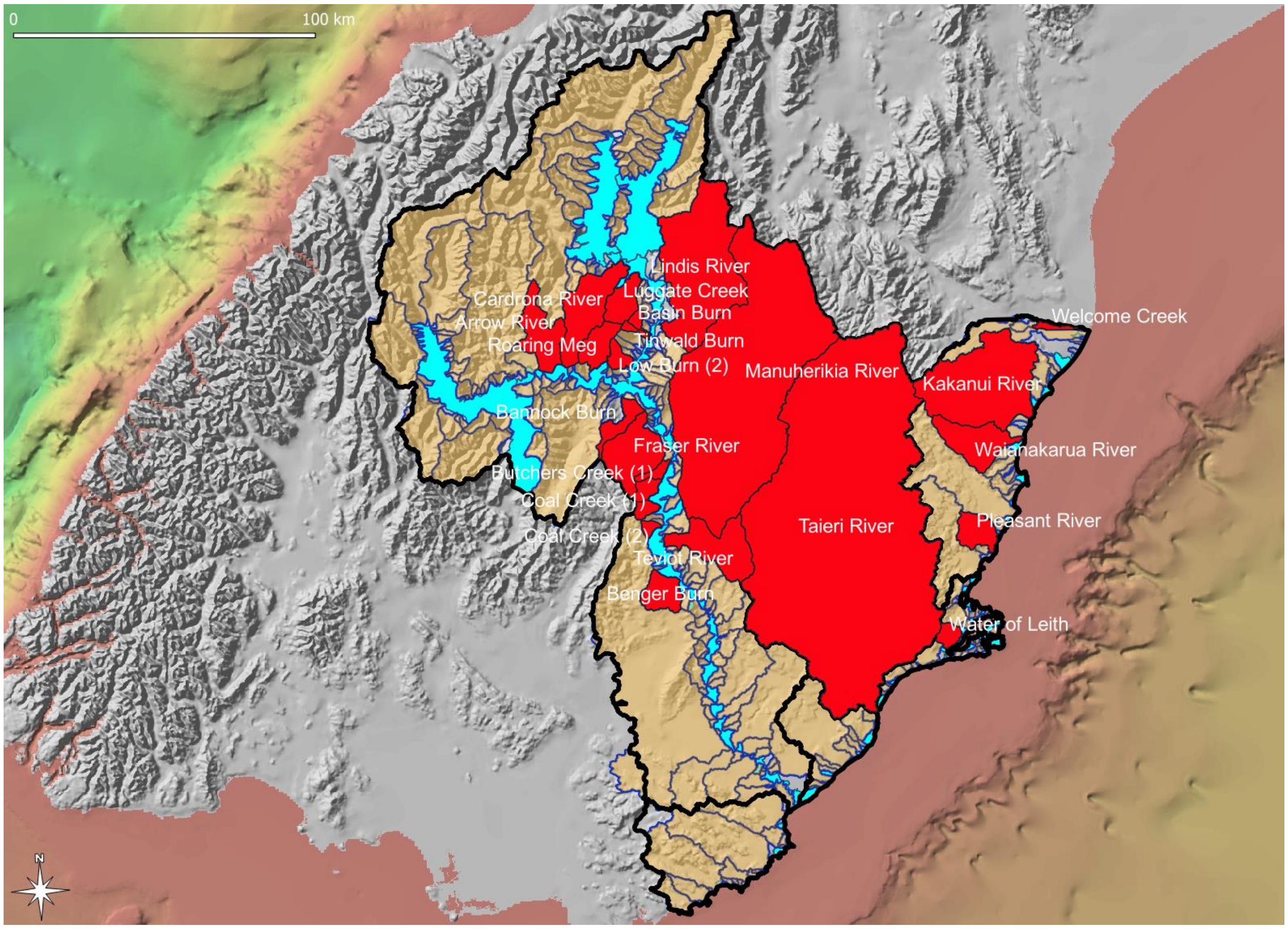
12(d)

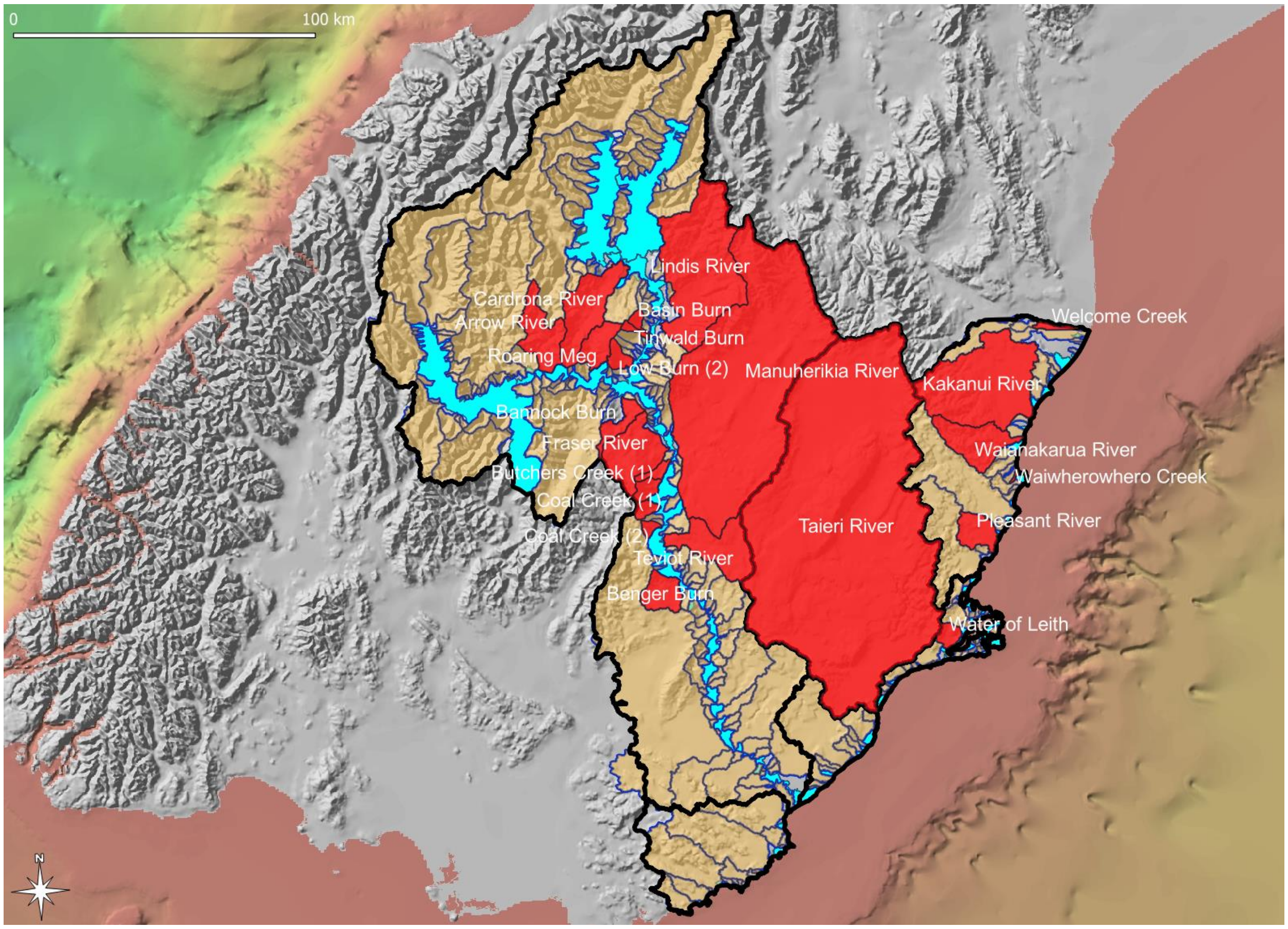

(e)

Fig. 12 Probable catchment status across the Otago Region: (a) 73 over-allocated catchments at the 5th percentile, (b) 57 over-allocated catchments at the 25th percentile, (c) 44 over-allocated catchments at the 50th percentile, (d) 23 over-allocated catchments at the 75th percentile, and (e) 22 over-allocated catchments at the 95th percentile. Over-allocated catchments are shown in red with names in white text.